\documentclass[aps,pra,twocolumn,groupedaddress,nofootinbib,notitlepage,showpacs,floatfix,10pt]{revtex4-1}

\usepackage{graphicx,graphics,epsfig,times,bm,bbm,amssymb,amsmath,amsfonts,mathrsfs,color,subfigure,hypernat}
\usepackage{braket}
\usepackage[pdfstartview=FitH]{hyperref}
\hypersetup{
    colorlinks=true,       
    linkcolor=red,          
    citecolor=magenta,        
    filecolor=magenta,      
    urlcolor=cyan,           
    runcolor=cyan
}

\usepackage{multirow}
\usepackage{rotating}

\def\s0{I}
\def\sx{\sigma^x}
\def\sy{\sigma^y}
\def\sz{\sigma^z}
\def\G{\mathcal{G}}
\def\eps{\epsilon}
\def\H{\bar{H}}
\def\lp{\left(}
\def\rp{\right)}
\def\Xb{\bar{X}}
\def\Yb{\bar{Y}}
\def\Zb{\bar{Z}}
\def\Pb{\bar{P}}
\def\O{\mathcal{O}}

\newcommand{\blue}[1]{\textcolor{blue}{#1}}
\newcommand{\bb}[1]{\boxed{#1}}
\newcommand{\ignore}[1]{}

\newcommand{\bes} {\begin{subequations}}
\newcommand{\ees} {\end{subequations}}
\newcommand{\bea} {\begin{eqnarray}}
\newcommand{\eea} {\end{eqnarray}}
\newcommand{\beq} {\begin{equation}}
\newcommand{\eeq} {\end{equation}}
\newcommand{\ba} {\begin{align}}
\newcommand{\ea} {\end{align}}

\begin{document}

\title{Optimized Dynamical Decoupling via Genetic Algorithms}
\author{Gregory Quiroz}
\affiliation{Department of Physics and Center for Quantum Information Science \&
Technology, University of Southern California, Los Angeles, California
90089, USA}
\author{Daniel A. Lidar}
\affiliation{Departments of Electrical Engineering, Chemistry, and Physics, and Center
for Quantum Information Science \& Technology, University of Southern
California, Los Angeles, California 90089, USA}
\begin{abstract}
We utilize genetic algorithms aided by simulated annealing to find optimal dynamical decoupling (DD) sequences for a single-qubit system subjected to a general decoherence model under a variety of control pulse conditions. We focus on the case of 
sequences with equal pulse-intervals and perform the optimization with respect to pulse type and order.
In this manner we obtain robust DD sequences, first in the limit of ideal pulses, 
then when including pulse imperfections such as finite pulse duration and qubit rotation (flip-angle) errors. Although our optimization is numerical, we identify
a deterministic structure that underlies the top-performing sequences.
We
use this structure to devise DD sequences which outperform previously designed concatenated DD (CDD) and quadratic DD (QDD) sequences in the presence of pulse errors. We explain our findings using time-dependent perturbation theory and provide a detailed scaling analysis of the optimal sequences.
\end{abstract}
\maketitle

\section{Introduction}
Quantum information processing (QIP) relies on the ability to implement high-fidelity quantum gate operations and successfully preserve quantum state coherence \cite{Nielsen:book}. One of the most challenging obstacles for reliable QIP is overcoming the inevitable interaction between a quantum system and its environment or bath. Unwanted interactions result in decoherence processes that 
cause quantum states to deviate from a desired evolution, consequently leading to computational errors and 
loss of 
coherence. In order for QIP to be realizable in the setting of 
open quantum systems, it is necessary to address the detrimental effects of decoherence.

Dynamical decoupling (DD) is one such method, which seeks to attenuate the effects of decoherence by applying 
strong and expeditious control pulses solely to the system \cite{ViolaKnillDD:98,ViolaKnillLloyd:99,Zanardi:1999:77,VitaliTombesi:99,Duan:1999:139,Byrd:2002:19}. Provided the pulses are applied over a time duration sufficiently shorter than the correlation time associated with the environment dynamics, DD effectively averages out undesirable interactions and preserves quantum states with a low probability of error, or fidelity loss. 
One advantage of DD
over quantum error correction (QEC) is that it is an open-loop technique, i.e., does not require feedback or measurement. Furthermore, DD 
has been widely experimentally studied in a number of systems, including ion traps \cite{Szwer-iontrapDD:10,BiercukUDD:09,Bermudez-iontrapDD:11}, nuclear magnetic resonance (NMR) \cite{AlvarezSuterUDD:11,SouzaAlvarezSuter,SouzaAlvarez:12}, solid state quantum dots \cite{Barthel-CDDexperimental:10}, and nitrogen vacancy (NV) centers in diamond \cite{LaraouiNVDD-11,Naydenov-NVDD:11,Wang-NVDD:12}.

The earliest known DD sequence constructions built upon the 
Hahn spin echo effect \cite{Hahn:50}, by applying a sequence of control pulses implementing $\pi$-rotations, each separated by a fixed time duration. 
Two particularly notable sequences 
include CPMG \cite{CPMG:58}, which utilizes cycles of two identical pulses to preserve spin magnetization along a single direction (useful for known quantum state preservation), and XY$_4$ \cite{XY4:69}, a four pulse multi-axis DD sequence that can increase coherence times 
isotropically (useful for unknown quantum state preservation). Both sequences were eventually extended to an ``XY-family" which incorporates 
longer pulse sequences to improve coherence even further and provide robustness against 
errors generated by experimental imperfections in the control pulses \cite{XYseqs:69}.

In the context of open quantum systems of interest to us here, the key feature responsible for
an increase in coherence time for CPMG and XY$_4$ is the suppression of the first order term in time-dependent perturbation theory for a system weakly coupled to a bath and subjected to dephasing or general decoherence, respectively. Concatenated DD (CDD) is a deterministic sequence design
that exploits this property by recursively embedding 
any base DD sequence (e.g., XY$_4$) into itself to successively suppress an additional order of the perturbation expansion at each level of concatenation \cite{KhodjastehLidar:04}. CDD has been extensively studied analytically \cite{KhodjastehLidar-CDD:07,LidarZanardi-DDdist:08,NLP:11}, numerically \cite{CDD-numericalWitzel:07,CDD-numericalZhang:07,CDD-numericalZhang:08}, and experimentally \cite{Barthel-CDDexperimental:10,Peng-CDD,Suter-CDD,WangDobrovitski-PulseErrorsSiDots:10,WangDobrovitski-PulseErrorsDD:10,AlvarezSuter:vCDD:12}, for a variety of 
systems, and its predicted ability to achieve high order suppression has been largely confirmed.

A CDD sequence of order $q$ using a base sequence of $K$ pulses uses $K^q$ equally spaced pulses to suppress the first $q$ orders of the perturbation expansion in the ideal pulses limit for a single qubit system subjected to general decoherence (e.g., $K=4$ for XY$_4$) \cite{KhodjastehLidar:04,KhodjastehLidar-CDD:07}. What distinguishes an XY$_4$-based CDD sequence of order $q$ from, e.g., a periodically repeated XY$_4$ sequence (PDD) with the same total number of pulses, is just the pulse order. Yet, (unsymmetrized) PDD achieves at most first order suppression, in contrast to $q$th order suppression for CDD. This begs the question of whether there exist other sequences that achieve even better performance than CDD, arrived at merely by optimizing the pulse order and type. This is the main question we address in this work, using numerical optimization based on genetic algorithms (GAs), supplemented with simulated annealing. DD optimization fits naturally within the constructs of GAs and has been previously utilized to compare the efficacy of optimal DD to randomized DD schemes in suppressing interactions between neighboring system qubits using selective ideal rotations \cite{randDD}.

Notwithstanding recent progress in pulse sequence optimization using unequal intervals, in particular the UDD sequence and its generalizations \cite{UDD:07,PasiniUhrig:10,YangLiuUDD:08,CUDD:09,Uhrig:2010:012301,WestFongLidar-QDD:10,QuirozLidarQDD:11,LidarKuoQDD:11,WangLiuNUDD:11,JiangImambekov-NUDD:11,Kuo:1207.1665}, we focus our optimization on the case of equal pulse-intervals, as this 
greatly simplifies the optimization, and is furthermore 
often the most convenient experimental situation.
Our first main result is to show that numerically optimal sequences 
comprising ideal, zero-width $\pi$-pulses can be located which exhibit the same error suppression characteristics 
as XY$_4$-based CDD while using $2^{r}$ times fewer pulses, where $r$ is the concatenation level. Furthermore, we identify in these 
sequences 
a deterministic structure which we use to construct additional, robust high-order sequences.


As the techniques of DD sequence construction have become increasingly sophisticated, the issue of robustness to control pulse imperfections has remained 
one of the prominent restrictions of sequence performance in experimental settings. Systematic errors such as rotation-angle or rotation-axis errors, and finite pulse duration errors brought about by
bandwidth constraints,
can generate additional decoherence that quickly destroys the decoupling efficiency of 
all known DD schemes. Robustness to such errors has been 
addressed by the XY-family of sequences and CDD, and in a systematic manner---for pulse-width errors---by Eulerian DD (EDD) \cite{EDD} and its generalization to logic gates, known as dynamically corrected gates (DCG) \cite{DCG,DCG1}. A concatenated version of DCG (CDCG) has been shown to be capable in principle of achieving arbitrarily accurate gates using finite width pulses \cite{CDCG}.
Protocols based on pulse-interval optimization have been shown experimentally to be highly sensitive to pulse imperfections, 
thus forfeiting their ideal-pulse decoupling efficiencies \cite{Xiao-UDDPulseErrors:11}. Certain numerical optimization techniques 
such as locally optimized DD (LODD) \cite{Biercuk-LODD:10}, bandwidth adapted DD (BADD) \cite{BADD:11}, optimized noise filtration DD (OFDD) \cite{ONFDD:09}, and Walsh function DD (WDD) \cite{WDD:11},
exhibit a degree of robustness to finite pulse duration, however, the relationship between sequence performance and 
rotation errors is unclear. 
{A more recent approach for combating pulse imperfections, known as Knill DD (KDD), utilizes a sequence of variable phase $\pi$-pulses separated by fixed pulse-intervals to generate an effective sequence of four $\pi$-pulses centered around a specified axis with an additional overall accumulated phase \cite{SouzaAlvarezSuter-RDD:11}. In contrast to the XY-family and XY$_4$-based CDD, KDD exhibits robustness to finite-width and flip-angle errors; however, this robustness is somewhat limited in the original construction as applying standard concatenation protocols to generate a hierarchy of KDD sequences does not appear to offer further improvement in sequence robustness. One important question which we seek to address here is whether it is possible to obtain a similar robustness to both forms of pulse imperfections utilizing $\pi$-pulses restricted to perpendicular axes, i.e., only X,Y, and Z pulses, while also attaining enhanced robustness with an increasing number of pulses.
}


To account for such errors, our numerical optimization is
extended to include pulse imperfections,
in particular finite pulse width, 
flip-angle errors, or both.
We show that robust DD sequences exist which perform considerably better than {the original} CDD sequence and 
sequences based on unequal pulse-intervals, such as UDD and its variants \cite{UDD:07,PasiniUhrig:10,YangLiuUDD:08,CUDD:09,Uhrig:2010:012301,WestFongLidar-QDD:10,QuirozLidarQDD:11,LidarKuoQDD:11,WangLiuNUDD:11,JiangImambekov-NUDD:11,Kuo:1207.1665}. However, sequence performance eventually saturates {with growing}
sequence length. 
Interestingly, we find that the deterministic structure identified in the ideal pulse limit provides robustness against both forms of pulse errors.

The structure of this paper is as follows. In Section~\ref{sec:2}, we supply background information and formal mathematical specifications for DD effectiveness in the context of 
general open quantum system {dynamics}. The particular error model utilized for this study is then discussed along with the details regarding the control Hamiltonian, which is responsible for the DD control fields. 
The results of our optimal sequence search are then given in Section~\ref{sec:4}, where we identify sequences that obtain a higher degree of DD efficiency than standard CDD sequences under the condition of ideal $\delta$-function pulses, and exhibit robustness to certain forms of pulse imperfections; namely, over and under-rotation errors. In Section~\ref{sec:5}, the search results are supplemented with a comparison between the GA optimal sequences, CDD, and QDD for each pulse specification defined in Sec.~\ref{sec:2}. Optimization proves to be most notably beneficial in the case of pulse imperfections where the GA sequences convey their superiority over known deterministic sequences. In Sec.~\ref{sec:6}, using the results of Sec.~\ref{sec:4} and \ref{sec:5}, we discuss the existence of a concatenation-based deterministic scheme built from optimal sequences that offers robustness to both finite-width and rotation errors. We present our conclusions in Section~\ref{sec:7}. Appendix~\ref{appsec:6qubitsearch} addresses the possibility of variations in our results due to bath specifications. Additional information regarding DD performance scaling and the application of GAs to DD optimization is given in Appendix~\ref{appsec:scalingD} and Appendix~\ref{appsec:algorithm}. Appendix~\ref{appsec:NumericScalingD} discusses how the effective error Hamiltonian is extracted numerically. The appendix is futher supplemented with additional results in Appendix~\ref{appsec:addresults} which support the main analysis presented in the paper.

\section{Background, Problem Setup, and Tools}
\label{sec:2}

In this section, we briefly review the pertinent
basic background and mathematical framework for DD. The general error model is then introduced and its numerical implementation specified.
The DD control Hamiltonian is discussed and 
specified for each of the four pulse conditions employed in this work, differing by the degree and nature of the pulse imperfections. Finally, we introduce 
a distance measure with which DD performance 
is quantified throughout the paper.

\subsection{Dynamical Decoupling}
\label{subsec:DDBackground}
Consider an open quantum system described by the Hamiltonian
\begin{equation}
H(t)= H_0+H_C(t).
\label{eq:genH}
\end{equation}
The time-independent term $H_0$ governs the internal dynamics of the system and environment, while $H_C(t)$ is responsible for the time-dependent DD control fields. The Hamiltonian $H_0$ is resolved further into
\begin{equation}
H_0\equiv H_S\otimes I_B+I_S\otimes H_B+H_{SB},
\label{eq:intH}
\end{equation}
where $H_S$ is the pure system Hamiltonian, $H_B$ is the pure environment Hamiltonian, 
$H_{SB}$ represents the system-environment interaction, and $I_{S(B)}$ is the identity operator on the system (bath).

For brevity, we denote
\begin{equation}
H_{\text{err}}\equiv H_S\otimes I_B+H_{SB}
\label{eq:genHe}
\end{equation}
as the \textit{error} Hamiltonian, where the pure system and system-environment interaction Hamiltonians constitute the sources of undesired system evolution and decoherence, respectively. 
Removal of undesired system evolution is particularly relevant when DD is utilized for high fidelity quantum memory storage \cite{ViolaKnillDD:98,KhodjastehLidar-CDD:07,UDD:07} or in a ``decouple-then-compute" approach to quantum gate construction \cite{NLP:11,ViolaKnillLloydBoundedDD:99,KhodjastehLidar:08}. Storing quantum memory requires initial state preservation, hence the desired system evolution 
is trivial action on the system. Any form of system dynamics present after the DD evolution would 
alter the initial state, resulting in storage errors. In a similar manner, gate errors can be acquired during the application of a nontrivial quantum gate if 
undesired system dynamics remain upon the completion of the gate operation. 
Undesired system evolution must also be removed in alternative DD-protected gate construction strategies, such as ``decouple-while-compute" \cite{ViolaKnillLloydBoundedDD:99,KhodjastehLidar:08,WestLidarFong:10}, in particular to prevent leakage errors when a decoherence-free subspace (DFS) or stabilizer code is used to enable computation while DD pulses are applied  \cite{WuLidar-DFS:02,ByrdLidar:01a}.

We assume that the control Hamiltonian $H_C(t)$ applies the DD 
pulses solely to the system. $H_C(t)$ consequently acts trivially on the pure environment Hamiltonian, $[H_C(t),H_B]=0\,\,\forall\,t$, and nontrivially on 
$H_{\text{err}}$. The manner in which $H_C(t)$ operates on $H_{\text{err}}$ ultimately determines the effectiveness of DD in suppressing the contributions of the error Hamiltonian to the system evolution. Demanding that each pulse operator 
anticommute with at least one term comprising $H_{\text{err}}$, the system is driven in such a way 
that $H_{\text{err}}$ can be effectively averaged out for sufficiently short time durations. 
How short can be elucidated in the interaction (``toggling") picture with respect to $H_C(t)$ \cite{ViolaKnillLloyd:99,ViolaKnillLloydBoundedDD:99,Haeberlen:book,ViolaKnillLloyd:03}, where
\bes
\bea
\tilde{H}_0(t)&=&U^{\dagger}_C(t)H_0 U_C(t)\\
&=&\tilde{H}_{\text{err}}(t)+I_S\otimes H_B
\eea
\ees
with the control unitary
\begin{equation}
U_C(t)=\mathcal{T}\exp\left(-i\int^{t}_0\,H_C(t)dt\right),
\end{equation}
where $\mathcal{T}$ denotes the time ordering operator. The unitary time evolution operator $\tilde{U}_0(t)$ satisfies the Schr\"odinger equation
\begin{equation}
i\frac{\partial}{\partial t}\tilde{U}_0(t)=\tilde{H}_0(t)\tilde{U}_0(t),\quad \tilde{U}_0(0)=I
\label{eq:U0}
\end{equation}
and $\tilde{U}_0(t)=U^{\dagger}_C(t)U(t)$, where $U(t)$ is the time evolution operator generated by Eq.~(\ref{eq:genH}). 

Employing time-dependent perturbation theory (TDPT) via the Magnus expansion \cite{Magnus:54,Blanes:08} to solve Eq.~(\ref{eq:U0}) we can write
\begin{equation}
\tilde{U}_0(\tau_c)=\exp\left(\sum^{\infty}_{n=1}\Omega^{(n)}(\tau_c)\right) ,
\end{equation}
with 
the anti-Hermitian operator $\Omega^{(n)}(\tau_c)$ representing the $n$th term in the Magnus operator expansion after a total DD cycle time $\tau_c$. The leading terms of the expansion are
\begin{eqnarray}
\Omega^{(1)}(\tau_c)&=&-i\int^{\tau_c}_{0}\tilde{H}_0(t_1)\,dt_1,\\
\label{eq:Omega1}
\Omega^{(2)}(\tau_c)&=&-\frac{1}{2}\int^{\tau_c}_0\,dt_1\int^{t_1}_0\,dt_2\left[\tilde{H}_0(t_1),\tilde{H}_0(t_2)\right],
\label{eq:Omega2}
\end{eqnarray}
while the $n$th order Magnus term is constructed recursively as a sum of $(n-1)$-fold commutators. A sufficient condition for convergence of the Magnus expansion is \cite{MoanOteoRos:99}
\begin{equation}
\int^{\tau_c}_0\|\tilde{H}_0(t)\|dt<\pi.
\label{eq:Magnus-cond}
\end{equation}

In agreement with average Hamiltonian theory (AHT) \cite{Haeberlen:book,ErnstBodenhausen-NMR:book}, the time-dependent evolution generated by $\tilde{H}_0(t)$ is 
formally identical to a time-independent evolution generated by the effective Hamiltonian 
\begin{equation}
\H_0=\frac{i}{\tau_c}\sum^{\infty}_{n=1}\Omega^{(n)}(\tau_c)=\H_B(\tau_c)+\H_{\text{err}}(\tau_c).
\label{eq:Heff}
\end{equation}
AHT 
applies here since we are only interested in the joint system-environment dynamics at the end of each stroboscopic DD period of $\tau_c$. 
We can further partition $\H_0$ into an effective pure environment term and a sum of effective error Hamiltonians
\begin{equation}
\H_{\text{err}}(\tau_c)\equiv\frac{i}{\tau_c}\sum^{\infty}_{n=1}\Omega^{(n)}_{\text{err}}(\tau_c),
\label{eq:H_err-expansion}
\end{equation}
where $\Omega^{(n)}_{\text{err}}(\tau_c)$ is the $n$th order Magnus operator containing non-trivial system operators, while $\H_B(\tau_c)$ contains only terms with trivial action on the system.

In light of 
Eq.~\eqref{eq:H_err-expansion} we find that DD facilitates an effective suppression of $H_{\text{err}}$ by suppressing $\H_{\text{err}}(\tau_c)$ up to some order in the Magnus expansion.
{When 
the first $N$ terms of the expansion of $\H_{\text{err}}(\tau_c)$ vanish} we speak of ``$N$th order decoupling." Assuming $N$th order {decoupling has been achieved}, the toggling frame evolution is given by
\bes
\begin{eqnarray}
\tilde{U}_0(\tau_c)&=&
e^{-i\tau_c [H_B(\tau_c)+ H_{\text{err}}(\tau_c)]}
\\
									 &=&e^{-i\tau_c \H_B(\tau_c)+\mathcal{O}[(\|\H'_{\text{err}}\|\tau_c)^{N+1}]},
\label{eq:Ueff}
\end{eqnarray}
\ees
where the evolution is predominately dictated by the effective pure environment Hamiltonian $\H_B(\tau_c)$ when 
$\|\bar{H}'_{\text{err}}\|\tau_c\ll \pi$
and $N\gg1$. Here, $\H'_{\text{err}}=\frac{i}{\tau_c}\sum^{\infty}_{n=N+1}\Omega^{(n)}_{\text{err}}(\tau_c)$ denotes 
the remaining effective error Hamiltonian and $\|A\|$ is the sup-operator norm of $A$ (largest singular value):
\begin{equation}
\|A\|=\sup_{\ket{\psi}}\frac{\|A\ket{\psi}\|}{\|\ket{\psi}\|}.
\end{equation}

Thus the effectiveness of DD is dependent upon intrinsic properties, 
in particular the strength of the interaction and pure environment Hamiltonians.
In situations where the internal dynamics are sufficiently fast, a short DD cycle time is desirable to maintain Eq.~\eqref{eq:Magnus-cond}. 
Furthermore, it is also desirable to achieve high order error suppression, $N\gg1$, so that the effects of $\H'_{\text{err}}$ are less consequential. We make use of both conditions to analyze DD in the presence of various strengths of internal dynamics, while determining the minimum number of control pulses required to obtain a given order of error suppression.

As we shall see when we present the analysis of optimal sequences, starting in Sec.~\ref{sec:4}, the effective error Hamiltonian~\eqref{eq:H_err-expansion} has significant explanatory power.

\subsection{Error Model}
\label{sec:errorModel}
The system of interest is a single-qubit system generically coupled to its environment. The internal dynamics are governed by 
\begin{equation}
H_0=\sum_{\mu\in\{I,x,y,z\}}\sigma^{\mu}\otimes B_\mu,
\label{eq:qubitHe}
\end{equation}
where $\sigma^{\mu}$ and $B_{\mu}$ are the spin-1/2 Pauli matrices and general bounded environment operators, respectively. Selecting a four-qubit spin bath to model the environment for the numerical search, the environment operators are given by
\begin{equation}
B_{\mu}=\sum_{i\neq j}\sum_{\alpha,\beta}c^{\mu}_{\alpha\beta}\left(\sigma^{\alpha}_i\otimes\sigma^{\beta}_j\right),
\label{eq:bathops}
\end{equation}
where $i,j$ index the bath qubits, $\alpha,\beta,\mu\in\{I,x,y,z\}$, and $c^{\mu}_{\alpha\beta}\in[0,1]$ are random coefficients chosen from a uniform probability distribution. The construction of $B_{\mu}$ permits at most two-body interactions between the environment qubits and three-body interactions between the system and environment. Note that Eq.~(\ref{eq:bathops}) contains terms proportional to the identity operator $\sigma^{I}_i\otimes \sigma^{I}_j$, which account for the pure system Hamiltonian described in Eqs.~(\ref{eq:intH}) and (\ref{eq:genHe}). 

Together, Eqs.~(\ref{eq:qubitHe}) and (\ref{eq:bathops}) encompass a wide range of experimentally relevant systems which suffer from system-environment interactions ranging from dephasing, longitudinal relaxation, or, more generally, the hyperfine interaction. Such systems, e.g., those outlined in the introduction: NMR , solid state quantum dots, etc., obviously contain a substantially larger number of bath spins than what we are considering in this study. The fact that we are able to validate our results using only four spins is based on the following: (1) increasing the number of bath spins does not appear to directly effect the convergence of the algorithm, as confirmed in \ref{appsec:6qubitsearch} for six bath spins, and (2) the error suppression properties of DD are well-characterized by the scaling of a relevant performance measure with the norm of the system-environment interaction and pure bath dynamics \cite{ViolaKnillDD:98,KhodjastehLidar:04}, both of which we specify prior to the search and vary over a wide range of values.

\subsection{Control Hamiltonian}
The general form of a single qubit control Hamiltonian is 
\begin{equation}
H_C(t)=\frac{1}{2}\sum_{\mu\in\{x,y,z\}}V_{\mu}(t)\sigma^{\mu},
\label{eq:H_C}
\end{equation}
where $V_{\mu}(t)$ is the control field {associated with} the $\sigma^{\mu}$ degree of freedom. All of the essential information regarding the DD sequence is contained within $V_{\mu}(t)$, i.e., pulse timings and amplitude profiles. In general, varying either quantity can result {in} drastically different optimal sequence constructions. Considering 
equal pulse-interval delay times of $\tau_d$ throughout the DD evolution, we examine how optimal sequence construction varies with amplitude profile. We consider the most customary pulse profiles: zero-width and rectangular, finite-width, and pursue an analysis of each profile with the addition of qubit rotation errors to model the existence of systematic errors brought about by faulty control fields.

\subsubsection{Ideal pulses}
The first type of pulse considered is an idealized, zero-width control field
\begin{equation}
V_{\mu}(t)=\sum_{j}\phi_0\,\delta(t-t^{\mu}_j),
\label{eq:idealpulse}
\end{equation}
where the Dirac delta function $\delta(t)$ constitutes the pulse profile. The pulses are applied at times $t^{\mu}_j$ and the angle of rotation is given by $\phi_0$. The control fields are restricted so that they act uni-axially for all time $t$. In terms of the single-qubit Bloch sphere, the condition can be visualized as allowing pulses solely along one of the three axes. We impose this constraint on all subsequent definitions of $V_{\mu}(t)$ as well.

\subsubsection{Finite-width pulses}
Since zero-width pulses are {experimentally} impossible, we relax the ideal pulse assumption and consider pulses of finite duration as well. We model the finite duration by
\begin{equation}
V_{\mu}(t)=\sum_{j}A\left[\Theta(t-t^{\mu}_j)-\Theta(t-t^{\mu}_j-\tau_p)\right],
\label{eq:FWpulse}
\end{equation}
representing a piecewise continuous control field with a rectangular profile \cite{ViolaKnillDD:98}. The pulse amplitude is denoted by $A$ and the pulse duration is $\tau_p$, {so that} $A\tau_p=\phi_0$. The Heaviside Theta function, {$\Theta(t)$}, dictates the pulse profile where it is assumed that the time to turn the pulse ``on" and ``off" is negligible, therefore, the pulse is well approximated by a square wave.

\subsubsection{Flip-angle errors}
An additional form of systematic error we consider is that of an over- or under-rotation in the angle $\phi_0$, commonly referred to as a flip-angle error \cite{Haeberlen:book}. This particular type of error is 
relevant, e.g., in nuclear magnetic resonance (NMR),
where inhomogeneity of the control field across the sample results in qubit rotation errors \cite{SouzaAlvarezSuter}. 
Flip-angle errors are also 
prevalent in other systems such as donor electron spins in Si systems \cite{WangDobrovitski-PulseErrorsSiDots:10,WangDobrovitski-PulseErrorsDD:10}.

In the case of zero-width pulses, the control field takes the form
\begin{equation}
V_{\mu}(t)=\sum_{j}\phi_0(1\pm\epsilon)\,\delta(t-t^{\mu}_j),
\label{eq:FAEpulse}
\end{equation}
where $\epsilon$ denotes the error in the rotation angle and the $+(-)$ refers to an over-(under-)rotation. By modeling the control field in this manner, it is assumed that the pulses are applied along their respective axes with zero or negligible error. The inclusion of rotation-axis errors has been previously studied for some common deterministic DD schemes \cite{WangDobrovitski-PulseErrorsUDD:10}, but is not included in our present study. 

\subsubsection{Finite-width flip-angle errors}
As a 
worst case scenario we also consider the 
combined effect of flip-angle errors for finite-width pulses. Assuming that the error in the pulse duration is negligible, a flip-angle error can be thought of as an error in the pulse amplitude. We model the combined error control field by
\begin{equation}
V_{\mu}(t)=\sum_{j}A(1\pm\epsilon)\left[\Theta(t-t^{\mu}_j)-\Theta(t-t^{\mu}_j-\tau_p)\right]
\label{eq:FWFAEpulse}
\end{equation}
and note that this particular form is one of the most prevalent pulse profiles encountered in experimental settings \cite{SouzaAlvarez:12,WangDobrovitski-PulseErrorsSiDots:10,Xiao-UDDPulseErrors:11,SouzaAlvarezSuter-RDD:11,WangDobrovitski-PulseErrorsUDD:10}.

\subsection{Distance Measure and Scaling}
Rather than the standard Uhlmann fidelity or trace-norm distance \cite{Nielsen:book} we use a state-independent distance measure, which significantly reduces the computational overhead.
Namely, we quantify DD performance 
using
\begin{equation}
D(U,G)=\frac{1}{\sqrt{2 d_S d_B}}\,\,\min_{\Phi}\|U-G\otimes\Phi\|_{F},
\label{eq:Dist}
\end{equation}
where $U$ represents the full evolution operator of the sequence [i.e., $U$ satisfies the Schr\"odinger equation~\eqref{eq:U0} with the Hamiltonian~\eqref{eq:genH}] , $G$ is the desired evolution of the system, the norm is the Frobenius norm $\|X\|_F = \sqrt{\text{Tr}(X^\dagger X)}$,
and $d_S$ and $d_B$ are the dimensions 
of the system and the environment Hilbert spaces $\mathcal{H}_S$ and $\mathcal{H}_B$, respectively \cite{KGBR:06}. The minimization problem can be solved analytically to obtain the closed form expression \cite{KGBR:06}:
\begin{equation}
D(U,G)=\sqrt{1-\frac{1}{d_S d_B}\|\Gamma\|_{\text{Tr}}},
\label{eq:DistClosedForm}
\end{equation}
where $\|\Gamma\|_{\text{Tr}}=\text{Tr}(\sqrt{\Gamma^{\dagger}\Gamma})$ represents the trace-norm and
\begin{equation}
\Gamma=\text{Tr}_S[U(G^{\dagger}\otimes I_B)],
\end{equation}
where $\text{Tr}_S$ denotes a partial trace over the system degrees of freedom.
In the subsequent analysis, $G\equiv I_S$ for the desired DD evolution and we denote $D\equiv D(U,I)$. 

We characterize optimal sequence performance with respect to two parameters associated with the internal dynamics: the ``strength" of the error Hamiltonian and pure environment dynamics given by
\begin{equation}
J=\|H_{\text{err}}\|,\quad \beta=\|H_B\|,
\label{eq:params}
\end{equation}
respectively, and three parameters whose relevance depends on the control Hamiltonian specifications: the pulse-interval $\tau_d$, the pulse duration $\tau_p$, and the rotation angle error $\eps$. Each of these parameters can be utilized to extract the scaling of the dominant term in the effective Hamiltonian by analyzing $D$
as a function of the parameter of interest. 

In the case of ideal pulses, 
the distance can be shown to be upper-bounded as
\begin{equation}
D\lesssim\mathcal{O}\left[(J+\beta)^{N+1}\tau^{N+1}_c\right],
\label{eq:scaling}
\end{equation}
where $N$ is the order of error suppression. See Appendix~\ref{appsec:scalingD} for {a proof}.
Analyzing $D$
as a function of $\tau_d$, $J$, and $\beta$, the order of error suppression and essentially the structure of the dominant effective error Hamiltonian operator can be determined for the relevant situations where the dynamics are dominated by system-environment interactions ($J\gg\beta$) or the bath dynamics ($J\ll\beta$). Similar studies can be performed for finite-width or flip-angle errors as well. We ultimately utilize this method in conjunction with AHT to characterize each optimal sequence and determine robustness to various pulse errors.

\section{Optimal sequences}
\label{sec:4}

In this section, we present numerically optimal $\pi$-pulse sequences obtained for the single-qubit system described by Eqs.~(\ref{eq:qubitHe}) and (\ref{eq:bathops}). Initially, we consider the case of ideal zero-width pulses.
Then, account for finite-width rectangular profiles, flip-angle errors, and finally the culmination of both types of errors. For all pulse profiles the number of pulses is varied from $K=1,2,\ldots,256$ with a pulse-interval $\tau_d=0.1$ns.\footnote{Our choice of units is arbitrary but is meant to be commensurate with electron spin qubits in, e.g., quantum dots.}

 Due to the piecewise continuous form of $H_C(t)$, the general structure of the sequences is described by 
\begin{equation}
U(\tau_c)=P_K\,f_{\tau_d}\,P_{K-1}\,f_{\tau_d}\cdots P_2\,f_{\tau_d}\,P_1\,f_{\tau_d},\\
\label{eq:seq}
\end{equation}
where $P_j$ is the unitary evolution operator achieved by the $j$th pulse and $f_{\tau_d}=e^{-i H_0 \tau_d}$ designates the "free evolution" propagator between successive pulses with pulse-interval $\tau_d$. The pulse operators are defined such that $P_j\in\G$, where $\G$ denotes a discrete set of allowable control pulses that depends upon the choice of $V_{\mu}(t)$. The total sequence time $\tau_c$ is also dictated by the choice of $V_{\mu}(t)$ since the finite duration of the pulse contributes when applicable, e.g., for a sequence of $m_d$ pulse delays and $m_p$ nontrivial pulses $\tau_c=m_d\tau_d+m_p\tau_p$. While Eq.~(\ref{eq:seq}) is not the most general DD evolution operator for fixed pulse-intervals, since it does not permit consecutive pulses without free evolution periods, it still captures a majority of the known sequences and additional highly robust sequence constructions. Further details regarding the algorithm can be found in Appendix~\ref{appsec:algorithm}.

The value of $K$ 
was varied over a significant range in our simulations, however, we found that only specific values of $K$ are 
relevant for successive error suppression. In particular, $K_{\text{opt}}=4,8,16,32,64,256$ correspond to the minimum number of pulses required to observe an increase in error suppression or significant improvement in performance. All of the remaining values of $K$ result in a sequence performance upper bounded by the performance of the previous $K_{\text{opt}}$. For example, the optimal sequences for $K=17,18,\ldots,31$ exhibit a performance proportional to that of $K=16$, if not worse. 

The values of $K_{\text{opt}}$ were obtained by analyzing the scaling of the performance of the optimal sequences identified  at each value of $K$ in the ideal pulse limit. For $K\leq 12$, optimal sequences were located by an exhaustive search, while $K> 12$ demands the use of the GA algorithm discussed in Appendix~\ref{appsec:algorithm}. Upon locating the optimal sequences, the performance measure $D$ is analyzed as a function of $\tau_c$. The order of error suppression, $N$, is then determined from Eq.~(\ref{eq:scaling}) by 
\begin{equation}
N=\frac{\log_{10}(D)}{\log_{10}[(J+\beta)\tau_c]}-1.
\end{equation}
Values of $K$ where $N$ is found to increase ultimately correspond to those identified as values of $K_{\text{opt}}$.

We find that the scaling method described above is a convenient numerical method for determining the structure of the dominant term in the effective error Hamiltonian for each of the optimal sequences obtained for a given $K_{\text{opt}}$. In order to fully characterize the scaling of $\H_{\text{err}}$ it is necessary to analyze the distance measure $D$ as a function of each relevant parameter. In the case of ideal pulses, this would correspond to analyzing the performance as a function of $\{J,\beta,\tau_d\}$ in the regimes of interaction-dominated dynamics ($J\gg\beta$) and environment-dominated dynamics ($J\ll\beta$) since each regime may exhibit different scalings. When finite pulse duration and flip-angle errors are included the number of parameters increases to either a subset of $\{J,\beta,\tau_d,\tau_p,\eps\}$ or the entire set if both forms of pulse errors are present. Futher details regarding this procedure can be found in Appendix~\ref{appsec:NumericScalingD}. It is then necessary to analyze the scaling of $D$ in each of the various parameter regimes in addition to the interaction or environment-dominated regimes, e.g., pulse-width dominated ($\tau_p \gg \tau_d$), free evolution dominated ($\tau_p \ll \tau_d$), flip-angle error dominated ($\eps \gg J\tau_d$), etc.
In the subsequent analysis, we make use of this technique in conjunction with direct calculation of the effective error Hamiltonian to fully characterize sequence performance for each pulse profile. The effective Hamiltonian calculation is utilized to provide additional insight into the structure of $\H_{\text{err}}$ that can not be observed from the scaling method, most notably in situations where multiple sequences exhibit identical performance scalings.

\subsection{Ideal pulses}
\label{subsec:idealpulse}

Here we examine the optimal sequence structure of ideal, zero-width pulses with respect to the strengths of the internal dynamics, $J$ and $\beta$, in the regime where {$J\tau_d,\beta\tau_d\in[10^{-10},10^2]$}. Under the condition of uni-axial pulses, the set of possible control pulses $\G=\{I,X,Y,Z\}$, where $I$ is the identity operator and
\begin{equation}
X(Y,Z)=-i\,\sigma^{x(y,z)}
\label{eq:idealopts}
\end{equation}
describe $\pi$-pulse unitary operators generated by 
Eq.~\eqref{eq:H_C}
when $H_C(t)$ is non-zero. Neither the error Hamiltonian, nor pure environment Hamiltonian, is present during the pulse evolution
 in the limit of zero-width (infinite amplitude) pulses.

Characteristics such as the dimension of the reduced search space $\mathcal{N}_R(K)$ are determined by the number of elements in the pulse set $\G$.
Under the conditions of ideal $\delta$-function pulses, $\mathcal{N}_R(K)=4^{K-1}$ at $l=l_{\max}$. Consequently, the initial search space only contains 16 possible sequence configurations for all $K$ using the complexity-increase procedure described in Appendix~\ref{subsubsec:ComplexReduce}. All 16 are chosen to represent the initial population at the commencement of the algorithm and the size of the population is kept constant throughout.

\subsubsection{Summary of Numerical Search}
Initially the algorithm is benchmarked at $K=4$, where it is confirmed that the well-known universal decoupling sequence \cite{XY4:69}
\begin{equation}
XY_4=Yf_{\tau_d}Xf_{\tau_d}Yf_{\tau_d}Xf_{\tau_d},
\label{eq:XY4}
\end{equation}
along with its obvious generalization
\begin{equation}
GA_4:= P_2f_{\tau_d}P_1f_{\tau_d}P_2f_{\tau_d}P_1f_{\tau_d},
\label{eq:GA_4}
\end{equation}
where $P_1\neq P_2\in \{X,Y,Z\}$, is indeed optimal over the range of $J,\beta$ specified above. The optimality of this particular sequence is attributed to its achievement of first order error suppression for general single-qubit decoherence \cite{ViolaKnillLloyd:99}, which can be confirmed numerically by analyzing the scaling of Eq.~(\ref{eq:scaling}) with respect to $\{J,\beta,\tau_d\}$, where
\begin{equation}
D\sim\left\{
\begin{array}{lcl}
\mathcal{O}(J\beta\tau^2_d) &:& J\ll\beta\cr
\mathcal{O}(J^2\tau^2_d)			&:& J\gg\beta
\end{array}\right..
\end{equation}
Alternatively, first order error suppression can be validated by calculating the effective error Hamiltonian for a specific choice of $P_j$, e.g.,
\begin{eqnarray}
\H^{XY_4}_{\text{err}}\approx &&-i\tau_d\sx\otimes[B_0,B_x]\nonumber\\
&+&\frac{i}{2}\tau_d\sz\otimes([B_0,B_z]-i\{B_x,B_y\}).
\end{eqnarray}

As mentioned above, the next interesting result occurs at $K=8$ where second order decoupling is first observed. The optimal sequences located at this particular value of $K$ can be partitioned into two general structures denoted as $a$-type and $b$-type sequences such that
\begin{eqnarray}
GA_{8a}&:=&IP_1P_2P_1IP_1P_2P_1, \label{eq:GA_8a}\\
GA_{8b}&:=&\lp P_3P_2\rp P_1 P_2 P_1\lp P_3P_2\rp P_1 P_2 P_1.
\end{eqnarray}
Note that we have dropped the free evolution periods for convenience of notation and highlighted the pulses which are not separated by free evolution periods with parentheses. In addition to the obvious structural differences between the two sequences, essentially described by whether $P_3=P_2$ is satisfied, a contrast is also observed from the standpoint of the effective error Hamiltonian. Since it is possible to effectively extract the dominant terms of the effective error Hamiltonian by examining the scaling of the performance measure, we turn to our numerical method and determine
\begin{eqnarray}
D_{8a}&\sim&\left\{
\begin{array}{lcl}
\mathcal{O}(J\beta^2\tau^3_d) &:& J\ll\beta\\
\mathcal{O}(J^3\tau^3_d)			&:& J\gg\beta
\end{array}\right.\\
D_{8b}&\sim&\mathcal{O}(J\beta^2\tau^3_d)\quad  \forall J,\beta.
\label{eq:GA_8}
\end{eqnarray}
Clearly, the difference occurs in the regime of interaction-dominated dynamics, where terms in the effective error Hamiltonian that solely comprise products of the interaction Hamiltonian begin to govern the scaling of $GA_{8a}$'s performance.

\begin{table}[t]
\centering
\begin{tabular}{cccc}
\multicolumn{2}{c}{Sequence} & \multirow{2}{*}{$J\ll\beta$} & \multirow{2}{*}{$J\gg\beta$}\\ \cline{1-2}

Name & Description &  & \\ \hline

$GA_4$ & $P_1P_2P_1P_2$ & $\O(J\beta\tau^{2}_d)$ & $\O(J^2\tau^{2}_d)$ \\ \hline

$GA_{8a}$ & $IP_1P_2P_1IP_1P_2P_1$ & $\bb{\O(J\beta^2\tau^{3}_d)}$ & $\O(J^3\tau^{3}_d)$ \\ 
$GA_{8b}$ & $P_3(GA_4)P_3(GA_4)$ & $\O(J\beta^2\tau^{3}_d)$ & $\bb{\O(J\beta^2\tau^{3}_d)}$ \\ 

$GA_{16a}$ & $P_3(GA_{8a})P_3 (GA_{8a})$ & $\bb{\O(J\beta^2\tau^{3}_d)}$ & $\O(J^3\tau^{3}_d)$ \\ 
$GA_{16b}$ & $GA_4[GA_4]$ & $\O(J\beta^2\tau^{3}_d)$ & $\bb{\O(J\beta^2\tau^{3}_d)}$ \\  \hline

$GA_{32a}$ & $GA_4[GA_{8a}]$ &$\bb{\O(J\beta^3\tau^{4}_d)}$ & $\O(J^2\beta^2\tau^{4}_d)$ \\ 
$GA_{32b}$ & $GA_{8a}[GA_{4}]$ &$\O(J\beta^3\tau^{4}_d)$ & $\bb{\O(J^2\beta^2\tau^{4}_d)}$ \\  \hline 

$GA_{64a}$ & $GA_{8a}[GA_{8a}]$ &$\bb{\O(J\beta^4\tau^{5}_d)}$ & $\O(J^3\beta^2\tau^{5}_d)$ \\ 
$GA_{64b}$ & $GA_{8b}[GA_{8b}]$ &$\O(J\beta^4\tau^{5}_d)$ & $\bb{\O(J^3\beta^2\tau^{5}_d)}$ \\ 
$GA_{64c}$ & $GA_4[GA_4[GA_4]]$ &$\O(J\beta^3\tau^{4}_d)$ & $\O(J^2\beta^2\tau^{4}_d)$ \\  \hline

$GA_{256a}$ & $GA_4[GA_{64a}]$ & $\bb{\O(J\beta^5\tau^{6}_d)}$ & $\O(J^3\beta^3\tau^{6}_d)$ \\
$GA_{256b}$ & $GA_{8b}[GA_{32a}]$  &   $\O(J\beta^5\tau^{6}_d)$ & $\bb{\O(J^3\beta^3\tau^{6}_d)}$ \\
$GA_{256c}$ & $GA_4[GA_{64c}]$ & $\O(J\beta^4\tau^{5}_d)$ & $\O(J^2\beta^3\tau^{5}_d)$ 
\end{tabular}
\caption{Summary of distance measure ($D$) scalings for each optimal $GA_K$ sequence identified in the ideal pulse limit, for a fixed 
pulse-interval of $\tau_d$. Boxed performance scalings highlight optimal sequences in each of the relevant $(J,\beta)$-regimes (columns) for each $K_{\text{opt}}$.}
\label{tbl:DScalingIdeal}
\end{table}

In addition to providing insight into the general structure of optimal sequences for $K=8$, the results also show an immediate correspondence with known $8$-pulse sequences, namely,
\begin{equation}
XY_8=IXYXIXYX.
\end{equation}
Known for providing second order error suppression for general single-qubit decoherence \cite{XYseqs:69}, $XY_8$ gains its decoupling attributes from its structure: the $XY_4$ sequence followed by a time-reversed copy. An alternative perspective of $XY_8$ is that of a concatenated sequence composed of $XY_4$ ($GA_4=XY_4$) and CPMG $=Pf_{\tau_d}Pf_{\tau_d}$  ($P=X$). In general, depending on the choice of the CPMG pulses two different variations arise: $GA_{8a}$ or $GA_{8b}$. Interestingly, the latter viewpoint is perhaps the most useful for sequence characterization since all remaining optimal sequences from $K=16$ to $K=256$ can be interpreted as concatenations of various combinations of CPMG, $XY_4$, and both $K=8$ optimal sequences. This result not only conveys the importance of these sequences as fundamental building blocks for arbitrary $K$ optimal sequences, but also the significance of concatenation in achieving high, perhaps even arbitrary, order error suppression in the regime of fixed-pulse intervals.

\begin{figure*}[t]
\centering
\subfigure[$K=4$]{\includegraphics[scale=0.024]{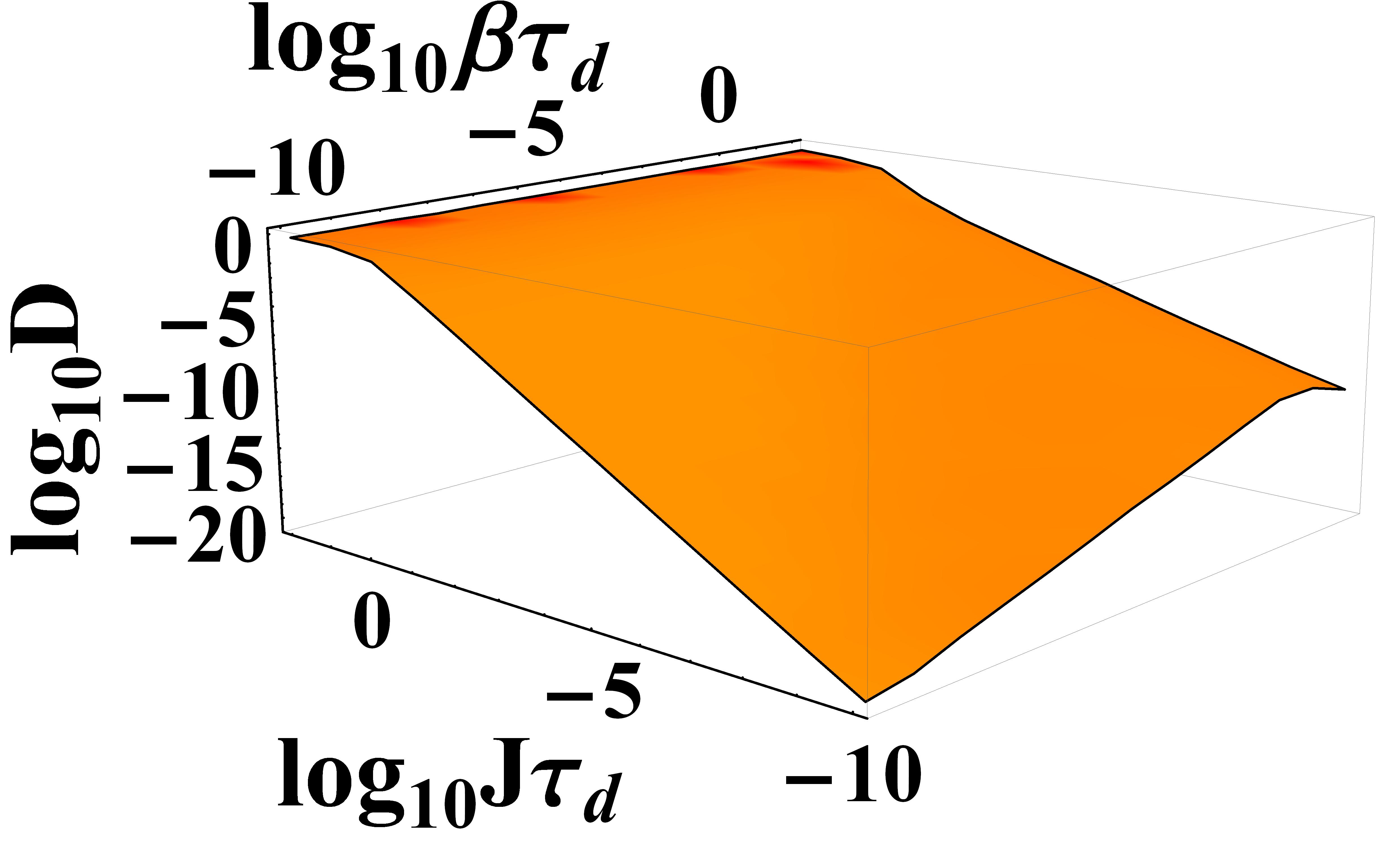}}\hspace{0.1cm}
\subfigure[ $K=8$]{\includegraphics[scale=0.024]{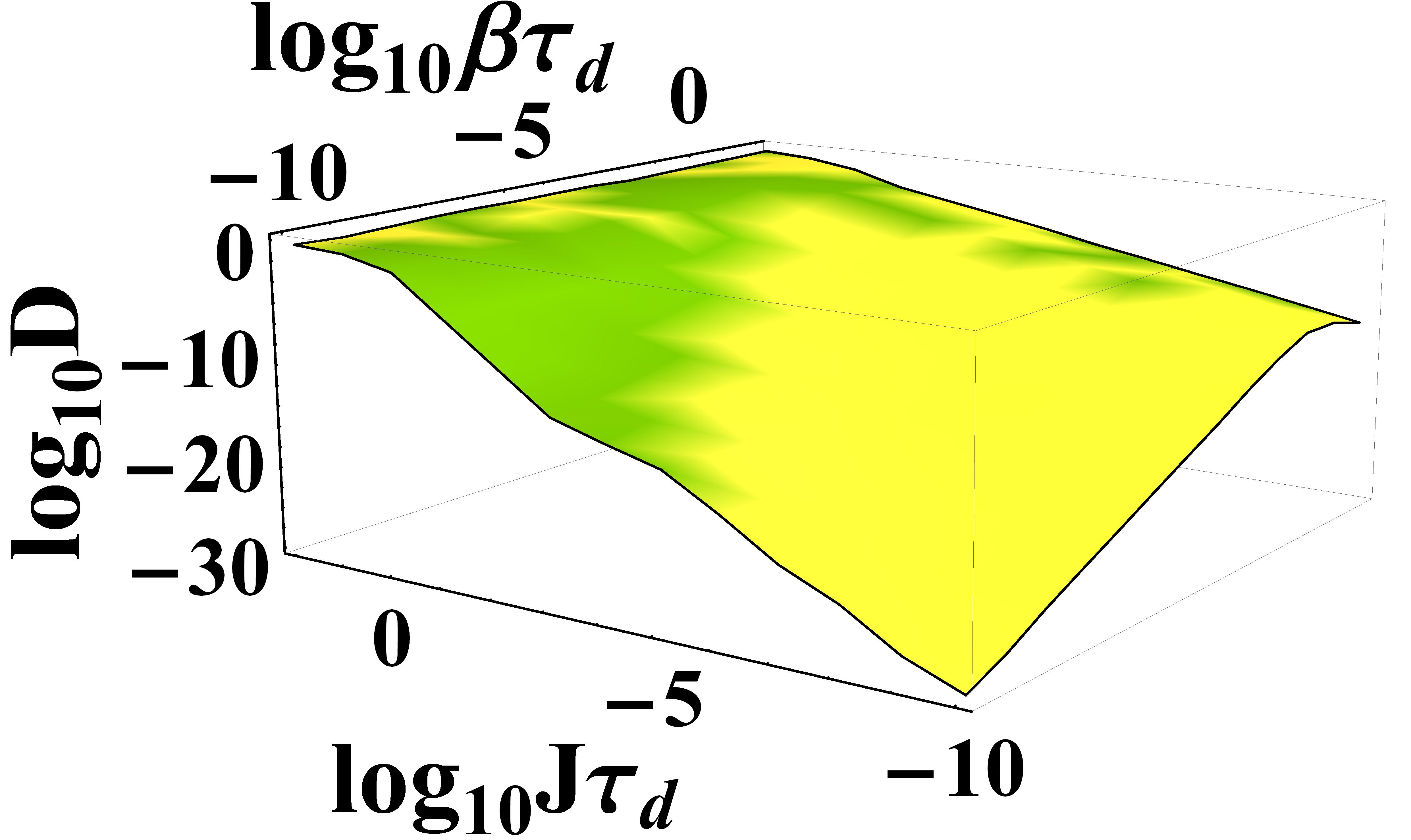}}\hspace{0.1cm}
\subfigure[ $K=16$]{\includegraphics[scale=0.024]{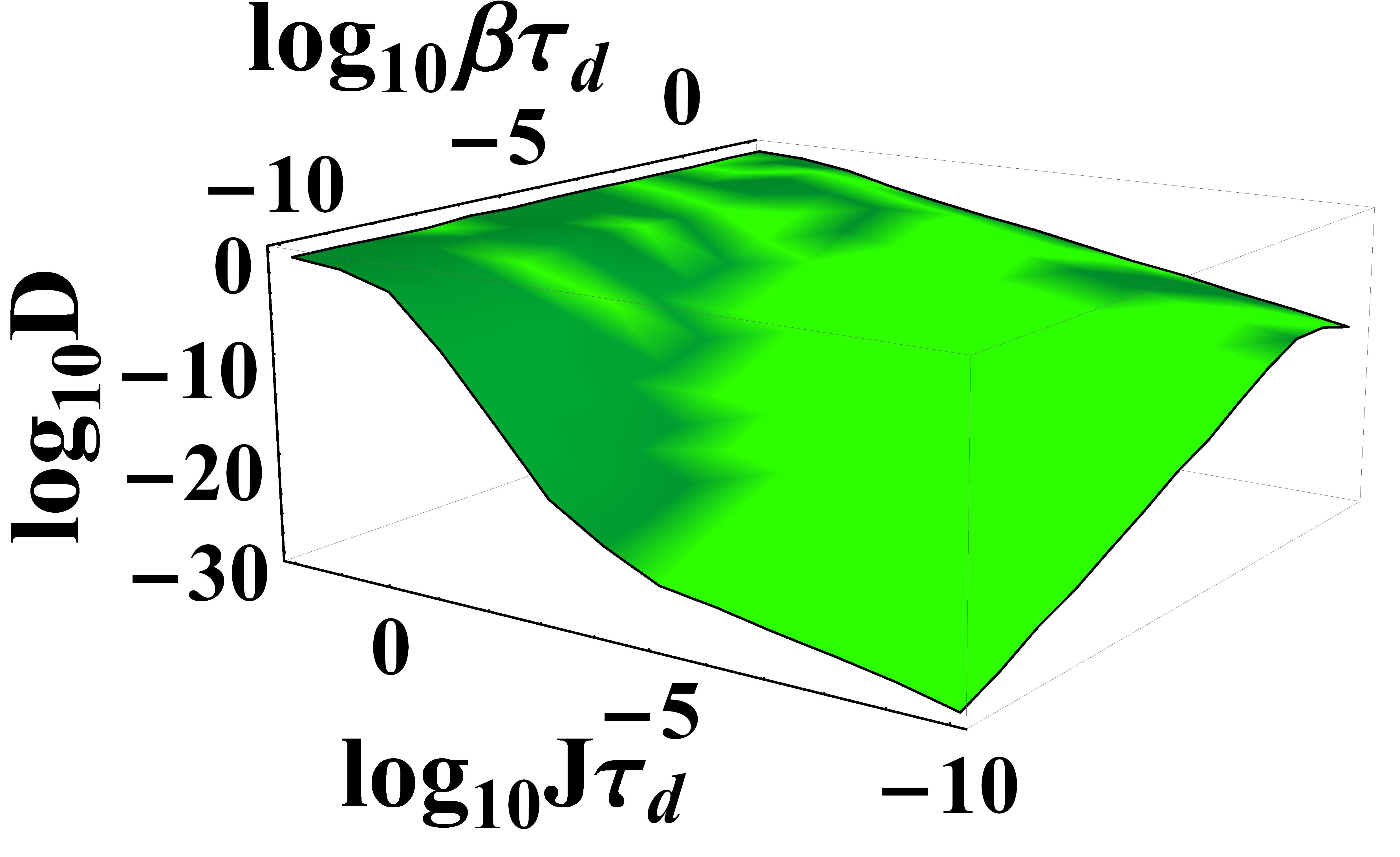}}\\
\subfigure[ $K=32$]{\includegraphics[scale=0.024]{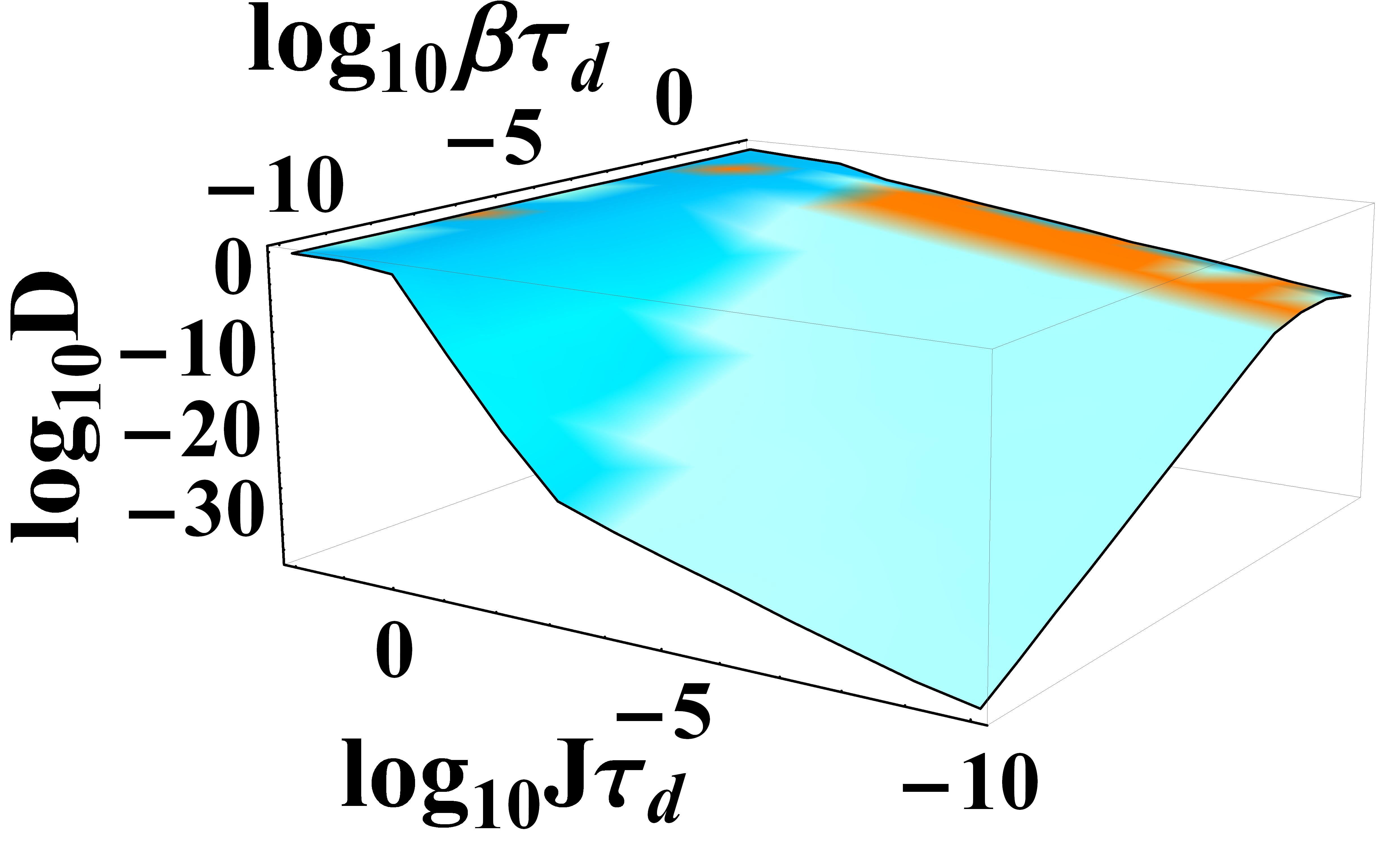}}\hspace{0.1cm}
\subfigure[ $K=64$]{\includegraphics[scale=0.024]{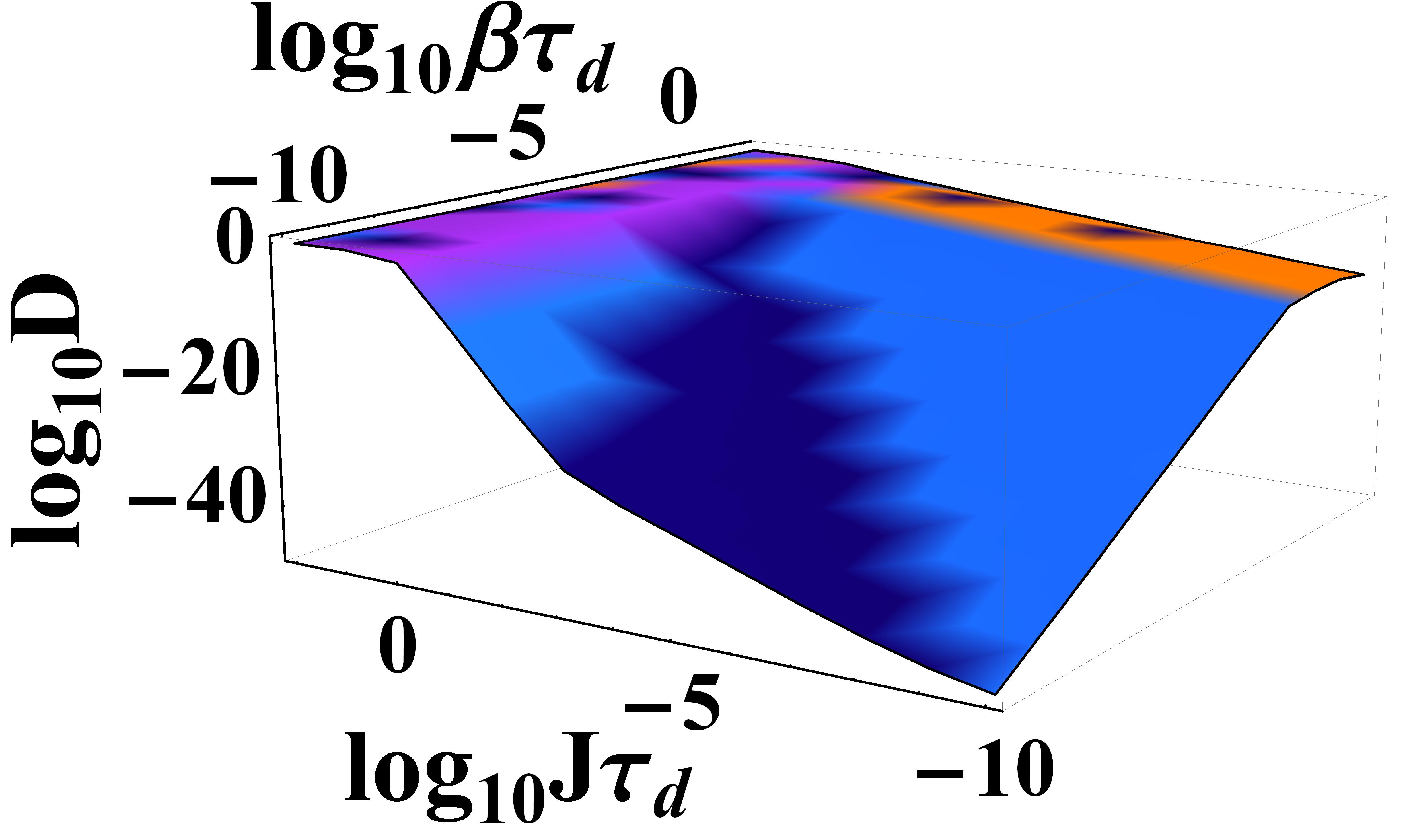}}\hspace{0.1cm}
\subfigure[ $K=256$]{\includegraphics[scale=0.024]{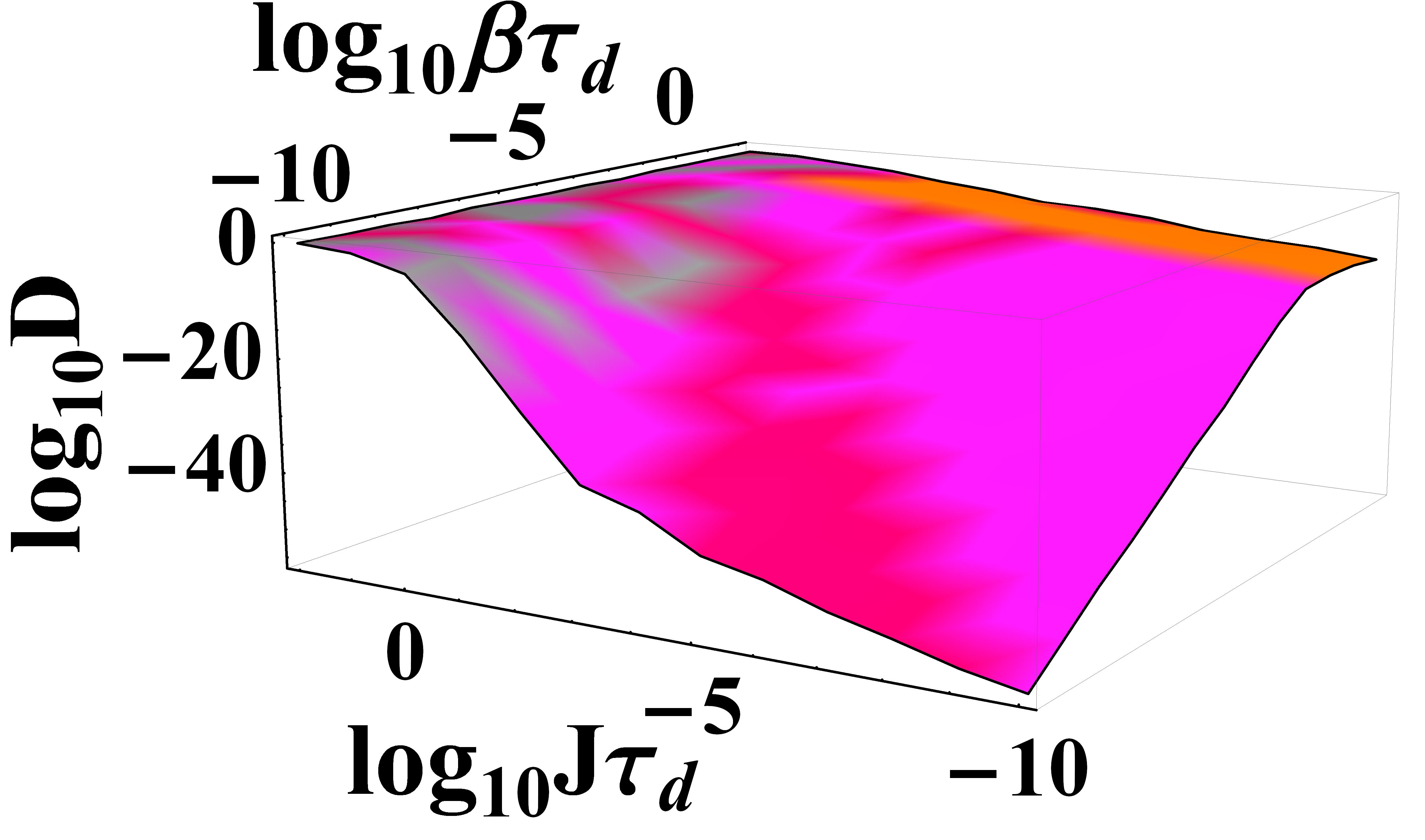}}\\
\subfigure{\includegraphics[scale=0.035]{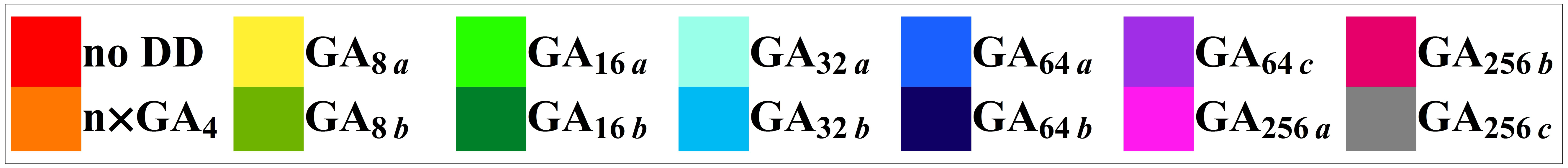}}
\caption{Performance of $GA_K$ sequences for $K=4,8,16,32,64,256$, as shown in (a)-(f), respectively, as a function of $J$ and $\beta$. The minimum pulse-interval is fixed at $\tau_d=0.1$ns and the results are averaged over 10 realizations of $B_{\mu}$. The $GA_{K\,a}$ sequences tend to be optimal in the range $J<\beta$, while $GA_{K\,b}$ are optimal for $J\geq \beta$. Sequences closely related to the deterministic structure of CDD, $GA_{K\,c}$, begin to appear as optimal sequences for $J\tau_d>1$ at $K=64$. The notation $n\times GA_4$ denotes the application of $n$ cycles of $GA_4$ and is dependent on the value of $K$. For the data presented above, $n=1,8,16,64$ 
for $K=4,32,64,256$, respectively.}
\label{fig:GALandscapePlots}
\end{figure*}

Let us now discuss the remaining sequences and focus on sequence lengths, and generalized sequence constructions, that yield additional orders of error suppression. We find that in order to achieve third, fourth, and fifth order decoupling a minimum of $32$, $64$, and $256$ pulses are required, respectively. The sequences responsible for these effects are:
\begin{equation}
\begin{array}{lcl}
GA_{32a}:=GA_4[GA_{8a}],                  & \quad &     GA_{32b}:=GA_{8a}[GA_{4}], \\
GA_{64a}:=GA_{8a}[GA_{8a}],             & \quad &    GA_{64b}:=GA_{8b}[GA_{8b}],\\
GA_{256a}:=GA_4[GA_{64a}],              & \quad &    GA_{256b}:=GA_{8b}[GA_{32a}],
\end{array}\nonumber
\end{equation}
where the brackets are used to denote a concatenated structure, e.g.,
\begin{eqnarray}
GA_{32a}&=&P_2(GA_{8a})P_1(GA_{8a})P_2(GA_{8a})P_1(GA_{8a})\nonumber\\
&=&GA_4[GA_{8a}].
\end{eqnarray}
The performance of both $32$-pulse sequences 
scales as $D_{32a,b}\sim\O(\tau^{4}_d)$. Similarly, the $64$ and $256$-pulse sequences obtain performance scalings of $D_{64a,b}\sim\O(\tau^{5}_d)$ and $D_{256a,b}\sim\O(\tau^{6}_d)$. A more detailed characterization of the performance scaling for each sequence is given in Table~\ref{tbl:DScalingIdeal}. Note that we also include $K=16$ sequences even though they do not achieve additional orders of error suppression. Their significance will become apparent in subsequent discussions on pulse imperfections presented in the main text and appendix. In regards to the remaining optimal sequences, one may notice the absence of $XY_4$-based CDD which contains sequences of $K=16,64,256$ pulses. The generalized sequences defined above in fact out perform CDD sequences and obtain an additional order of decoupling at each of the $K=16,64,256$ sequence lengths.

\subsubsection{Characterization of $GA_K$ Sequences in $(J,\beta)$-space}
In Figure \ref{fig:GALandscapePlots}, the space of optimal sequences is characterized as a function of $J\tau_d$ and $\beta\tau_d$ for each of the optimal sequence lengths discussed above. A general trend is observed for each sequence type, where a-type sequences tend to be optimal when $J<\beta$ and b-type sequences dominate the $J>\beta$ regime. Note that when the bath dynamics are dominant such that $\beta\tau_d\gg1$, repeated cycles of  $GA_4$ become the preferred sequence. Here, bath self-averaging effects are more prominent and sequence effectiveness is reduced dramatically for more sophisticated sequence structures which possess cycle times that are longer than $\beta^{-1}$. It is important to note that additional sequences beyond those discussed above also appear: $GA_{64c}=GA_4[GA_4[GA_4]]$ and $GA_{256c}=GA_4[GA_{64c}]$. These structures correspond to generalized XY$_4$-based CDD sequences which only appear to be optimal when $J\tau_d\gg1$. This result is consistent with the performance scaling equations presented in Table \ref{tbl:DScalingIdeal}, where the quadratic scaling in $J$ for c-type (as opposed to cubic scaling for a,b-type) sequences is clearly more favorable when $J\tau_d\gg1$.

\begin{table*}[t]
\centering
\begin{tabular}{cccc}
\multicolumn{2}{c}{Sequence} & \multirow{2}{*}{$J\tau_p\ll \eps$} & \multirow{2}{*}{$J\tau_p\gg\eps$}\\ \cline{1-2}

Name & Description &  & \\ \hline

$RGA_2$ & $\Pb P$ & $\bb{\O(J\tau_d)}$ & $\O(J\tau_p)$ \\ 

$RGA_4$ & $\Pb_2 P_1 \Pb_2 P1$ & $\O(J^2\tau^{2}_d,\eps^2)$ & $\bb{\O(J^2\tau^{2}_d,J\tau_p)}$ \\
$RGA_{4'}$ & $\Pb_2 \Pb_1 \Pb_2 P1$ & N/A & N/A\\ \hline

$RGA_{8a}$ & $I \Pb_1 P_2 \Pb_1 I P_1 \Pb_2 P_1$ & $\bb{\O(\eps J\tau_d)}$ & $\O(J\tau_p)$\\ 
$RGA_{8c}$ & $P_1 P_2 P_1 P_2 P_2 P_1 P_2 P_1$ & $\O(\eps J\tau_d,\eps^2)$ & $\bb{\O(J^2\tau^{2}_d,J^2\tau_d\tau_p)}$\\ \hline

$RGA_{16a}$ & $\Pb_3(RGA_{8a}) P_3 (RGA_{8a})$ & $\bb{\O(J\tau_p)}$ & $\O(J\tau_p)$\\ 
$RGA_{16b''}$ & $RGA_{4'}[RGA_{4'}]$ & $\O(\eps^2)$ & $\bb{\O(J\tau_p)}$\\ \hline

$RGA_{32a}$ & $RGA_4[RGA_{8a}]$ & $\O(\eps^2)$ & $\O(J\tau_p)$\\
$RGA_{32c}$ & $RGA_{8c}[RGA_4]$ & $\O(\eps^2)$ & $\O(\eps J\tau_p)$\\ \hline

$RGA_{64a}$ & $RGA_{8a}[RGA_{8a}]$ & $\bb{\O(J\tau_p)}$ & $\O(J\tau_p)$\\
$RGA_{64c}$ & $RGA_{8c}[RGA_{8c}]$ & $\O(\eps^2)$ & $\bb{\O(\eps J\tau_p)}$\\ \hline

$RGA_{256a}$ & $RGA_{4}[RGA_{64a}]$ & $\O(\eps^2)$ & $\O(J\tau_p)$\\

\end{tabular}

\caption{Summary of distance measure $D$ scalings for optimal $RGA_K$ sequences identified for DD evolution subjected to finite-width pulses of duration $\tau_p$ and flip-angle errors with rotation error $\eps$ in the regime of system-environment interaction-dominated ($J\gg\beta$) dynamics. The sequences with the best performance scaling in each parameter regime (column) for each $K_{\text{opt}}$ are boxed. 
Note specifically the case of strong pulses dominated by flip-angle errors ($J\tau_p\ll \eps$) where $RGA_{8a}$, $RGA_{16a}$, and $RGA_{64a}$ obtain the more favorable performance scaling.}
\label{tbl:scaling-FWPE}
\end{table*}

\subsection{Finite-Width and Flip-Angle Errors}
\label{subsec:FiniteAndFlip}
Flip-angle and finite-width pulse errors are prevalent in a variety of experimental settings. Therefore, it is necessary for DD sequences to be robust against both types of errors simultaneously if reliable computation is to be implemented under the protection of DD in realistic setups. Designs based on (C)DCG \cite{DCG,DCG1,CDCG} do not apply in this case as they do not address flip-angle errors. The KDD sequence is applicable, but unlike the present study, it employs $\pi$ pulses which are not uni-axial \cite{SouzaAlvarezSuter-RDD:11}. As a final consideration, we investigate the inclusion of the two forms of pulse errors and search for optimal sequences at each $K_{\text{opt}}$. By performing this search, we are essentially addressing the possibility of constructing fault-tolerant DD in the most convenient possible arrangement: fixed pulse-interval and 
rectangular pulse shape. 

The control pulse set is now $\G=\{I,X,Y,Z,\Xb,\Yb,\Zb\}$, where each pulse operator is defined by
\begin{equation}
X(Y,Z)=e^{-i\tau_p[A(1+\eps)\sigma^{x(y,z)}+H_0]}
\end{equation}
with $\{\Xb,\Yb,\Zb\}$ corresponding to 180-degree phase rotations ($A\rightarrow -A$) and $I\equiv \exp(-i\tau_pH_0)$. The identity operator is chosen in this manner so that the cycle time for a given $K$ is equivalent for all sequences. Note that pulse operators for non-trivial pulses now include the internal Hamiltonian $H_0$ to effectively model pulses which are finite in duration and amplitude. This form reduces to the ideal pulse operators given in Eq.~(\ref{eq:idealpulse}) by considering infinite pulse amplitude ($A\rightarrow\infty$) and a pulse duration $\tau_p\rightarrow0$ so that $\|H_C(t)\|\tau_p=\pi/2$ is maintained and $\|H_0\|\tau_p\rightarrow 0$. 

Of course we must still make the strong pulse assumption $\|H_C(t)\|\gg\|H_0\|$ in order to reduce the additional errors generated by $H_0$ during the pulse. We enforce this assumption in the following section and utilize it to calculate effective pulse dynamics where the contributions of $H_0$ are essentially perturbations to $H_C(t)$. The effective pulse operators are then used to calculate effective error Hamiltonians for each sequence.

The
amplitudes of the error and bath Hamiltonians 
are chosen as $J=1$MHz and $\beta=1$kHz, respectively. Optimal sequence performance is analyzed with respect to $\eps$ and $\tau_p$ in the regime where $J\gg\beta$ and $J\tau_d\ll 1$ to characterize sequence robustness as a function of errors generated by the DD pulses.

\subsubsection{Results of Numerical Search}
Here, we present a summary of our numerical search for pulse error-optimized sequences, which we will refer to as robust $GA_K$ ($RGA_K$) sequences. The first case to examine is that of $K=4$, where we locate two optimal sequences:
\begin{equation}
RGA_4:=\Pb_2 f_{\tau_d}P_1f_{\tau_d} \Pb_2f_{\tau_d} P_1f_{\tau_d}
\label{eq:RGA_4}
\end{equation}
and two cycles of
\begin{equation}
RGA_2:=\Pb f_{\tau_d}P f_{\tau_d},
\end{equation}
which we will denote as $2\times RGA_2$. It is perhaps not surprising that a robust version of $GA_4$ appears as an optimal sequence given the results of the ideal pulse analysis. The more interesting result is the emergence of $2\times RGA_2$ as an optimal four-pulse sequence since it does not provide complete first order error suppression. Its presence is clearly attributed to its robustness against pulse imperfections, rather than errors generated by free evolution.

Determining which form of pulse error is addressed most effectively by $RGA_2$, and $RGA_4$ for that matter, is best accomplished via direct calculation of the effective error Hamiltonian. In the case of $RGA_4$, the effective dynamics are governed by
\begin{equation}
\H^{RGA_4}_{\text{err}}\approx-\frac{\pi^2\eps^2}{2\tau_c}\sz-\frac{4\tau_p}{\pi\tau_c}\sz B_{x-y}+\O(\eps J\tau_p,\eps J\tau_d),
\label{eq:RGA4Heff_FWPE}
\end{equation}
with $B_{x-y}\equiv B_x-B_y$, while for $RGA_2$,
\begin{equation}
\H^{RGA_2}_{\text{err}}\approx  \sx B_x-\frac{4\tau_p}{\pi\tau_c}(1-\eps)\sz B_y+\O(\eps J\tau_p,\eps J\tau_d).
\label{eq:Heff-FWPE2}
\end{equation}
Here, the pulses are taken as $\{P_1,P_2\}=\{X,Y\}$ and $P=X$ for $RGA_4$ and $RGA_2$, respectively. Note that $RGA_4$ produces first order decoupling in $\tau_p$ and $\eps$, yet does not effectively address errors generated by the finite pulse duration. The effective error Hamiltonian for $RGA_2$ confirms the presence of $\O(J\tau_d)$ terms and also displays a similar lack of first order decoupling in $\tau_p$. The primary distinction between the two sequences is the suppression of $\O(\eps^2)$ errors provided solely by $RGA_2$. In summary, the regimes of optimal performance for each sequence can be characterized as follows:  $RGA_4$ is most advantageous when the pulse imperfections are relatively small and the free evolution periods ascribe to the primary source of decoherence ($J\tau_d>J\tau_p$ and $J\tau_p> \eps$), while $RGA_2$ is most effective when flip-angle errors are dominant ($\eps>J\tau_p$ and $\eps> J\tau_d$).

Continuing the search to $K=8$, we find that all $8$-pulse optimal sequences which exhibit robustness to both finite pulse-width and flip-angle errors can be described by 
\begin{eqnarray}
RGA_{8a}&:=&I \Pb_1 P_2 \Pb_1 I P_1 \Pb_2 P_1,\label{eq:RGA_8a}\\
RGA_{8c}&:=&P_1P_2P_1P_2P_2P_1P_2P_1.\label{eq:RGA_8c}
\end{eqnarray}
Again, a correspondence between $RGA_K$ and $GA_K$ sequences is observed with $RGA_{8a}$, as it is essentially $GA_{8a}$ with \blue{$\pi$} phase adjustments required for additional robustness against pulse imperfections. The latter sequence did not appear as an optimal sequence in the ideal pulse analysis; however, it is identified as a generalized version of the well-known Eulerian DD (EDD) sequence which is known for producing first order decoupling in the pulse duration \cite{EDD}. See Appendix~\ref{subsec:finitepulse} for a detailed discussion of EDD and additional $RGA_K$ sequences optimized excusively for finite pulse-width.

\begin{figure*}[t]
\centering
\subfigure[\ $K=4$]{\includegraphics[scale=0.024]{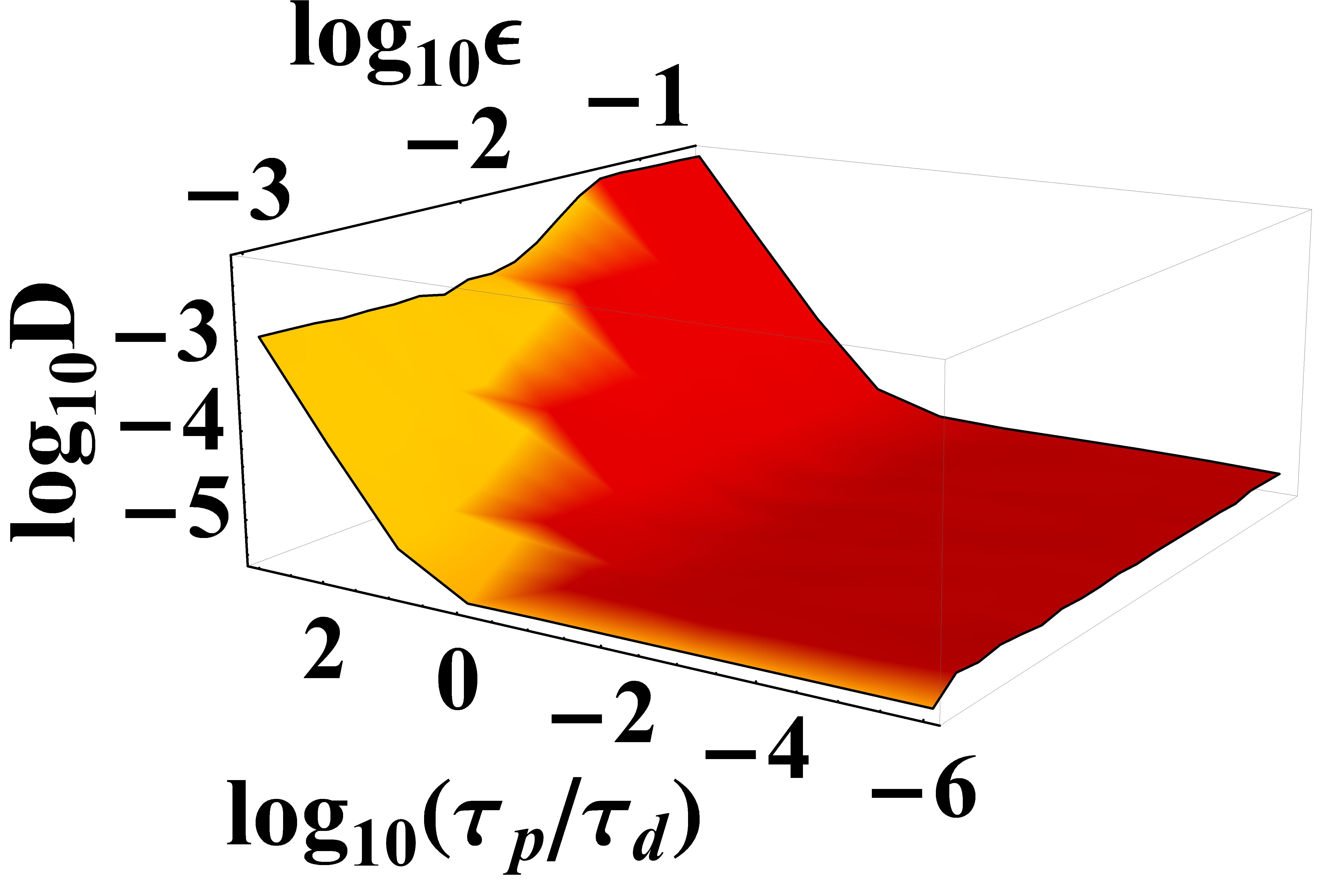}}\hspace{0.1cm}
\subfigure[\ $K=8$]{\includegraphics[scale=0.024]{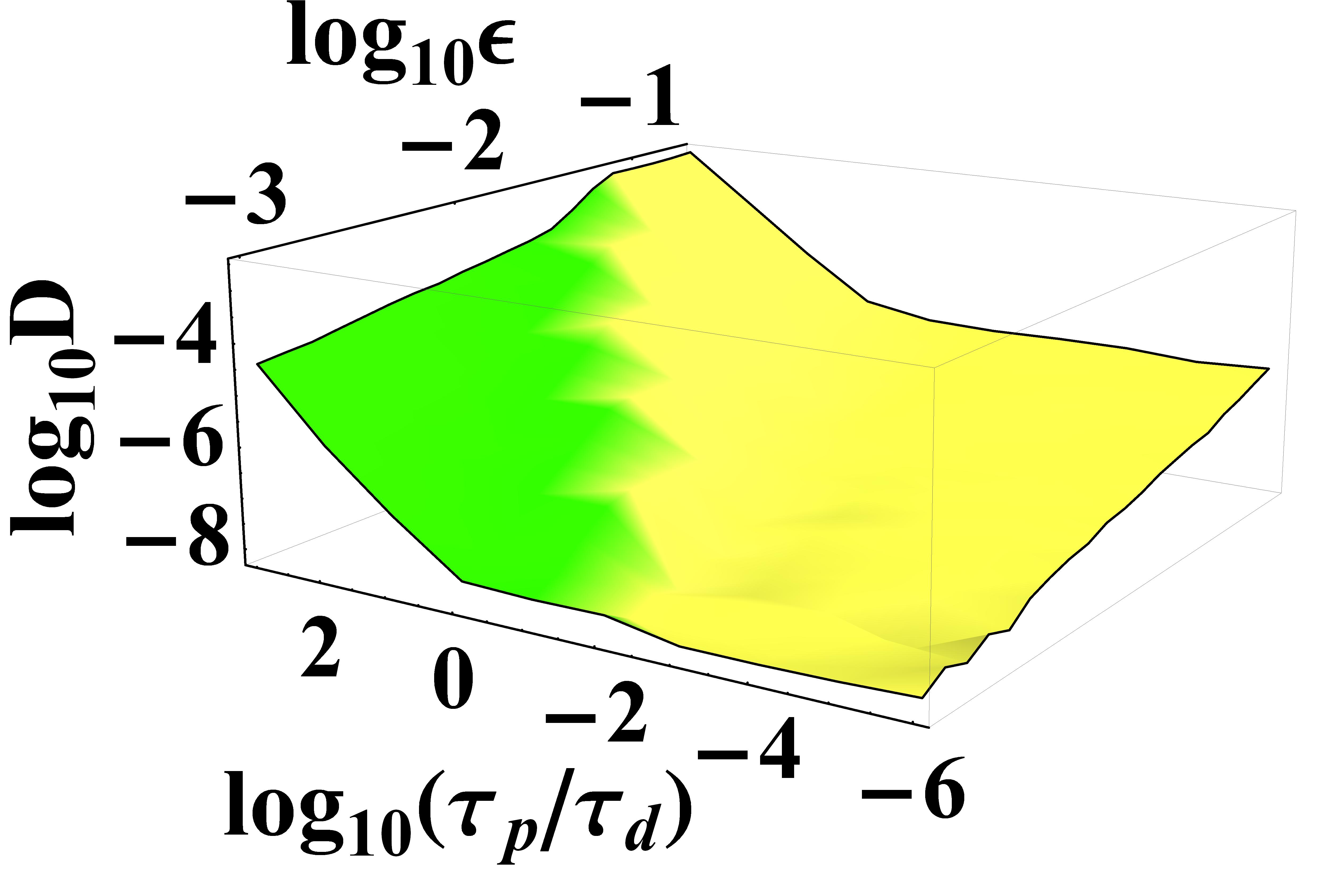}}\hspace{0.1cm}
\subfigure[\ $K=16$]{\includegraphics[scale=0.024]{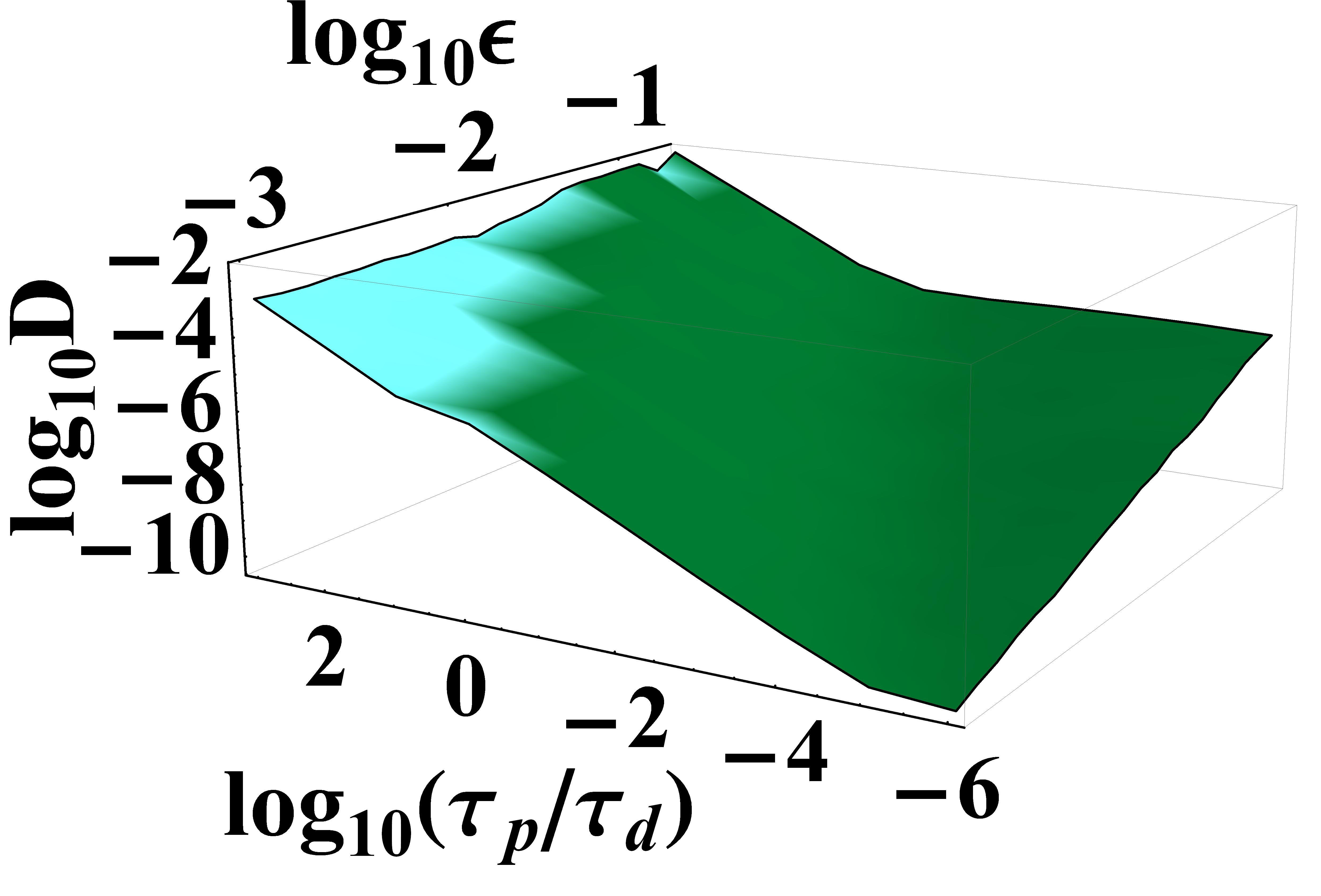}}\\
\subfigure[\ $K=32$]{\includegraphics[scale=0.024]{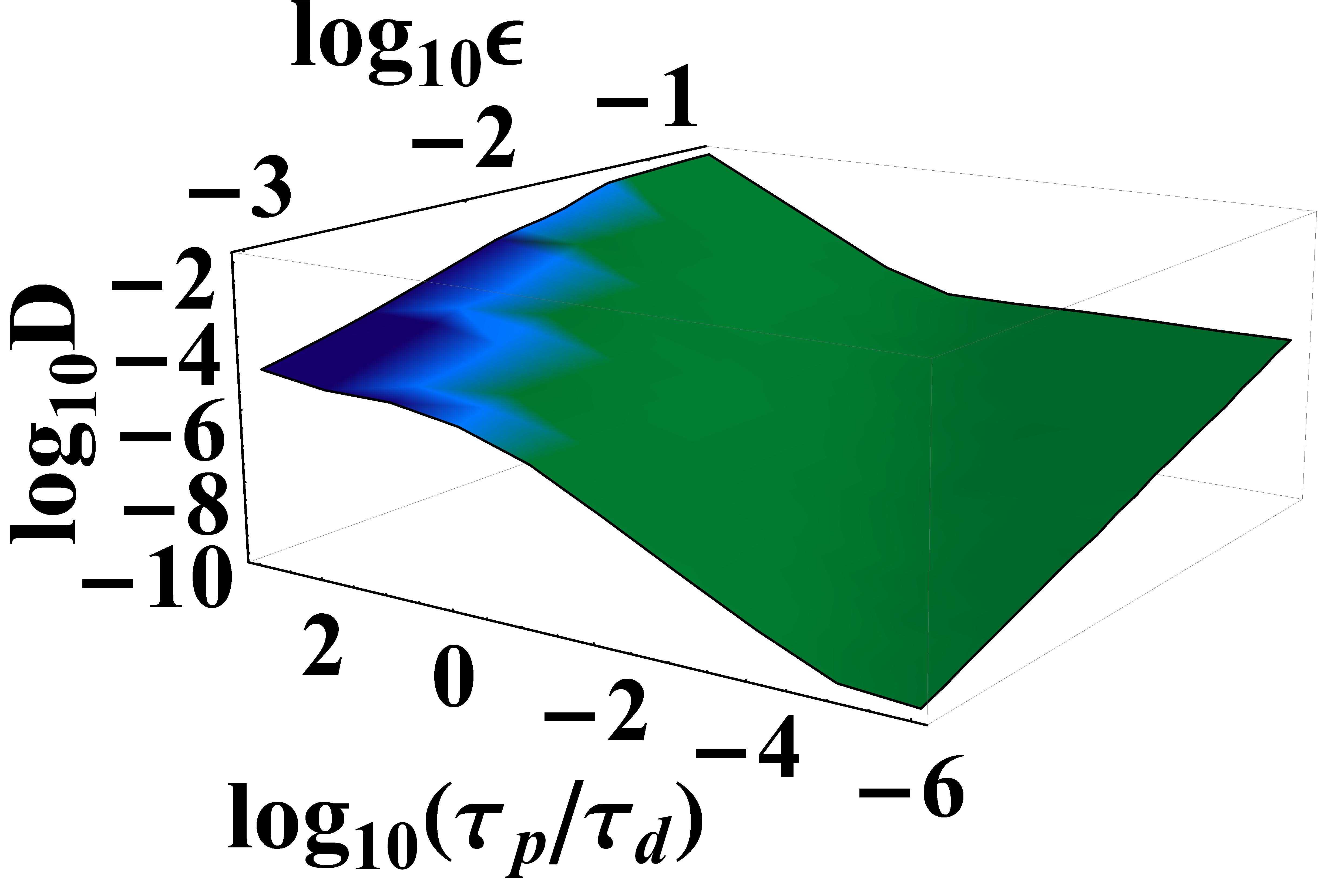}}\hspace{0.1cm}
\subfigure[\ $K=64$]{\includegraphics[scale=0.024]{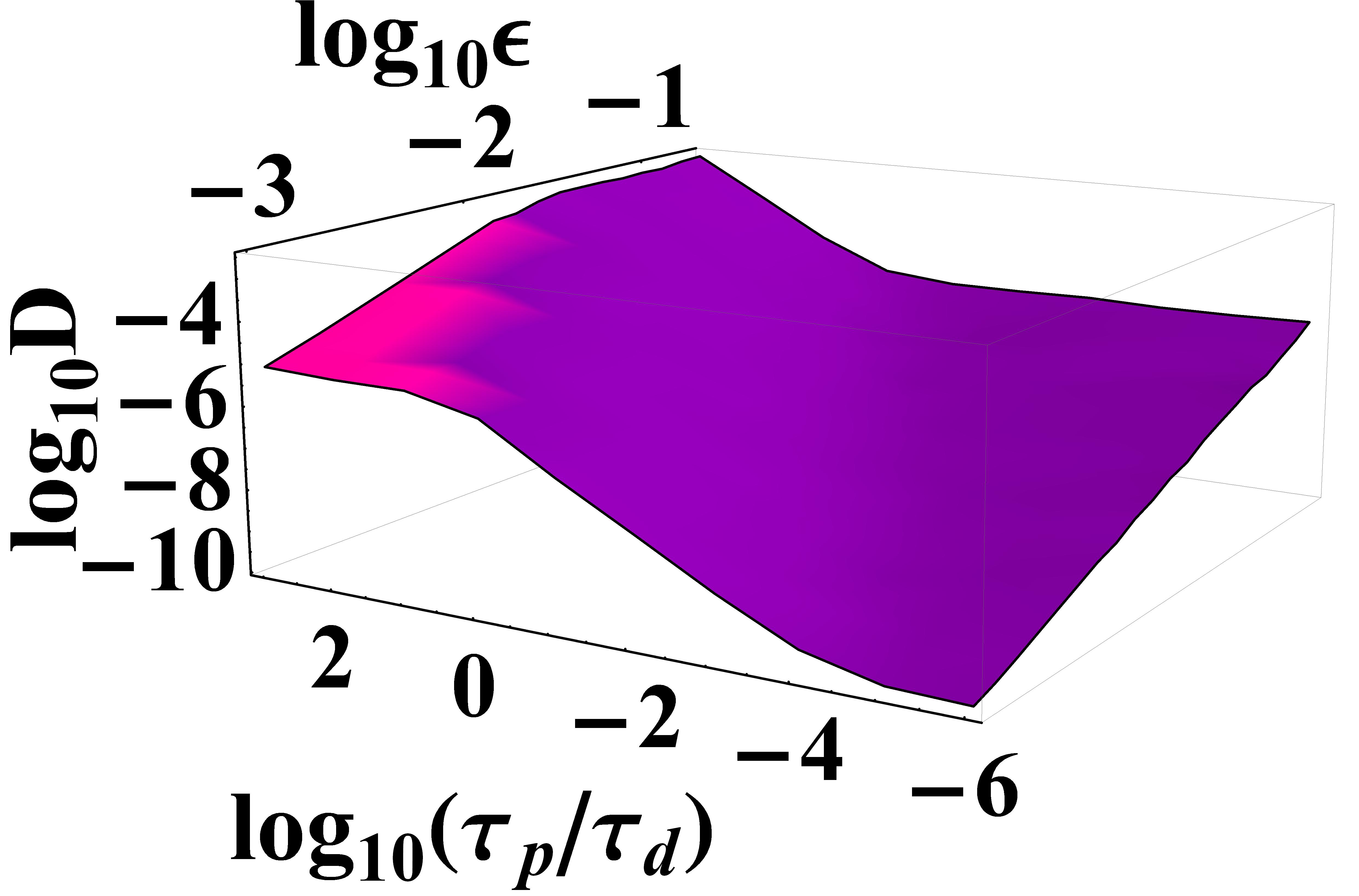}}\hspace{0.1cm}
\subfigure[\ $K=256$]{\includegraphics[scale=0.024]{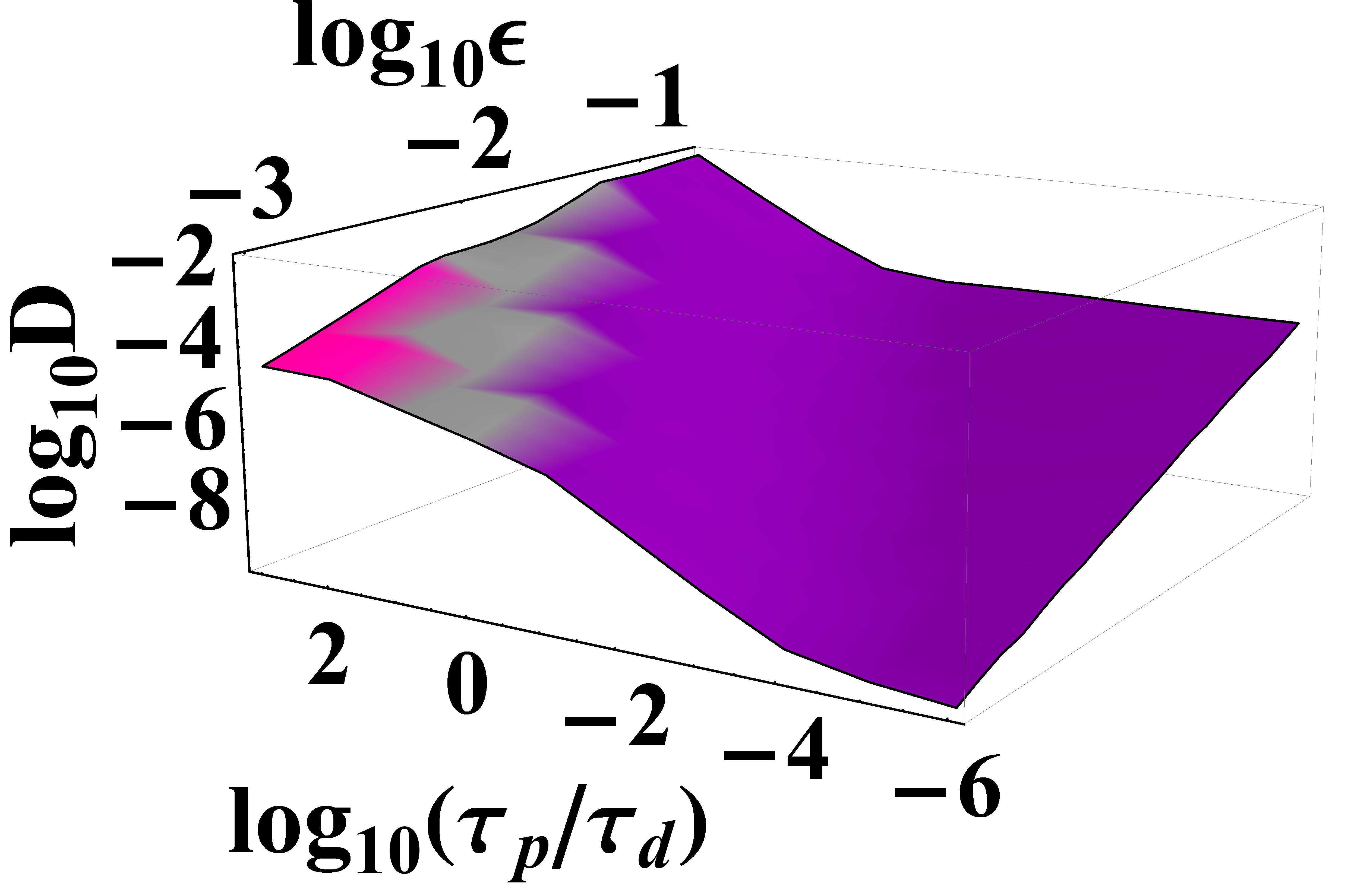}}\\
\subfigure{\includegraphics[scale=0.0285]{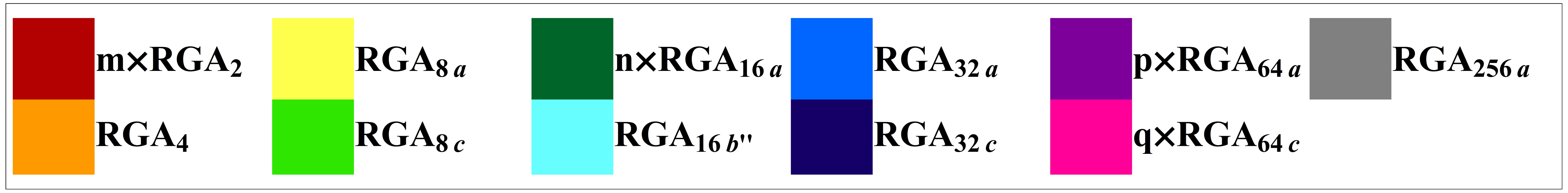}}
\caption{Performance of $RGA_K$ sequences for $K=4,8,16,32,64,256$, as shown in (a)-(f), respectively, as a function of $\eps$ and $\tau_p/\tau_d$. The minimum pulse-interval is fixed at $\tau_d=0.1$ns, while $J=1$MHz and $\beta=1$kHz. The flip-angle error is varied from a $1\%$ to a $10\%$ error and $\tau_p$ is varied throughout a wide range of values so that $\tau_p\ll\tau_d$, and $\tau_p\gg\tau_d$, is explored. The majority of the parameter space is dominated by the $RGA_{K\,a}$ sequences, even in the finite-width error-dominant regime. The robustness of $RGA_{K\,a}$ to pulse imperfections ultimately saturates at $K=64$. All simulations are averaged over 10 realizations of $B_{\mu}$. The optimal sequences that are used in multiple $K_{\text{opt}}$ values require the following number of cycles: {(a) $m=2$, (c) $n=1$, (d) $n=2$, (e) $p=q=1$, (f) $p=q=4$.}}
\label{fig:GALandscapePlotsFWPE}
\end{figure*}

The effective error Hamiltonian for $RGA_{8a}$,
\begin{equation}
\H^{RGA_{8a}}_{\text{err}}\approx\frac{4\tau_p}{\pi\tau_c}\left[\sx B_z+\sz(B_x-2B_z)\right]-\frac{4\pi\eps\tau_d}{\tau_c}\sz B_{x-y},
\label{eq:HeffRGA8a_FWPE}
\end{equation}
displays an improvement in performance over both $K=4$ optimal sequences in its ability to produce second order decoupling in both $\tau_d$ and $\eps$. Further improvement in robustness is observed for $RGA_{8c}$, where
\begin{eqnarray}
\H^{RGA_{8c}}_{\text{err}}\approx&&(\sx+\sy)\left[\frac{\pi^3\eps^3}{2\tau_c}-\frac{4\eps\tau_p}{\tau_c}(B_x+B_y)\right]\nonumber\\
&+&\O(\eps J\tau_d,J^2\tau^{2}_d)
\end{eqnarray}
confirms the first order decoupling in $\tau_p$ expected from EDD sequences and, interestingly, displays a second order decoupling in $\eps$ as well. While $RGA_{8c}$ appears to address pulse imperfections more effectively that $RGA_{8a}$, its optimality is limited to the regime where $\tau_p>\tau_d$ and $J\tau_p>\eps$. Below, we will illustrate this result by analyzing performance numerically as a function of $\eps$ and $\tau_p$.

The remaining optimal sequences obtained for $K=16,32,64,256$ do not appear to offer any additional robustness. Each sequence is either robust against finite pulse-width or flip-angle errors, but not both forms of pulse errors simultaneously. The effective error Hamiltonians are generally defined as 
\begin{equation}
\H^{a}_{\text{err}}\approx \frac{\tau_p}{\tau_c}\sum_{\mu,\nu}\gamma_{\mu\nu} \,\sigma^{\mu}B_\nu +\O(\eps J\tau_p,\eps J\tau_d)
\end{equation}
for $a$-type sequences and
\begin{equation}
\H^{c}_{\text{err}}\approx \frac{\eps^2}{\tau_c}\sum_{\mu}\zeta_\mu\, \sigma^{\mu} +\O(\eps J\tau_p,\eps J\tau_d)
\end{equation}
for $c$-type sequences, with the worse possible case of $\H^{b}_{\text{err}}=\H^{a}_{\text{err}}+\H^{c}_{\text{err}}$ appearing primarily for $b$-type sequences. Table \ref{tbl:scaling-FWPE} presents all remaining sequences we have located, and outlines the scaling of the effective error Hamiltonians for each sequence.

In summary, our results essentially indicate that attaining robustness for a wide range of $\tau_p$, $\tau_d$, and $\eps$ solely by manipulating the sequence configuration is insufficient. However, this does not invalidate the sequences we have obtained here since if a particular parameter regime is achievable, namely, the strong-pulse regime where flip-angle errors are dominant ($J\tau_p\ll J\tau_d\ll \eps$), sequences such as $RGA_{8a}$, $RGA_{16a}$ and $RGA_{64a}$ still exhibit a high level of robustness. This is not only evident from their ability to suppress errors solely proportional to $\eps$ and $\tau_d$ [see Appendix~\ref{subsec:flipErrors} for a discussion of $RGA_K$ excusively for flip-angle errors], but also from the fact that error accumulation for $\O(J\tau_p)$ terms  is not observed as the number of pulses is increased.
 Therefore, if the pulse duration can be made small relative to the pulse-interval then the finite-width errors are less consequential and it is still possible to maintain some form of robustness without the need for additional techniques. When such a regime is not attainable it may be necessary to utilize pulse shaping techniques to aid flip-angle error robust sequences in the suppression of finite-width pulses, or to exploit composite pulses to suppress flip-angle errors for sequences highly robust to finite-width pulses. Ultimately, a combination of sequence configuration, pulse shaping, and composite pulses is most likely the path forward to constructing fault-tolerant DD sequences for a wide range of parameter regimes.

\subsubsection{Characterization of $RGA_K$ Sequences in $(\eps,J\tau_p)$-space}
In Figure~\ref{fig:GALandscapePlotsFWPE}, the regions of optimal performance for the $RGA_K$ sequences are characterized with respect to magnitude of the flip-angle error $\eps$ and the ratio of the pulse duration $\tau_p$ to pulse delay $\tau_d$. For each $K$, an evident partitioning in the space is observed depending on the form of the prevailing pulse error indicating that it is not possible to combat both forms of pulse imperfections simultaneously by solely manipulating sequence configuration. We find that the a-type sequences, along with $RGA_2$, are more effective against flip-angle errors and offer increasing performance for $K=16,32,64,256$ as $\tau_p/\tau_d\rightarrow0$. [See Appendix~\ref{subsec:flipErrors} for implications and details regarding this result.]  Upon reaching $\tau_p=\tau_d$, the optimal sequence structure is highly dependent upon the relationship between $J\tau_p$ and $\eps$. Sequences of a-type continue to be the preferred structure for $\eps>J\tau_p$, while b-type and additionally defined (c-type) sequences are optimal when $\eps<J\tau_p$. These result are essentially illustrations of the analysis given in the previous section where sequence effectiveness is described in terms of the effective error Hamiltonian.

\section{Comparison with Known Deterministic Schemes}
\label{sec:5}

\label{sec:CDDandQDDCompare}
In this section, $GA_K$ and $RGA_K$ optimal sequences are compared to two known deterministic DD schemes: CDD and QDD, for each pulse profile. 

CDD represents a fair comparison to the numerically optimal sequences since in both cases the pulse-intervals are fixed and the error suppression properties are dictated only by the sequence structure. Schemes which rely on optimized pulse delays to gain decoupling efficiency, such as QDD (see below), 
have much better scaling and error suppression properties than fixed delay schemes in the ideal pulse limit. However, the $GA_K$ optimal sequences can be expected to prevail in the case of non-ideal pulse profiles, where no robust version of QDD currently exists. Optimal control theory can also be used to generate robust DD sequences \cite{Tabuchi:12}.

The foundation of single-qubit CDD rests on the universal decoupling group (modulo irrelevant phases)
\begin{equation}
\mathcal{S}=\{I,X,Y,Z\},
\label{eq:DDgroup}
\end{equation}
 i.e, the single-qubit Pauli group, which implements first order error suppression by symmetrizing $H_{\text{err}}$ over $\mathcal{S}$. Higher order error suppression can be achieved by continuing the symmetrization recursively, such that each additional level averages out the leading term in the Magnus expansion of each sub-level. At the $r$th level of recursion, 
\begin{eqnarray}
\text{CDD}_{r}&=&\Yb\,\text{CDD}_{r-1}\Xb\,\text{CDD}_{r-1}\Yb\, \text{CDD}_{r-1}X\, \text{CDD}_{r-1}\quad\quad
\end{eqnarray}
where CDD$_{0}=f_{\tau_d}$ and the first level of symmetrization, $r=1$, corresponds to $RGA_4'$ with $\{P_1,P_2\}=\{X,Y\}$:
\begin{eqnarray}
CDD_1&=&RGA_{4'}\nonumber\\
&=&(Yf_{\tau_d}Y)(Zf_{\tau_d}Z)(Xf_{\tau_d}X)(If_{\tau_d}I).
\end{eqnarray}
Note that $\mathcal{S}$-based CDD requires $4^r$ pulses to accomplish $r$th order error suppression \cite{KhodjastehLidar-CDD:07}.

\begin{figure*}[t]
\subfigure[]{\includegraphics[width=0.417\textwidth]{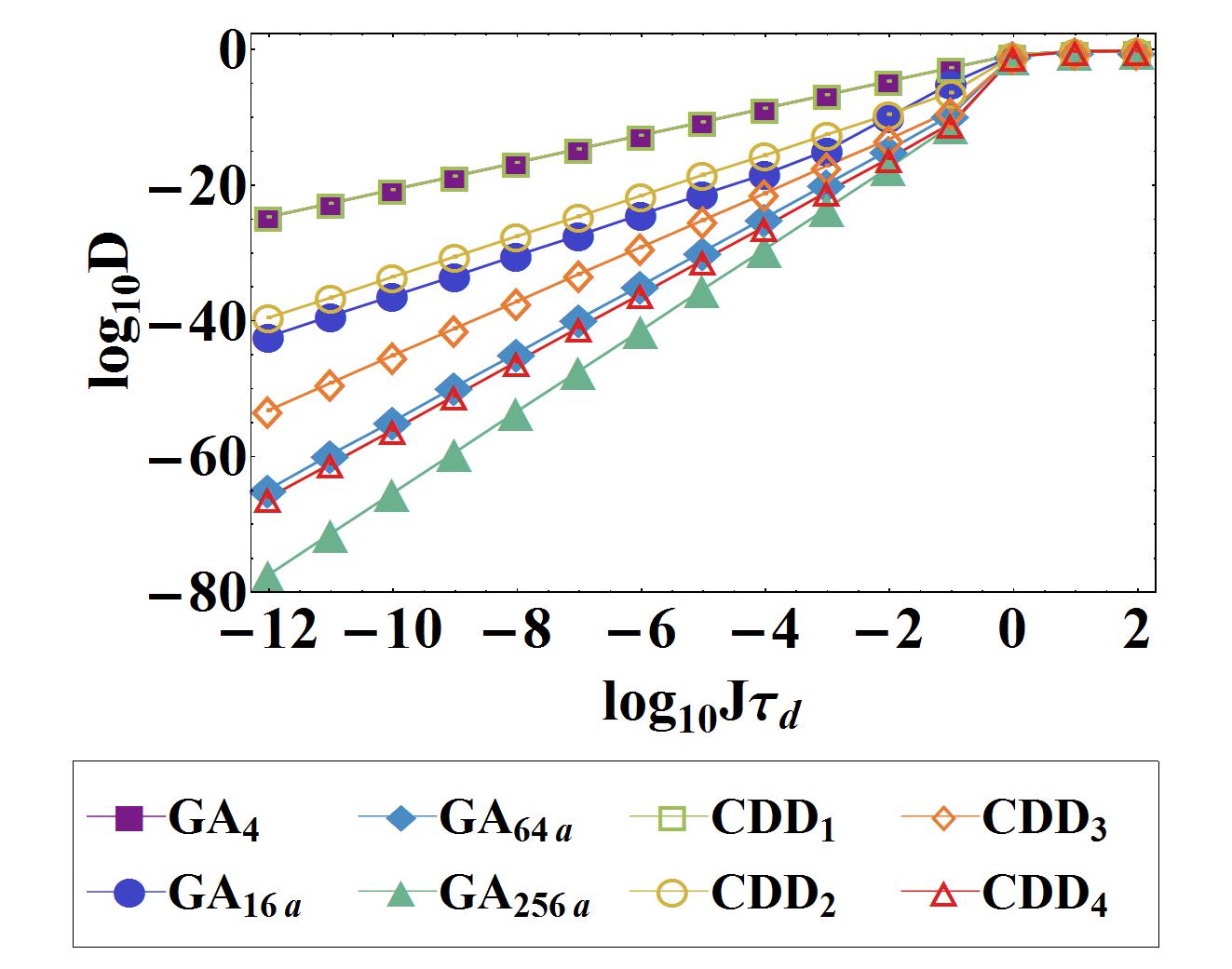}}
\subfigure[]{\includegraphics[width=0.417\textwidth]{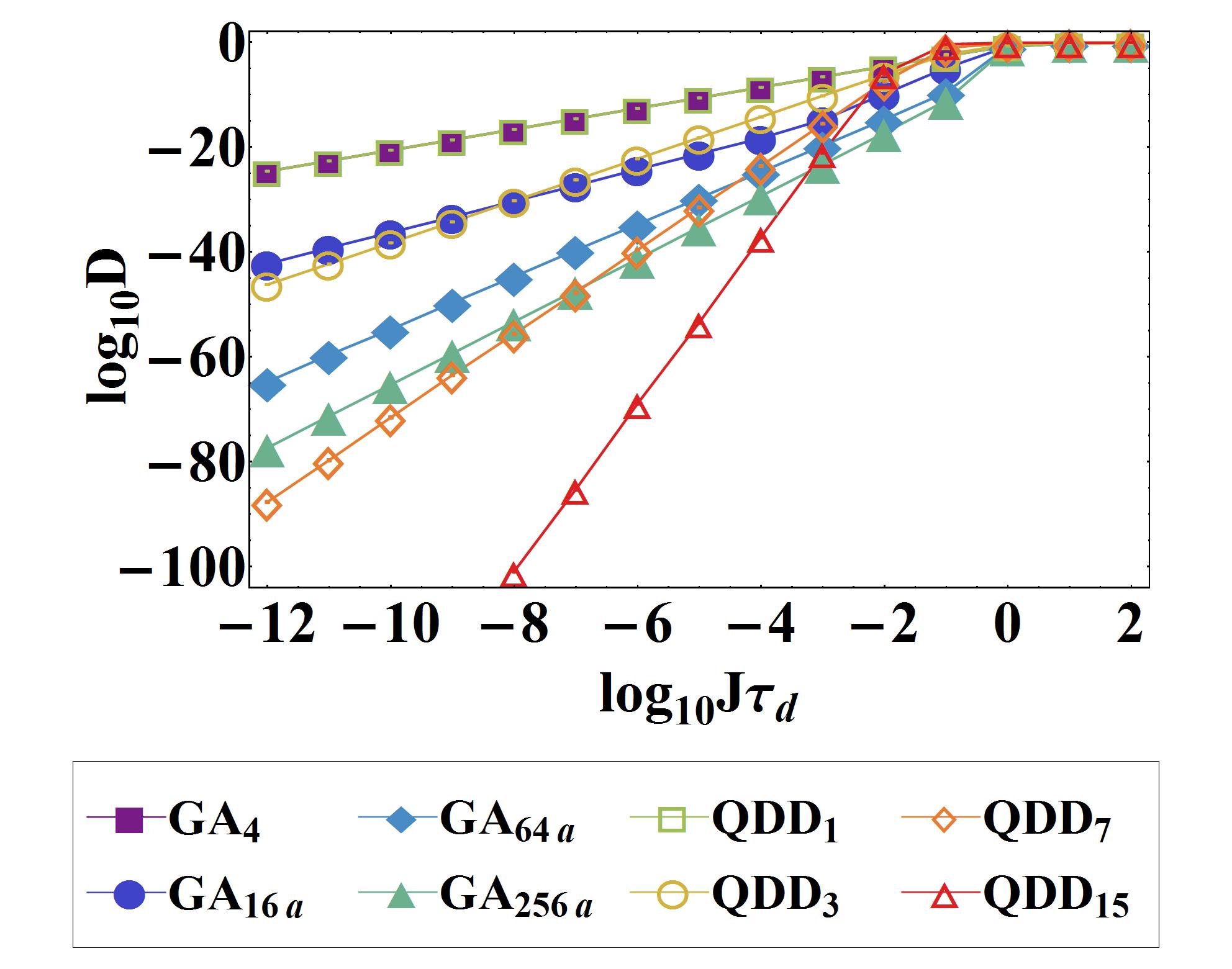}}
\caption{Comparison of performance for (a) CDD (empty symbols) and (b) QDD (empty symbols) versus $GA_K$ (filled symbols) as a function of the pulse-interval $\tau_d$ in the ideal pulse limit. The strength of the error Hamiltonian is chosen as $J=1$MHz and the strength of the pure-environment dynamics $\beta=1$kHz. Optimal GA sequences achieve a higher of order error suppression than CDD as the number of pulses increases; note $GA_{64a}$ and $GA_{256a}$ as compared to CDD$_3$ and CDD$_4$, respectively. In contrast, QDD outperforms $GA_K$ for all sequence lengths, consistent with the expected superiority of interval-optimized schemes in the ideal pulse limit.}
\label{fig:GAvsCDDvsQDD}
\end{figure*}

A large body of work now exists concerning DD sequences with non-uniform pulse-intervals, a technique which enables 
a drastic improvement over the exponential scaling of CDD.
Uhrig DD (UDD) is one such method, which applies DD pulses 
separated by unequal time intervals, at instances determined
by a closed form expression originally developed in the context of the spin-boson model \cite{UDD:07}. 
Using $N$ control pulses, 
UDD 
suppresses the first $N$ orders of the time-dependent perturbation theory 
expansion along the directions which do not commute with the pulses, provided
the bath spectral density contains a sharp high-frequency cutoff \cite{UDD:07,PasiniUhrig:10}, or for 
generic bounded bath operators \cite{YangLiuUDD:08}. For an analysis of the scaling properties and corresponding distance measure performance of UDD see Ref.~\cite{Uhrig:2010:012301}. UDD can be extended to combat general single-qubit decoherence by nesting two anti-commuting UDD sequences. 
The resulting quadratic DD (QDD) scheme
\cite{WestFongLidar-QDD:10}, 
achieves $\min(M_1,M_2)$th order decoupling using $M_1\times M_2$ pulses in the ideal pulse limit, where $M_j$, $j=1,2$, refers to the number of pulses in each nested UDD sequence. The decoupling efficiency of QDD has been extensively studied and confirmed numerically \cite{QuirozLidarQDD:11} and analytically \cite{LidarKuoQDD:11}.

Most relevant to us, because it decouples a single qubit from a general system-bath interaction, is the QDD sequence, generated by nesting two UDD sequences as
\begin{eqnarray}
\text{QDD}_{M_1,M_2}&=&\Gamma^{M_2+1}_2\prod^{M_2+1}_{j=1}\Gamma_2\,\text{UDD}^{\Gamma_1}_{M_1}(\lambda^{(M_2)}_j\tau_d)\nonumber\\
&=&\text{UDD}^{\Gamma_2}_{M_2}[\text{UDD}^{\Gamma_1}_{M_1}(\tau_d)],
\end{eqnarray}
where $\Gamma_1\neq\Gamma_2$ are the generators of $\mathcal{S}$ and
\begin{equation}
\text{UDD}^{\Gamma}_{M}(\tau_d)=\Gamma^{M+1}\prod^{M+1}_{k=1}\Gamma f_{\lambda^{(M)}_k\tau_d}
\end{equation}
The free evolution periods for each UDD sequence are dictated by the normalized pulse-intervals 
\begin{equation}
\lambda^{(M)}_k=\frac{t^{(M)}_k-t^{(M)}_{k-1}}{t^{(M)}_1-t^{(M)}_0},
\end{equation} 
where
\begin{equation}
t^{(M)}_k=\tau_c\sin^2\left(\frac{k\pi}{2M+2}\right),\quad j=1,2,\ldots,M+1,
\end{equation}
and (minimum) pulse delay $\tau_d$. 
In the following analysis, we will focus on $M_j=M$, $j=1,2$, to account for the effectively uniform decoherence model of Eq.~(\ref{eq:qubitHe}) and arbitrarily choose $\Gamma_j\in\{Z,X\}$ as the generators of $\mathcal{S}$.

\subsection{Ideal Pulses}
In Figure~\ref{fig:GAvsCDDvsQDD}(a), we compare the performance of CDD$_l$ to GA$_K$ with respect to $J\tau_d\in[10^{-12},10^2]$ for $J=1$MHz and $\beta=1$kHz in the ideal pulse limit. Numerically optimal sequences first coincide with CDD$_l$ at $K=16$, where both achieve second order error suppression. The main advantage of $GA_{16a}$ over CDD$_2$ is a reduction in the error amplitude by a factor of approximately $10^3$. Significant improvement in error suppression is observed for $K=64$ and $K=256$, where GA optimal sequences offer an additional order of error suppression over corresponding CDD sequences, $r=3,4$. The results indicate that CDD does not constitute an optimal deterministic sequence structure for fixed pulse delays. In Sec.~\ref{sec:6}, we elaborate on this fact and explore the possibility of designing a deterministic scheme to correctly describe the optimal GA sequences.

Pulse interval optimized sequences have been shown to far surpass the decoupling efficiency of any known fixed pulse-interval scheme in the ideal pulse limit, requiring only a quadratic increase in the number of pulses to suppress an additional order of the Magnus expansion. We validate the above statement by comparing QDD$_M$ to the optimal GA sequences in Fig.~\ref{fig:GAvsCDDvsQDD}(b) for $M=1,3,7,15$; an equivalent number of pulses to $K=4,16,64,256$, respectively. For each set of sequence orders $\{M,M\}$ QDD$_M$ attains a decoupling order of $M$, which is beyond the ability of the GA$_K$ for an equivalent number of pulses. QDD$_M$ superiority is an indication that optimizing with respect to pulse-interval, in addition to pulse configuration, is necessary to obtain higher decoupling efficiency. However, these results depend heavily on the ideal pulse limit. As we show in the next subsection, the story is very different when finite width and flip-angle error effects are accounted for.

\subsection{Finite-width and Flip-angle Errors}
As a final comparison, we examine the performance and robustness of $RGA_K$, CDD$_r$, and QDD$_M$ in the presence of both finite-width and flip-angle pulse errors. In particular, we focus on the performance as a function of $\eps$ and $\tau_p$ relative to the size of the pulse-interval $\tau_d=0.1$ns. The flip-angle error is varied from zero to a $15\%$ rotation error, the ratio of pulse duration to interpulse delay $\tau_p/\tau_d\in[10^{-5},10^4]$, and $J=1$MHz with $\beta=1$kHz. 
We average over $10$ realizations of $B_\mu$.

\begin{figure*}[t]
\centering
\includegraphics[width=0.8\linewidth]{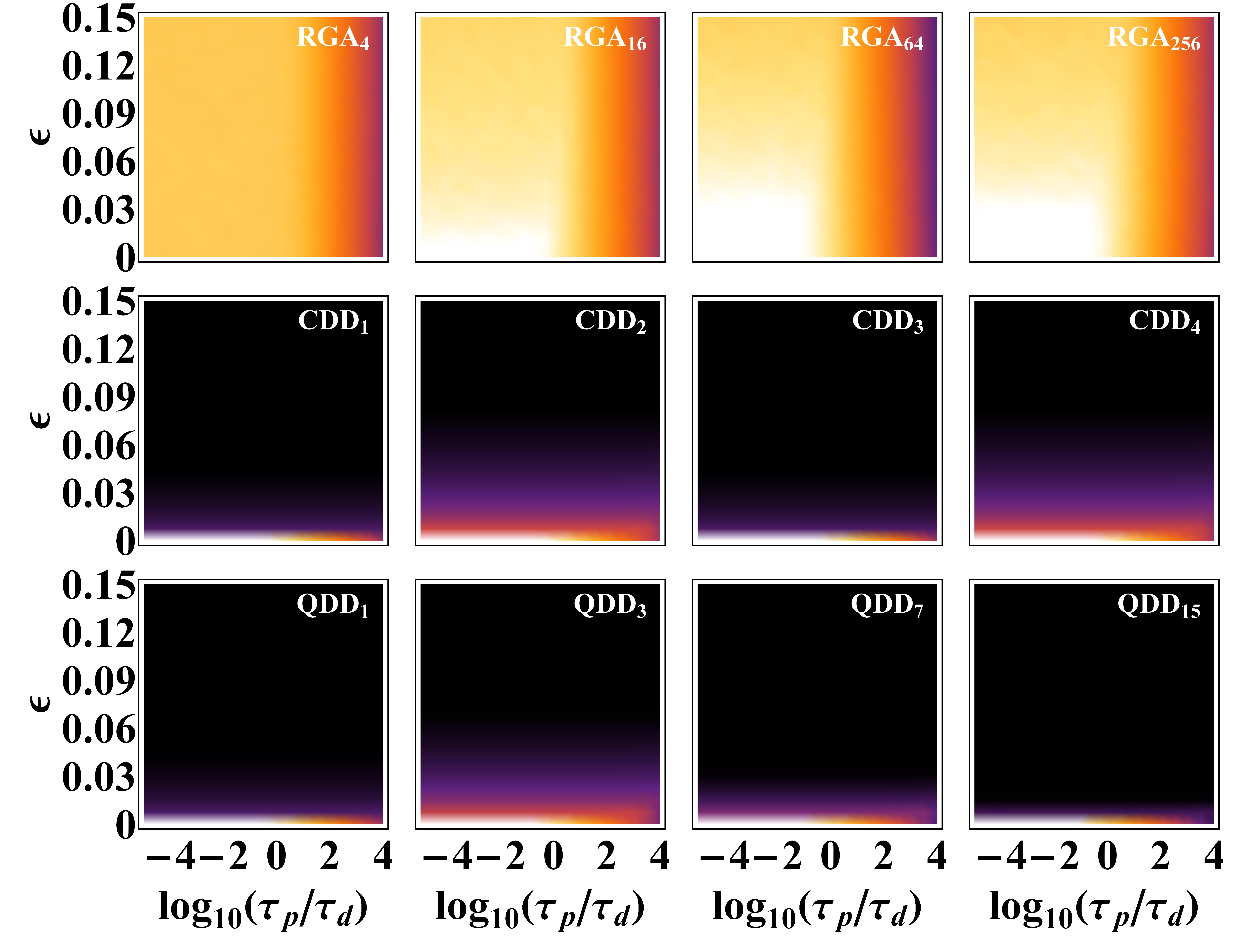}
\raisebox{0.18\height}{\centering\includegraphics[scale=0.09]{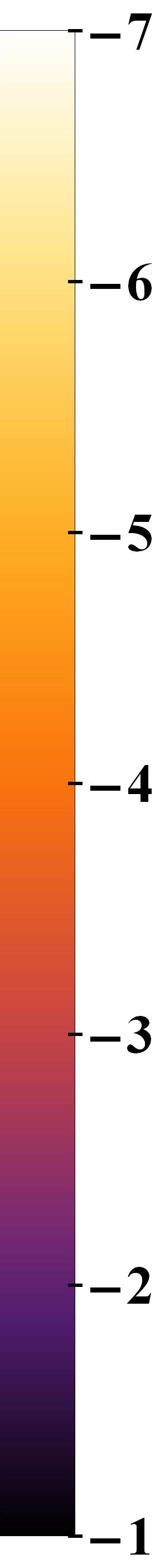}}
\caption{Performance of $RGA_K$, CDD$_r$, and QDD$_M$ when subjected to finite pulse duration and flip-angle errors. The pulse-interval is fixed at $\tau_d=0.1$ns, while $J=1$MHz and $\beta=1$kHz. $RGA_K$ sequences significantly outperform both CDD$_r$ and QDD$_M$
for $K=4,16,64,256$. The most notable region of robustness exists for $\eps<0.04$ 
and $\tau_p<\tau_d$ for $K=16,64,256$.}
\label{fig:GAvsQDDvsCDDFWPE}
\end{figure*}

The performance of $RGA_K$ is shown in Fig.~\ref{fig:GAvsQDDvsCDDFWPE} in panels (a)-(d) for $K=4,16a,64a,256a$, respectively. Although additional optimal configurations were identified at other sequence lengths, we focus on these particular values of $K$ to compare them directly to CDD$_r$, $r=1,2,3,4$, and QDD$_M$, $M=1,3,7,15$. $RGA_K$ performance is found to be primarily dependent upon $\tau_p$, only showing significant $\eps$-dependence when $\eps\geq0.04$. Below $\eps=0.04$, specifically in the region where $\tau_p<\tau_d$, robustness increases as the number of pulses is increased from $K=16a$ to $K=256a$. Error accumulation within this particular range of values appears not to be an issue. CDD$_r$ performance is displayed in panels (e)-(h) of Fig.~\ref{fig:GAvsQDDvsCDDFWPE} for $r=1,2,3,4$, respectively. In contrast to $RGA_K$, CDD$_r$ performance exhibits a strong dependence on flip-angle errors rather than finite pulse duration. Optimal performance is heavily concentrated around small values of $\eps$ due to the low level of robustness exhibited by CDD$_r$ for flip-angle errors. Robustness to finite pulse duration appears to be most noticable for $r=2,4$, although performance is still 
rather poor compared to $RGA_K$.

Lastly, we examine QDD$_M$ in panels (i)-(l) in Fig.~\ref{fig:GAvsQDDvsCDDFWPE}, where $M=1,3,7,15$, respectively. The lowest sequence order, $M=1$, generates the exact same sequence as $RGA_4$, therefore, performance is identical. The remaining sequence orders result in continual error accumulation, which is evident from the steady decline in performance from $M=3$ to $M=15$. As in the case of CDD$_r$, QDD$_M$ performance is primarily $\eps$-dependent and degrades rapidly with increasing $\eps$. Similarly to the additional comparisons of numerically optimized sequences and deterministic schemes for faulty DD pulses given above, $RGA_K$ sequences significantly outperform CDD$_r$ and QDD$_M$.

\section{Existence of Deterministic Structure}
\label{sec:6}

\begin{figure}[t]
\centering
\includegraphics[width=\columnwidth]{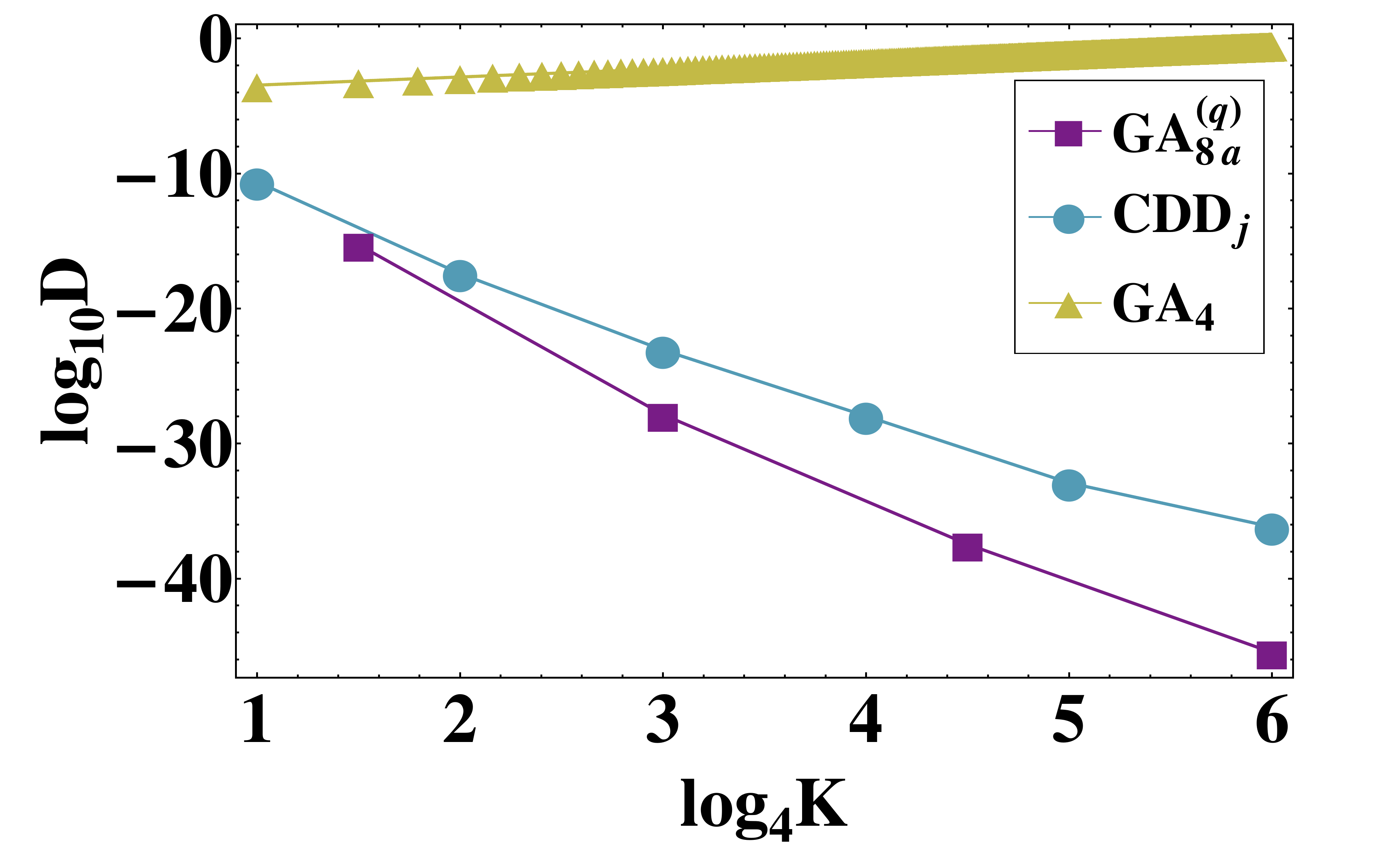}
\caption{Comparison of $GA^{(q)}_{8a}$, CDD$_r$, and $XY_4$ performance after one cycle as a function of the number of pulses $K$ for ideal, zero-width pulses. The strength of the error Hamiltonian and environment dynamics are set to $J=1$MHz and $\beta=1$kHz, respectively, and the minimum pulse-interval $\tau_d=0.1$ns. Results are averaged over 25 random realizations of $B_{\mu}$, where the error bars are shown, but quite small. As expected by the results of the GA search $GA^{(q)}_{8a}$, $q=1,\ldots,4$ is indeed superior to CDD$_r$, $r=1,\ldots,6$, and $XY_4$}
\label{fig:CXY8vsCDD}
\end{figure}

Concatenation appears to play an important role in the construction of optimal DD in the case of fixed pulse-intervals. One of the goals of this study is to determine whether the optimal sequences provided above can be generalized into a deterministic concatenation scheme for arbitrary order decoupling. As suggested by the ideal pulse analysis summarized in Table~\ref{tbl:DScalingIdeal}, such a scheme is possible by utilizing $GA_{8a}$ as the fundamental unit of concatenation in
\begin{equation}
GA^{(q)}_{8a}=GA_{8a}[GA^{(q-1)}_{8a}],
\label{eq:CXY8}
\end{equation}
where $GA^{(0)}_{8a}\equiv f_{\tau_d}$. Requiring $8^{q}$ pulses, $GA^{(q)}_{8a}$ achieves $2q$th order error suppression; a quadratic improvement over the decoupling efficiency of CDD, which requires $4^{2q}$ pulses to achieve an equivalent decoupling order. The increased decoupling efficiency is facilitated by second order error suppression provided by $GA_{8a}$, which essentially boosts the efficiency by a factor of 2 at each level of concatenation.

In Fig.~\ref{fig:CXY8vsCDD}, we compare the performance of $GA^{(q)}_{8a}$ to that of CDD$_r$ and $GA_4$ in the case of zero-width pulses as a function of the number of pulses $K$. The results are averaged over 25 random realizations of $B_{\mu}$, where the pulse operators are designated by $\{P_1,P_2\}=\{X,Y\}$ for each generalized sequence. The strengths of the error Hamiltonian and bath dynamics are given by $J=1$MHz and $\beta=1$kHz, respectively, and the minimum pulse delay $\tau_d=0.1$ns. In comparing $q=1,2,\ldots,4$ and $l=1,2,\ldots,6$, the performance of $GA^{(q)}_{8a}$ improves dramatically as the level of concatenation increases, far exceeding that of CDD$_r$ and $GA_4$.

{However, the truly meaningful test of sequence performance is in the presence of pulse errors.} Defining 
\begin{equation}
RGA^{(q)}_{8a}=RGA_{8a}[RGA^{(q-1)}_{8a}],
\end{equation}
to effectively combat the inclusion of flip-angle and finite-width pulse errors, we compare the performance of $RGA^{(q)}_{8a}$ to CDD$_r$ and $RGA_4$ in Fig.~\ref{fig:CXY8vsCDDFWPE} for $\{P_1,P_2\}=\{Y,X\}$. The relevant Hamiltonian parameters are equivalent to those chosen for the ideal case, while the flip-angle error $\eps=0.01$. As opposed to fixing $\tau_d$, we select a fixed cycle time $\tau_c$ to analyze the relationship between $\tau_p$ and $\tau_d$ as the number of pulses grows. In particular, we consider $\tau_c=1$ns and $\tau_p/\tau_c=10^{-10}$. We expect robustness to be 
most noticeable in the strong pulse regime, $\tau_p\gg\tau_d$, where the primary form of pulse error is due to the flip-angle errors. Performing the analysis with a fixed cycle time allows us to examine robustness as a function of concatenation level and as 
$\tau_p\rightarrow\tau_d$, simultaneously. As expected by direct calculation of the effective Hamiltonian for $RGA_4$ [see Eq.~(\ref{eq:RGA4Heff_FWPE})], increasing the number of pulses via multiple DD cycles does not offer an enhancement in sequence performance due to an immediate accumulation of error proportional to $\tau_p$ and $\eps^2$. {In contrast, CDD$_r$ performance remains fairly consistent and oscillates between two values, seemingly dependent on the parity of the concatenation level.} In Refs.~\cite{KhodjastehLidar-CDD:07,NLP:11}, a similar study of CDD performance conveyed that effective cancellation of pulse-width errors occurs for a range of concatenation levels if the pulse width is much smaller than the {pulse-interval; eventual saturation in performance as $r$ increases occurs once this condition is violated. The performance characteristics of CDD$_r$ differ here due to the presence of flip-angle errors, which were not accounted for in Refs.~\cite{KhodjastehLidar-CDD:07,NLP:11}. Combined errors are most effectively addressed by $RGA^{(q)}_{8a}$, as can be seen by the improvements in sequence performance as concatenation level increases. Note that the performance eventually begins to show signs of saturation as the concatenation level increases. Essentially, the pulse-interval is approaching a value comparable to the pulse duration which leads to finite-width errors becoming a more significant decoherence mechanism than flip-angle errors or the error Hamiltonian. $RGA_{8a}$ and its concatenated versions do not provide protection against finite-width errors, and therefore their performance becomes hindered by the presence of terms that are first order in $\tau_p$.

\begin{figure}[t]
\centering
\includegraphics[width=\columnwidth]{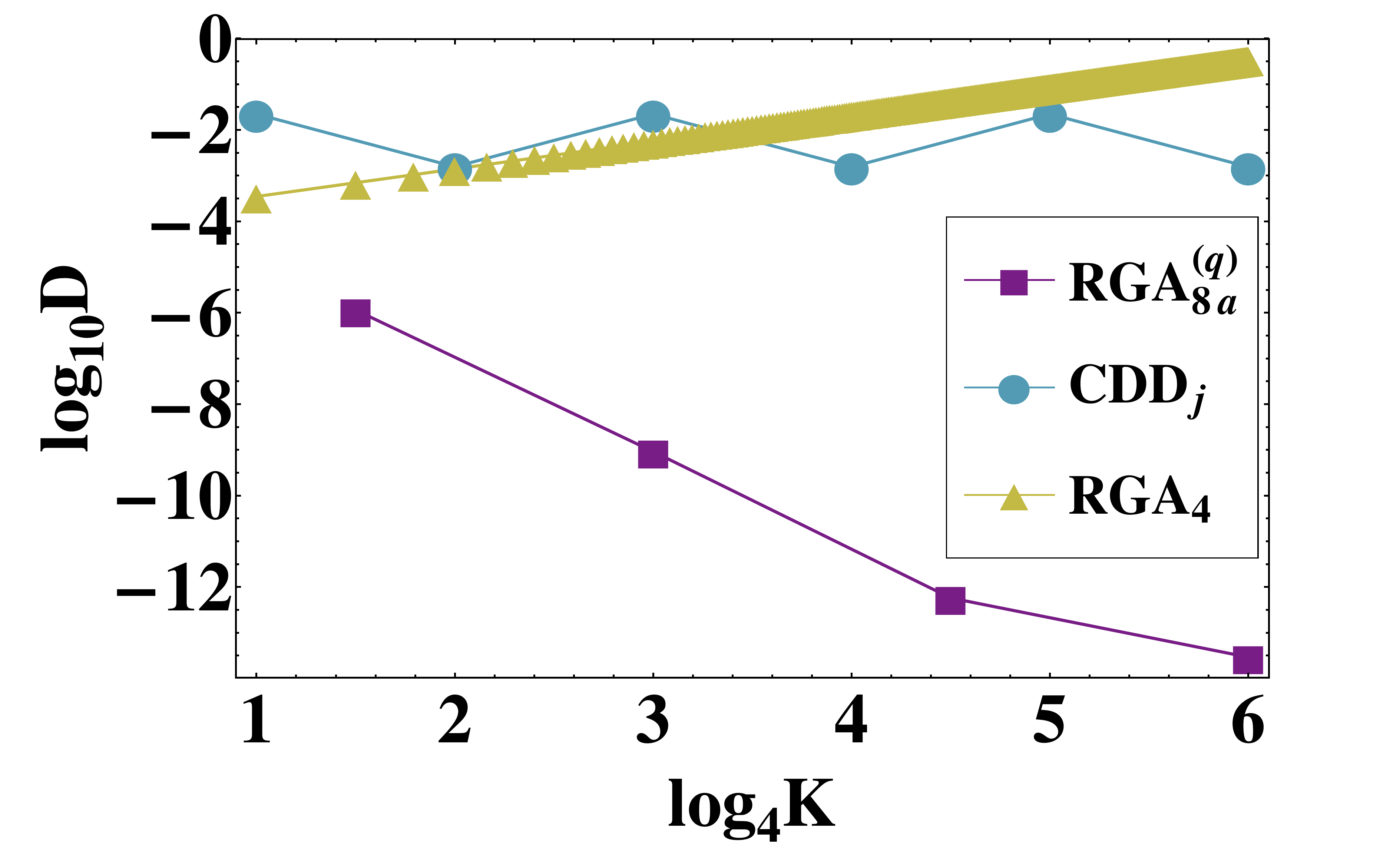}
\caption{(color online) Comparison of $RGA^{(q)}_{8a}$, CDD$_r$, and 
RGA$_4$ performance versus the number of pulses $K$ for combined pulse errors (flip-angle and finite-width) after one cycle. Hamiltonian parameters, $J$ and $\beta$, are 
the same as those in Fig.~\ref{fig:CXY8vsCDD} with the total cycle time fixed at $\tau_c=1$ns, as opposed to $\tau_d$. Results are averaged over 25 realizations of $B_{\mu}$ with $\eps=0.01$ and $\tau_p/\tau_c=10^{-10}$. The performance of $RGA^{(q)}_{8a}$ improves as the number of pulses is increased within a given cycle time $\tau_c$. For the specified parameters, CDD$_r$ does not exhibit enhanced performance with increasing concatenation level.}
\label{fig:CXY8vsCDDFWPE}
\end{figure}

\section{Conclusions}
\label{sec:7}

In 
this work we showed that numerically optimal DD sequences can be constructed using a genetic algorithm in conjunction with a simulated annealing convergence accelerator and a novel complexity-reduction technique. The search focused on sequences containing $K=1,2,\ldots,256$ pulses, however, we identified optimal performance at $K=4,8,16,32,64,256$ and compared each sequence to known deterministic schemes, such as CDD and QDD. The ideal-pulse analysis showed that, under the constraint of a fixed pulse interval, optimal sequences can be constructed which outperform CDD, yet fall short of the decoupling efficiency realized by QDD. Optimization proved to be quite beneficial in the case of finite-width and flip-angle errors, where  numerically optimal sequences obtained a level of robustness that could not be reached by either 
CDD or QDD.

The culmination of our study is centered around the identification of a deterministic sequence structure that obtains a high decoupling efficiency in the ideal limit and robustness to errors generated by pulse imperfections. We determined that $RGA_{8a}$ [Eq.~\eqref{eq:RGA_8a}] is the 
favored generating sequence for a majority of the pulse profiles considered. Concatenating this particular sequence, it is possible to suppress the first $2q$ terms in the Magnus expansion using $8^q$ pulses. Compared to the $4^q$ 
pulses required for the widely used original version of CDD \cite{KhodjastehLidar:04}, the concatenated version of 
$RGA_{8a}$
utilizes quadratically fewer pulses to obtain the same decoupling order. Although it is not possible to obtain such high levels of decoupling in the presence of pulse imperfections, 
$RGA_{8a}$ contains an inherent robustness that continues to aid the error suppression process as the level of concatenation increases. This result is most apparent in our final study of faulty DD pulses, which includes both finite duration and flip-angle errors; we found the 
concatenated 
$RGA_{8a}$ construction 
to be the most robust scheme available for fixed pulse-interval DD sequences.

The importance of optimizing over pulse-intervals was clearly displayed in the ideal pulse analysis, where sequence configuration optimization alone could not supply the decoupling efficiency achieved by QDD. However, in agreement with previous work, we have shown that pulse-interval optimized sequences fail to be robust against additional errors generated by faulty DD pulses. Future work should focus on
extending the search algorithm to incorporate multi-qubit systems and unequal pulse delays, to obtain robust DD sequences in the presence of various pulse errors. 

\section{Acknowledgements}
We thank Dr. Joaquin Cerda Boluda for his 
valuable contribution towards constructing a highly efficient GA algorithm. We would also like to thank Gerardo Paz Silva, Gonzalo A. Alvarez, Dieter Suter, and Wan-Jung Kuo for useful discussions and valuable input.
This research was supported by the ARO MURI grant
W911NF-11-1-0268, by the Department of Defense, by the
Intelligence Advanced Research Projects Activity (IARPA)
via Department of Interior National Business Center contract
number D11PC20165, and by NSF grants No. CHE-924318
and CHE-1037992. The U.S. Government is authorized to
reproduce and distribute reprints for Governmental purposes
notwithstanding any copyright annotation thereon. The views
and conclusions contained herein are those of the authors and
should not be interpreted as necessarily representing the official policies or endorsements, either expressed or implied, of
IARPA, DoI/NBC, or the U.S. Government.

\appendix

\section{Independence of Optimal Sequences on Bath Hilbert Space Dimension}
\label{appsec:6qubitsearch}
In order to support our claim in Sec.~\ref{sec:errorModel} regarding the independence of optimal sequence configuration on the dimension of the bath Hilbert space, we consider the case of six bath spins and search for ideal pulse optimal sequences specifically for $K=4$. The results obtained from the search are identical to those presented in Sec.~\ref{subsec:idealpulse}, where $RGA_{4}$ is the dominant optimal sequence. In Fig.~\ref{fig:GA4landscape_6qb}, the results are summarized for $J\tau_d\in[10^{-10},10^3]$ and $\beta\tau_d\in[10^{-10},10^3]$. Note that, as in the case of four bath spins, $RGA_4$ remains optimal for most values of $J,\beta$.

\begin{figure}[t]
\centering
\includegraphics[width=0.5\textwidth]{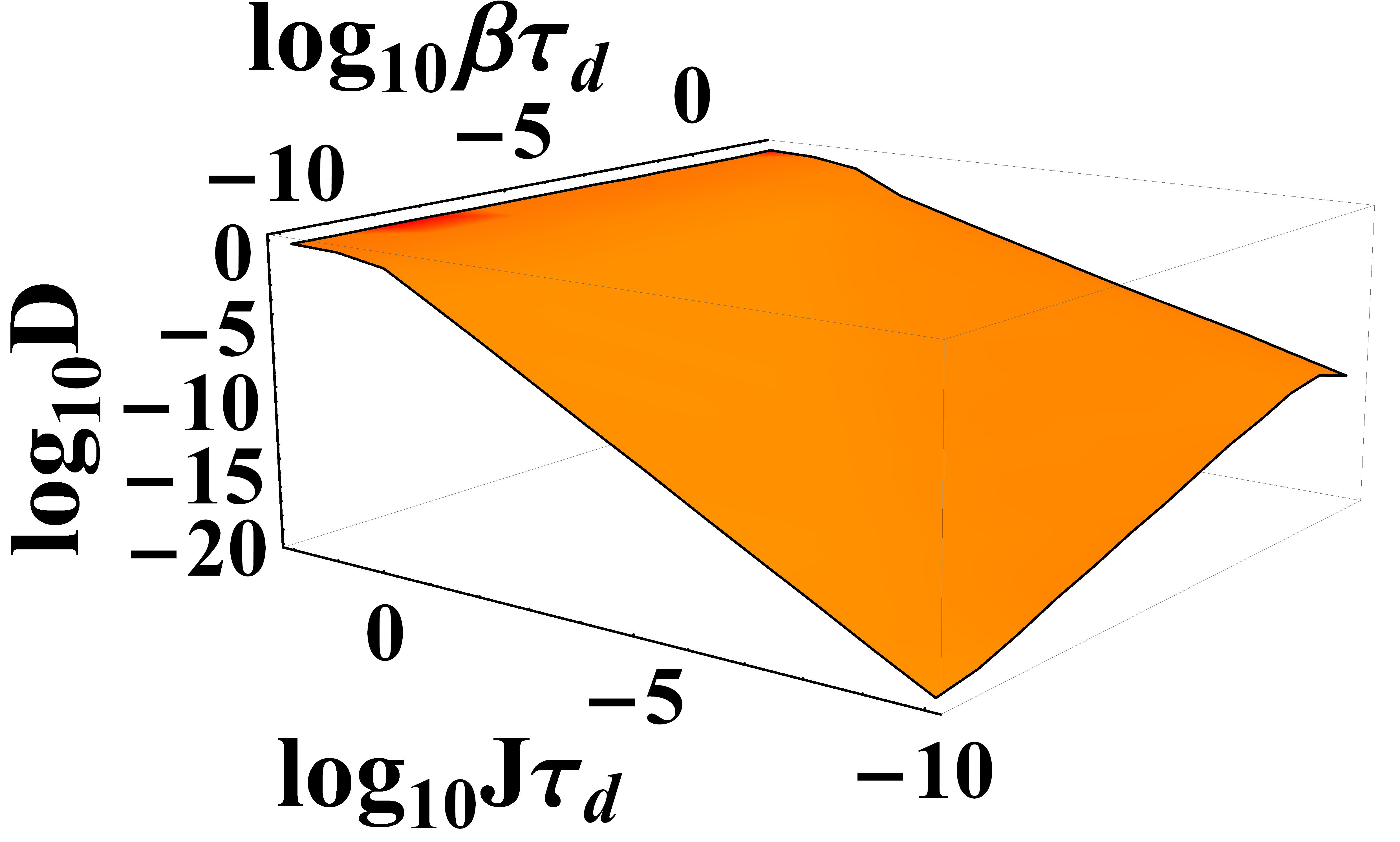}
\caption{Summary of results for ideal pulse GA search for six bath spins. Regions of optimal performance for each sequence are identical to those obtained for the four bath spin case.}
\label{fig:GA4landscape_6qb}
\end{figure}


\section{Scaling of Performance}
\label{appsec:scalingD}
Here, we prove the scaling of distance measure $D(U,I_S)$ described in Eq.~(\ref{eq:scaling}) for the ideal pulse limit. First, we prove that the general distance measure $D(U,G)$ [see Eq.~(\ref{eq:Dist})] satisfies
\begin{equation}
D(U,G)\leq \frac{1}{\sqrt{2}}\|U-G\otimes I_B\|.
\label{eq:DBound}
\end{equation}
The upper-bound is obtained by utilizing the steps originally taken in Ref.~\cite{KGBR:06} to obtain the closed form expression of $D(U,G)$ given in Eq.~(\ref{eq:DistClosedForm}), where initially it is shown that, for $\Phi$ satisfying $\Phi^\dagger \Phi = I_B$,
\begin{eqnarray}
D(U,G)&=&\frac{1}{\sqrt{2d_Sd_B}}\min_\Phi\|U-G\otimes \Phi\|_F\nonumber\\
&=&\min_\Phi \sqrt{1-\frac{1}{d_S d_B}\text{Re}\{\text{Tr}\left[U(G^{\dagger}\otimes \Phi^{\dagger})\right]\}}\nonumber\\
&=&\min_\Phi \sqrt{ 1-\frac{1}{d_S d_B}\text{Re}\{\text{Tr}_B\left[\text{Tr}_S[U (G^{\dagger}\otimes I_B)]\Phi^{\dagger}\right]\}}\nonumber\\
&=&\min_\Phi \sqrt{1-\frac{1}{d_S d_B}\text{Re}[\text{Tr}(\Gamma\Phi^{\dagger})]}.
\label{eq:Dmin}
\end{eqnarray}
It is then noticed that computing the minimization problem of Eq.~(\ref{eq:Dist}) is equivalent to finding the maximum value of $\text{Re}[\text{Tr}(\Gamma\Phi^{\dagger})]$ over all unitary $\Phi$. In order to complete the proof, the singular value decomposition (SVD) $\Gamma=W \Sigma V^{\dagger}$ is invoked, where $W,V$ are unitary and $\Sigma=\text{diag}(s_1,\ldots,s_{d_B})$ is a real diagonal matrix containing the singular values $s_1\geq s_2\geq\cdots\geq s_{d_B}{\geq 0}$. The relevant expression
becomes
\begin{equation}
\text{Tr}(\Gamma\Phi^{\dagger})=\text{Tr}(W\Sigma V^{\dagger}\Phi^{\dagger})=\text{Tr}[\Sigma (V^{\dagger}\Phi^{\dagger}W)]
\end{equation}
and the final details of the proof involve showing that $\text{Re}\{\text{Tr}[\Sigma (V^{\dagger}\Phi^{\dagger}W)]\}$ is essentially maximized if and only if $V^{\dagger}\Phi^{\dagger}W=I_B$, or equivalently when $\Phi=WV^{\dagger}$. Note that 
\begin{equation}
\text{Re}\{\text{Tr}[\Sigma (V^{\dagger}\Phi^{\dagger}W)]\}\leq\text{Tr}(\Sigma)
\label{eq:GBound}
\end{equation}
holds for all $\Phi$ that differ from $WV^{\dagger}$.

It is 
now straightforward to obtain the bound expressed in Eq.~(\ref{eq:DBound}) since by choosing $\Phi=I_B$ we are satisfying the lower bound expression of Eq.~(\ref{eq:GBound}) and generating the upper-bound 
\begin{eqnarray}
\label{eq:FinalBound1}
D(U,G)&\leq&\frac{1}{\sqrt{2 d_S d_B}}\|U-G\otimes I_B\|_F\\
&\leq&\frac{1}{\sqrt{2}}\|U-G\otimes I_B\|.
\label{eq:FinalBound2}
\end{eqnarray}
Note that we have used the inequality $\|T\|_F\leq \sqrt{d_Sd_B}\|T\|$ in order to transform from the Frobenius norm [Eq.~(\ref{eq:FinalBound1})] to the sup-operator norm [Eq.~(\ref{eq:FinalBound2})].

Choosing $G=I_S$, as specified by the desired action of the DD evolution on the system, and the toggling frame evolution operator $U=\tilde{U}_0(\tau_c)$, the distance measure upper-bound becomes
\begin{eqnarray}
D\leq\frac{1}{\sqrt{2}}\|\tilde{U}_0(\tau_c)-I_{S}\otimes I_B\|.
\end{eqnarray}
Expressing the unitary evolution operator $\tilde{U}_0(\tau_c)$ as a time-dependent perturbation expansion
\begin{equation}
\tilde{U}_0(\tau_c)=I_{S}\otimes I_B+\sum^{\infty}_{n=1}\tilde{U}^{(n)}_0(\tau_c),
\end{equation}
with Dyson operators
\begin{equation}
\tilde{U}^{(n)}_0(\tau_c)=
(-i)^n\int^{\tau_c}_{0}dt_1\cdots\int^{t_{n-1}}_{0}dt_n\prod^{n}_{j=1}\tilde{H}_0(t_j),
\label{eq:DysonOps}
\end{equation}
we obtain, using the triangle inequality,
\begin{eqnarray}
D\leq\frac{1}{\sqrt{2}}\|\sum^{\infty}_{n=1}\tilde{U}^{(n)}_0(\tau_c)\|
{\leq}\frac{1}{\sqrt{2}}\sum^{\infty}_{n=1}\|\tilde{U}^{(n)}_0(\tau_c)\| .
\label{eq:DUn}
\end{eqnarray}
The final step is to show that $\|U_n(\tau_c)\|$ obtains the scaling claimed by Eq.~(\ref{eq:scaling}). This is accomplished by using $(1)$ the triangle inequality, $(2)$ sub-multiplicativity, and $(3)$ unitary invariance to obtain an upper-bound on Eq.~(\ref{eq:DysonOps}) as follows:
\begin{eqnarray}
\|\tilde{U}^{(n)}_0(\tau_c)\|&=&\|
(-i)^n\int^{\tau_c}_{0}dt_1\cdots\int^{t_{n-1}}_{0}dt_n\prod^{n}_{j=1}\tilde{H}_0(t_j)\|\nonumber\\
&\stackrel{(1)}{\leq}&
\int^{\tau_c}_{0}dt_1\cdots\int^{t_{n-1}}_{0}dt_n\|\prod^{n}_{j=1}\tilde{H}_0(t_j)\|\nonumber\\
&\stackrel{(2)}{\leq}&
\int^{\tau_c}_{0}dt_1\cdots\int^{t_{n-1}}_{0}dt_n\prod^{n}_{j=1}\|\tilde{H}_0(t_j)\|\nonumber\\
&\stackrel{(3)}{=}&
\int^{\tau_c}_{0}dt_1\cdots\int^{t_{n-1}}_{0}dt_n\prod^{n}_{j=1}\|H_0\|\nonumber\\
&=&\frac{\tau^{n}_c}{n!}
\|H_0\|^n
\stackrel{(1)}{\leq}
\frac{\tau^{n}_c}{n!}(J+\beta)^n.
\end{eqnarray}
As a result, Eq.~(\ref{eq:DUn}) achieves the upper-bound
\begin{equation}
D\leq\frac{1}{\sqrt{2}}\sum^{\infty}_{n=1}\frac{\tau^{n}_c}{
n!}(J+\beta)^n =\frac{1}{\sqrt{2}}\left(e^{\tau_c(J+\beta)}-1\right).
\end{equation}
Let us now consider the case of $N$th order error suppression, where all $\tilde{U}^{(n)}_0(\tau_c)$ for $n\leq N$ 
vanish. 
The bound then truncates to
\begin{equation}
D\leq\frac{1}{\sqrt{2}}\sum^{\infty}_{n=N+1}\frac{\tau^{n}_c}{
{n!}}(J+\beta)^n,
\end{equation}
which scalings accordingly as
\begin{equation}
D\lesssim\O\left[\tau^{N+1}_c(J+\beta)^{N+1}\right],
\end{equation}
when $J\tau_c\ll1$ and $\beta\tau_c\ll1$ are satisfied.

\section{Algorithm}
\label{appsec:algorithm}

Genetic Algorithms represent an approach to optimization problems based on the properties of natural evolution. Given an initial population and a definition of fitness, the algorithm simulates the processes of selection, reproduction, and mutation in an attempt to locate the member in the population with the highest probability of survival. In regards to DD, the population can be thought of as a subset of all possible sequence configurations, where a configuration is specified by the order and types of pulses, for a given sequence length $K$. 
The member with the highest probability of survival 
is the sequence which maximally suppresses system-bath interactions with respect to a particular distance measure. In the following subsections we outline the representation of the population and discuss how selection, reproduction, and mutations are implemented in the setting of DD optimization.

\subsection{Chromosome structure}
The canonical approach to GAs is to define a member of the population by a set of genes, loosely referred to as a chromosome. Each gene can be thought of as a parameter in the optimization problem which contributes in some way to the fitness, and therefore the probability of selection, of the member. Defining a member in the population as a DD sequence, Eq.~(\ref{eq:seq}) is translated directly into its corresponding chromosome
\begin{equation}
C^{(\alpha)}_{j}=\{P_1,P_2,\ldots,P_{K-1},P_{K}\},
\label{eq:chrom1}
\end{equation}
representing the $j$th member in the $\alpha$th generation. The genes are given by the pulses in the sequence, therefore the number of genes increases with increasing sequence length. In general, a sequence and its corresponding chromosome do not have to be structurally equivalent. Later we will elaborate on why the naive translation of Eq.~(\ref{eq:chrom1}) is not favorable for DD optimization and discuss how it can be refined; however, for now Eq.~(\ref{eq:chrom1}) is adequate to describe each aspect of the algorithm outlined in the subsequent subsections.

The population is given by the set of chromosomes $\{C^{(\alpha)}_{j}\}^{Q}_{j=1}$, where each $C^{(\alpha)}_{j}$ corresponds to a sequence $U^{(\alpha)}_{j}(\tau_c)$ and $Q$ is the population size. The total number of possible sequence configurations, $\mathcal{N}(K)$, is determined by both the length of the sequence and the number of pulse types in $\G$. The size of the sequence space grows exponentially with the length of the sequence, $\mathcal{N}(K)=|\G|^K$, where $|\G|$ is  the number elements in $\G$.

The search space can be reduced by imposing the cyclic DD condition $U_C(\tau_c)=I_S$, which is applicable for our focus on quantum memory preservation. The condition can be recast in the context of the search problem as
\begin{equation}
\prod^{K}_{j=1}P^{\text{ideal}}_j\propto I_S
\label{eq:DDcond}
\end{equation}
on all $C^{(\alpha)}_{j}$, where only the ideal, zero-width version of the pulse is used when finite-width or flip-angle error pulse profiles define $V_{\mu}(t)$. Applying Eq.~(\ref{eq:DDcond}), the search space is reduced to $\mathcal{N}_{R}(K)=|\G|^{K-1}$, where only $|\G|^{-1}$ of the original search space accounts for viable DD sequences.

The initial population is chosen at random from the reduced search space, such that $Q\ll \mathcal{N}_R(K)$. In general, the size of $Q$ is somewhat arbitrary and expected to vary depending on the number of degrees of freedom specified by the problem. In the context of DD optimization, the size of the initial population will ultimately end up fixed for all sequence lengths due to the structure of the initial chromosomes; see Section \ref{subsubsec:ComplexReduce} for additional details.

\subsection{Selection}
Associated with each chromosome $C^{(\alpha)}_{j}$ is a selection probability $p^{(\alpha)}_j$. This quantity defines the probability of being selected for reproduction in generation $\alpha$ and is given by 
\begin{equation}
p^{(\alpha)}_j=\frac{q^{(\alpha)}_j}{\sum_{i}q^{(\alpha)}_i},
\label{eq:probselect}
\end{equation} 
where $q^{(\alpha)}_j$$=$$-\log_{10}D^{(\alpha)}_j$ represents the performance, or fitness, of the $j$th sequence. Here, we impose the cyclic DD condition as well, $G=I_S$, and denote $D^{(\alpha)}_j\equiv D^{(\alpha)}_j(U(\tau_c),I_S)$. The logarithm is included in the definition of the fitness due to complications with the selection probability that are attributed to the extreme sensitivity of Eq.~(\ref{eq:Dist}), and any distance measure for that matter, to sequence variations. The exchange of a single pulse in a sequence with any other member of the decoupling set can result in a change in performance up to many orders of magnitude. Since the reduced search space does not eliminate all poorly performing sequences, the fitness can vary greatly in any generation. As a result, there is a reduced contribution of high performance sequences in the selection probability distribution. The logarithm counteracts this issue by increasing the resolution of the selection probability.

\subsection{Crossover}
In each generation $2Q$ offspring are produced from the current population. Members of the population are chosen for reproduction based on their probability of selection. The selection process is constrained such that 
the crossover procedure only occurs between two distinct members of the population. Members with a high probability of selection not only possess a higher likelihood of reproduction, but also have a higher probability of reproducing with multiple members in a single generation since each crossover is an independent event.

Reproduction is implemented by a crossover between two members in the population, yielding two offspring. To best illustrate the crossover, consider the two chromosomes
\begin{eqnarray}
C^{(\alpha)}_j&=&\{P_1,\ldots,P_i,\ldots,P_k\},\\
C^{(\alpha)}_{j'}&=&\{R_1,\ldots,R_i,\ldots,R_k\},
\label{eq:crossover_I}
\end{eqnarray}
where $P_i,R_i\in\G$. The offspring are created by splicing the parent chromosomes at a location chosen at random, where each pulse location has an equal probability of being chosen. Taking the splice point to be the $i$th pulse site, the resulting offspring are
\begin{eqnarray}
\tilde{C}^{(\alpha)}_{j}&=&\{P_1,\ldots,P_i,R_{i+1},\ldots,R_k\},\\
\tilde{C}^{(\alpha)}_{j'}&=&\{R_1,\ldots,R_i,P_{i+1},\ldots,P_k\}.
\label{eq:crossover_F}
\end{eqnarray}
It is essential that the offspring still satisfy Eq.~(\ref{eq:DDcond}), however it is not necessarily true that each is guarenteed to do so. If the DD condition is not satisfied, the pulse located at the splice point is manipulated until the condition is satisfied. For example, if $\tilde{C}^{(\alpha)}_{j}$ does not fulfill the DD condition then it is transformed to
\begin{equation}
\tilde{\tilde{C}}^{(\alpha)}_{j}=\{P_1,\ldots,P'_i,R_{i+1},\ldots,R_k\},
\end{equation}
where it now is in agreement with Eq.~(\ref{eq:DDcond}) and $P'_i\in\G$. In the situation that $\tilde{\tilde{C}}^{(\alpha)}_{j}$ cannot be found, the splice point is chosen again and the process is repeated until the proper offspring are created.

The condition set forth by Eq.~(\ref{eq:DDcond}) restricts the crossover process and in some cases does not allow it at all. By permitting the manipulation of the pulse at the splice point, it is ensured that only the probability of selection dictates reproduction. It is always possible to construct offspring from the above process, since there is no constraint on yielding offspring which are identical to the parent chromosomes. Thus, every set of parent chromosomes is guaranteed to produce some form of offspring.

Upon producing the $2Q$ offspring, the best $Q/4$ parents and $3Q/4$ offspring are taken to be the new population. The partitioning was chosen based on what appeared to be the most beneficial to the convergence of the algorithm. No duplicate sequences are allowed in the new population, however if the updated population size is less than $Q$ then new members are generated at random from within the reduced search space.

\subsection{Mutation}
After reproduction, the new population composed of $Q/4$ parents and $3Q/4$ offspring is used to create $2Q$ mutated sequences, $Q$ single-site and $Q$ double-site. Every sequence in the population participates in both mutation processes, however only a portion of the mutated sequences is retained for the succeeding generation.

Single-site mutations are performed by choosing a pulse site at random and altering the pulse until Eq.~(\ref{eq:DDcond}) is again satisfied. If the DD condition is unsatisfiable then the original pulse is replaced and a different pulse site is chosen. It is possible that only the original configuration satisfies Eq.~(\ref{eq:DDcond}). In this situation the mutated member is simply a duplicate sequence, therefore it is discarded.

Double-site mutations correspond to linked single-site mutations. The process begins in a similar manner by choosing a pulse site at random, say the $i$th site with pulse $P_i$. An additional pulse site is now chosen at random from the set of pulse sites which have pulse types equivalent to $P_i$, e.g., the $j$th site. Both $P_i$ and $P_j$ are updated simultaneously until the DD condition is again satisfied. If an additional pulse site does not exist, then the initial site is re-selected and the double-site mutation process is repeated. As in the case of the single-site mutation, if the DD condition cannot be satisfied then the mutated sequence is accepted as the original configuration and discarded.

At the conclusion of the two mutations, a portion of the parent, offspring, and mutated sequences will comprise the new population. Only the sequences that have the highest fitness with respect to Eq.~(\ref{eq:Dist}) are desired from each division of the population. We find that the best $Q/8$ parent, $5Q/8$ offspring, $Q/8$ single-site mutated, and $Q/8$ double-site mutated sequences comprise a favorable distribution for the new population. Other distributions were considered such as taking the best $Q/4$ of all mutated sequences, as well as different proportions of the offspring. However, no distribution appeared to yield a higher probability of optimal sequence convergence.

\subsection{Necessary Convergence Accelerators}
As noted above, single-site perturbations may result in large deviations in sequence performance. Hence, the logarithm was introduced to decrease the performance gap between poor- and well-performing sequences, thereby increasing the resolution of the selection probability. However, this adjustment only proves to aid in optimal convergence for sequences comprised of $K<16$ pulses. This is evident from a simple comparison between CDD and numerically located sequences at $K=16,64,256$, where the numerically ``optimal" sequences perform far worse than CDD.

We suspect that the local minima convergence is ultimately attributed to significant deviations in sequence performance that result in relatively large local minima traps. We alleviate this complication by introducing two convergence accelerators which act to reduce the size and presence of large traps, thereby smoothening what we refer to as the fitness landscape. Both accelerators are crucial for the algorithm to converge on global optima as the number of pulses increases beyond $K=16$.

\subsubsection{Reducing Local Traps via Complexity}
\label{subsubsec:ComplexReduce}
Although the size of $\mathcal{N}(K)$ is decreased by imposing the cyclic DD condition, the resulting reduced search space, $\mathcal{N}_R(K)$, maintains its exponential scaling in the number of control pulses. Hence, there is still a high probability of the subspace containing low-performance sequences that lead to large local traps. This issue is resolved by reducing the search space further and systematically increasing its size as the algorithm iterates, such that the search space at the termination of the algorithm is $\mathcal{N}_R(K)$. The additional reduction is achieved by constraining the complexity of the chromosome, thereby moderating the possible sequence configurations.

\begin{figure}[t]
\centering
\includegraphics[width=\columnwidth]{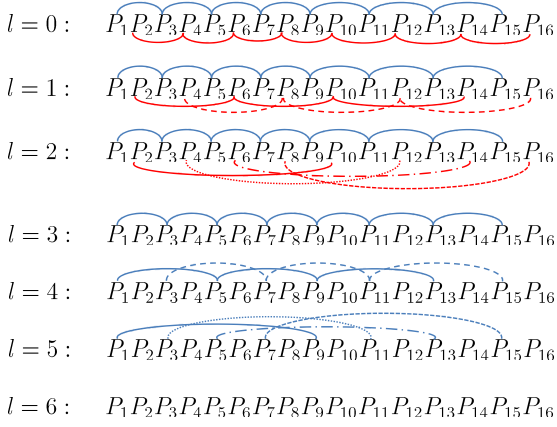}
\vspace{-.75cm}
\caption{Complexity-reduction protocol for $K=16$. The upper (blue) curves denote the linking between odd pulse sites and the lower (red) curves denote even pulse site linking. The process begins with all odd and even pulses linked, then continues by removing links between every other even pulse site. Links are removed until all even sites are uncorrelated, thereafter the odd sites undergo the same process.}
\label{fig:Linking}
\end{figure}

Initially, we choose each chromosome to represent the most elementary two-dimensional sequence
\begin{equation}
C^{(l=0,\alpha=0)}_{j}=\{P_{s_1},P_{s_2}\},
\end{equation}
where $l$$=$$0$ is the initial complexity index. The notation $P_{s_j}$ denotes a pulse $P_j$ applied at the locations specified by the set $s_j$. We choose $s_1$ and $s_2$ to contain only the odd and even pulse sites, respectively, for the initial population. This is a 
relevant construction since all known deterministic DD schemes utilizing fixed free intervals contain the same pulse at either every even or odd site \cite{XY4:69,KhodjastehLidar-CDD:07}. Moreover, it conveniently reduces the space to only $|\G|^2$ sequence configurations for all $K$.

For a general complexity index $l$, the chromosome is defined as
\begin{equation}
C^{(l,\alpha)}_{j}=\{P_{s_1},P_{s_2},\ldots,P_{s_{\tilde{K}(l)}}\},
\label{eq:chrom2}
\end{equation}
where we require that
\begin{equation}
\bigcup^{\tilde{K}(l)}_{i=1}s_i=\text{all sites} \quad \text{and} \quad \bigcap^{\tilde{K}(l)}_{i=1}s_i=\emptyset
\end{equation}
be satisfied so that only one control pulse is applied at each pulse site. The number of sets $\{s_j\}$ is determined by
\begin{equation}
\tilde{K}(l)=\left\{
\begin{array}{cc}
\frac{3}{2}(l+\frac{4}{3}) :&  l\,\text{even}\\
\\
\frac{3}{2}(l+1):& l\,\text{odd}
\end{array}\right.
\end{equation}
for $l=0,1,2,\ldots,l_{\max}$. At maximum complexity, $l_{\max}$, the most general sequence within $\mathcal{N}_R(K)$ is permitted. Hence, each $s_j$ is a single element set containing only the $j$th pulse site. An example of the complexity-increase procedure is illustrated in Figure \ref{fig:Linking} for $K=16$. Note that at each level of complexity-increase we have chosen to remove constraints only pertaining to odd or even sites. The constraint between every other even site is removed until each even pulse site is independent, after which the same is performed on the odd sites.

In contrast to Eq.~(\ref{eq:chrom1}), the number of elements in Eq.~(\ref{eq:chrom2}) increases as the algorithm iterates. It is important to note that this aspect does not imply an increase in the number of pulses, rather an increase in the permissible search space. This is an attractive feature since it not only diminishes the presence of local traps, but also yields an initial set of sequence configurations which only scales quadratically in $|\G|$.
Choosing sequences for the initial population is obviously much more favorable here since the space is drastically smaller than $\mathcal{N}_R(K)$. In principle, it may even be possible to choose the entire set as the initial population. For a single-qubit system subjected to ideal $\pi$-pulses, we find that the complete initial set of configurations is indeed computationally convenient, consisting of only 16 possible configurations. Other pulse profiles lead to larger initial sets, but, remarkably, optimal sequence convergence is possible for initial populations of only 16 sequences.

\subsubsection{Fitness annealing}
Substantial differences in sequence fitness manifest local traps in the fitness landscape. By decreasing the complexity of the chromosome only the probability of generating local traps is diminished. In order to control the relative differences between high- and low-performance sequences, we introduce an annealing process into the selection probability. Adopted from Ref. \cite{PetiteauGAPaper:10}, the selection probability is redefined as
\begin{equation}
p^{(l,\alpha)}_j(T)=\frac{\tilde{q}^{(l,\alpha)}_j(T)}{\sum_i \tilde{q}^{(l,\alpha)}_i(T)},
\end{equation}
[compare with Eq.~(\ref{eq:probselect})] such that
\begin{equation}
\tilde{q}^{(l,\alpha)}_j(T)=\exp\left(\frac{q^{(\alpha)}_j-q^{(\alpha)}_{\text{best}}}{T(\alpha)}\right).
\label{eq:TempFit}
\end{equation}
The performance of the most fit member in the $\alpha$th generation is denoted by $q^{(\alpha)}_{\text{best}}$ and the temperature function is given by
\begin{equation}
T(\alpha)=T_{0}\left(\frac{T_f}{T_{0}}\right)^{\alpha/\alpha_c}\left[1-\eta\sin\left(\frac{\lambda\pi}{\alpha_c}\alpha\right)\right].
\label{eq:Temp}
\end{equation}
The temperature function utilized here is a modified version of the one introduced in Ref.~\cite{PetiteauGAPaper:10}, where we have included the sinusoidal function to reduce the probability of local minima convergence as $T(\alpha)$ decreases from the initial temperature $T_{0}$ to the final temperature $T_f$. The number of generations between these temperatures is dictated by the cutoff generation $\alpha_c$, {which is chosen based on the value of $K$.  For large $K$, we pick $\alpha_c$ to be large as well since the annealing process is to accelerate global minimum convergence while reducing the probability of local minimum convergence.} The remaining parameters $\eta$ and $\lambda$ are related to the amplitude and frequency of the oscillations, respectively. {Upon increasing the complexity index $l$, the annealing process resets with an initial temperature $T_0$ chosen so that all sequences in the current population have an equal likelihood of being chosen for reproduction.}


\section{Extracting effective error Hamiltonian scaling numerically}
\label{appsec:NumericScalingD}
In Sec.~\ref{subsec:idealpulse}, we discussed the scaling of the distance measure $D$ for each optimal sequence in the ideal pulse limit without direct calculation of the effective Hamiltonian. We obtain this scaling by assuming that $D$ has the form
\begin{equation}
D\sim \O(J^{n_{J}}\beta^{n_{\beta}}\tau^{N+1}_{d}),
\label{eq:NumericScalingIdeal}
\end{equation}
where $N$ is the decoupling order of the sequence and $n_{J}+n_{\beta}=N+1$. First, the decoupling order is determined by examining $\log_{10}D$  as a function of $\tau_{d}$, as this quantity scales linearly in $\tau_{d}$ with a slope of $N+1$. The scaling of $\tau_{d}$ is only dependent upon the decoupling order $N$, and therefore is independent of the relative magnitudes of $J$ and $\beta$. The only constraint we consider is $J\tau_{d}\ll 1$ and $\beta \tau_{d}\ll1$ in order to satisfy the condition $\Vert H^{\prime}_{\text{err}}\tau_{d}\Vert\ll 1$ discussed in Sec.~\ref{subsec:DDBackground} for effective error suppression.

The scaling of the remaining quantities, $J$ and $\beta$, is determined for each relevant parameter regime ($J\ll \beta$ and $J\gg \beta$) by analyzing $\log_{10}D$ as a function of $J\tau_{d}$ by varying $J$ for fixed $\tau_{d}$ and $\beta$. The logarithm of the performance {measure} can again be expected to scale linearly in $J\tau_d$, now with a slope of $n_J$. The value of $n_J$ will ultimately depend on the 
{magnitude} of $J$ relative to $\beta$, however in either case $n_J$ is well-defined. The scaling of $\beta$ is now 
determined from $n_J+n_\beta=N+1$, where $n_\beta$ is the only unknown quantity. Note that this method can be easily extended to include finite-width pulses, flip-angle errors, or both, by assuming
\begin{eqnarray}
D&\sim\O(J^{n_J}\beta^{n_\beta}\tau^{n_p}_p\tau^{n_d}_d),\\
D&\sim\O(\eps^{n_\eps}J^{n_J}\beta^{n_\beta}\tau^{N+1}_d),\\
D&\sim\O(\eps^{n_\eps}J^{n_J}\beta^{n_\beta}\tau^{n_p}_p\tau^{n_d}_d),
\end{eqnarray}
respectively, where $n_p+n_d=n_J+n_\beta$.

\section{Additional Results}
\label{appsec:addresults}

\subsection{Finite-width Pulses}
\label{subsec:finitepulse}
In this section, we account for errors excusively due to finite-width rectangular pulses of duration $\tau_p$. We note that in this case EDD is a known way to achieve first order pulse-width error suppression, with the added assumption that pulse shaping is possible \cite{EDD}. Each optimal sequence construction is examined with respect to {$J/\beta\in[10^{-15},10^{3}]$} for $\beta=1$kHz and $\tau_p/\tau_d\in[10^{-6},10^3]$ for $\tau_d=0.1$ns. The set of allowable control pulses is given by $\G=\{I,X,Y,Z,\Xb,\Yb,\Zb\}$, where the unitary pulse operators 
\begin{equation}
X(Y,Z)=e^{-i\tau_p(A\sigma^{x(y,z)}+H_0)},
\label{eq:pulseOperatorFW}
\end{equation}
are generated from Eq.~(\ref{eq:FWpulse}). Throughout the following section we will again enforce the strong pulse assumption [see Sec.~\ref{subsec:FiniteAndFlip}] to calculate the effective error Hamiltonian for various optimal sequences.

\subsubsection{Summary of Numerical Search}
In optimizing over finite-width flip-angle errors, we consider a situation in which both forms of pulse imperfections are essentially equally prevalent. Here, we focus on a case where the flip-angle errors are neglible and the finite-width duration of the pulse completely determines the errors due to pulse imperfections. While one may expect a considerable overlap with the finite-width flip-angle error results, many of the optimal sequences located in this section will differ due to their inability to supply any form of flip-angle error suppression.

The first case we consider is $K=4$, where we find that $RGA_4$ [see Eq.~(\ref{eq:RGA_4})] and $RGA_{4'}$ [see Table \ref{tbl:scaling-FWPE}] are the only optimal sequence configurations. The two sequences result in similar effective error Hamiltonians:
\begin{eqnarray}
\label{eq:HeffFW4}
\H^{RGA_4}_{\text{err}}&\approx&\frac{4\tau_p}{\pi\tau_c}\sz B_{x-y}+\bar{H}^{GA_4}_{\text{err}}\\
\H^{RGA_{4'}}_{\text{err}}&\approx&\frac{4\tau_p}{\pi\tau_c}\left(\sy B_z+\sz B_x\right)+\bar{H}^{GA_4}_{\text{err}},
\label{eq:HeffFW4p}
\end{eqnarray}
which scale linearly in $\tau_p$ and, therefore, do not provide first order error suppression in the pulse duration. Note the difference in error distribution between the effective error Hamiltonians generated simply by reversing the phase of a single pulse. Pulse imperfections generate errors along the $\sz$-channel for $RGA_4$ and along the $\sy$ and $\sz$ channels for $RGA_{4'}$. Depending on the form of the system-environment interaction, the difference in sequence performance could be quite drastic. For example, consider the case of uniform decoherence in the xy-plane ($B_x=B_y$) which results in complete first order decoupling in $\tau_p$ for $RGA_4$ only.

The above results indicate that robustness to pulse imperfections can be extremely sensitive to variations in pulse phases, even in the simplest case of uni-axial pulses. This statement continues to hold true for
\begin{eqnarray}
RGA_{8a^{\prime}}&:=&{I P_1 \Pb_2 P_1 I  P_1 \Pb_2 P_1}\\
RGA_{8b}&:=&RGA_2[RGA_{4}]
\end{eqnarray}
and $RGA_{8c}$ [see Eq.~(\ref{eq:RGA_8c})], the finite-width pulse error-optimized sequences for $K=8$. Although obvious similarities between $RGA_{8a}$ [see Eq.~(\ref{eq:RGA_8a})] and $RGA_{8a'}$ exist, the effect of altering pulse phases is quite significant. $RGA_{8a}$ does not produce first order decoupling in $\tau_p$, while the effective error Hamiltonian for $RGA_{8a'}$,
\begin{equation}
H^{RGA_{8a^{\prime}}}_{\text{err}}\approx\frac{16\tau_d\tau_p}{\pi\tau_c}\sy B^{2}_x-\frac{8\tau_d\tau_p}{\pi\tau_c}\sy\{B_x,B_y\}+\O(J\beta\tau_d\tau_p),
\label{eq:Heff-RGA8ap}
\end{equation}
conveys complete first order suppression in the pulse duration. $RGA_{8a'}$ is the primary optimal sequence located at $K=8$, but is also accompanied by less-robust sequences such as $RGA_{8b}$ and $RGA_{8c}$. $RGA_{8b}$ is a robust form of $GA_{8b}$ [see Eq.~\ref{eq:GA_8a}] whose lack of first order decoupling in $\tau_p$ is particularly favorable when $J\gg\beta$. The remaining sequence, $RGA_{8c}$, denotes a generic version of the Eulerian DD (EDD) sequence and attains first order error suppression in $\tau_p$ by traversing the Cayley graph $\Gamma=\Gamma(\mathcal{S},\G)$, where $\mathcal{S}$ denotes the single-qubit Pauli group with elements denoting the graph vertices and $\G=\{I,X,Y,Z\}$ is the generating set comprising the edges \cite{EDD}. The original construction (captured by $RGA_{8c}$) utilized two closed Eulerian cycles on $\Gamma$ such that the second is completed by returning along the first path. However, additional paths exist which do not require closed cycles to obtain first order suppression in $\tau_p$. One such case is $RGA_{8a^\prime}$, where the initial path and its inversion are both \emph{open} Eulerian paths, as illustrated in Fig.~\ref{fig:path}. In comparing $RGA_{8a^\prime}$ and $RGA_{8c}$, the most important aspect appears to be the manner in which the paths are traversed rather than their closure. 
Note that variations in pulse phases may aid in the error suppression process, but are not necessarily required to obtain first order decoupling in $\tau_p$ as $GA_{8a}$ also perform the task. In terms of performance, $RGA_{8c}$ does not match the second order error suppression in $\tau_d$ found for $RGA_{8a'}$, as indicated by
\begin{eqnarray}
\H^{RGA_{8c}}_{\text{err}}&\approx&-\frac{4\tau_d\tau_p}{\pi\tau_c}\left[\sx (B^{2}_y+\{B_x,B_y\}) \right.\nonumber\\
&&+\left.\sy (B^{2}_x+\{B_x,B_y\})\right]+\O(J^2\tau^{2}_d).
\label{eq:Heff-FW8c}
\end{eqnarray}
This attribute of $RGA_{8c}$ is ultimately the cause for the overall advantageous performance of $RGA_{8a'}$.

\begin{figure}
\centering
\vspace{-0.4cm}
\includegraphics[scale=0.25]{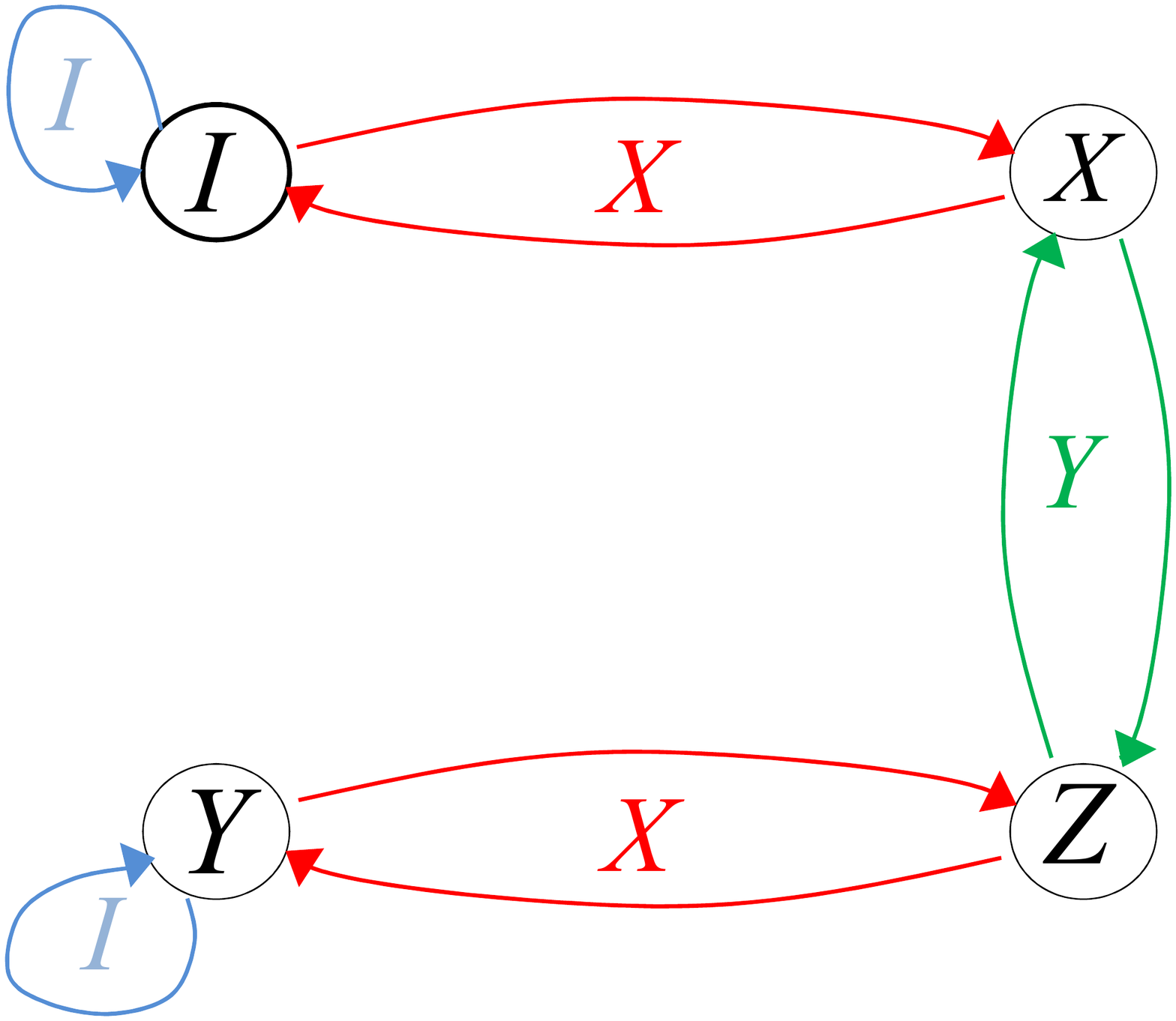}\\
\includegraphics[scale=0.15]{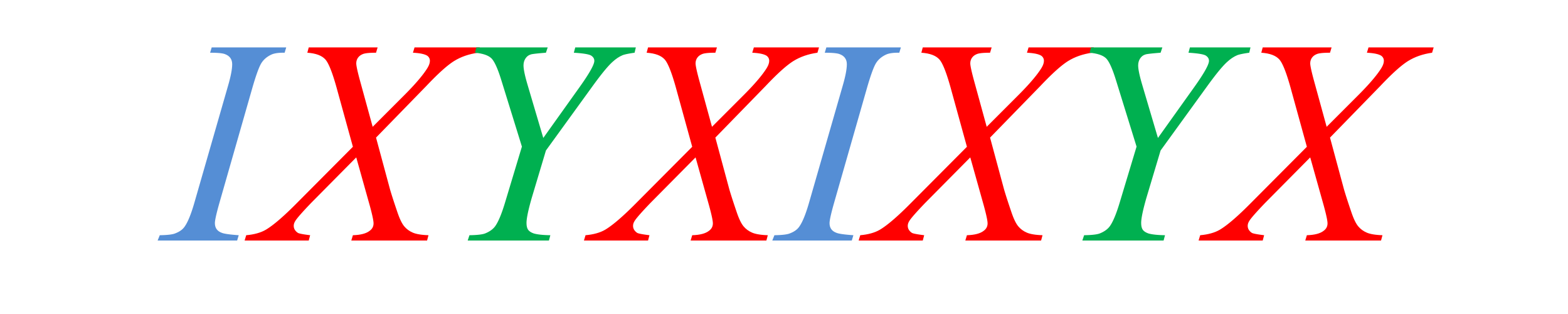}
\caption{Pictorial depiction for the action of $RGA_{8a^\prime}$ as an Eulerian path along the Cayley graph $\Gamma(\mathcal{S},\mathcal{G})$ with vertices $\mathcal{S}=\{I,X,Y,Z\}$ and generating set $\mathcal{G}=\{I,X,Y,Z\}$. Note that unlike the EDD construction [Eq.~(\ref{eq:RGA_8c})], $RGA_{8a^\prime}$ is generated by Eulerian paths, rather than cycles.}
\label{fig:path}
\end{figure}

Beyond $K=8$, optimal sequence configurations are generally characterized by two specific sequences:
\begin{eqnarray}
RGA_{16a'}&:=&P_3(RGA_{8a'})P_3(RGA_{8a'}),\\
RGA_{64c}&:=&RGA_{8c}[RGA_{8c}].
\end{eqnarray}
The former emerges at $K=16,32,64,256$, where cycles of $RGA_{16a'}$ are utilized to generate the correct number of pulses for each corresponding $K$ value. Additional sequences such as $RGA_{16b'}:=RGA_4[RGA_{4'}]$ and $RGA_{32c}:=RGA_{8c}[RGA_4]$ appear at $K=16$ and $K=32$, respectively, yet neither generate the effective symmetrization of error along all three decoherence channels achieved by $RGA_{16a'}$. Effective dynamics for $RGA_{16b'}$ and $RGA_{32c}$ are essentially described by
\begin{eqnarray}
\H^{RGA_{16a^\prime}}_{\text{err}}&\approx& \frac{8i\tau_d\tau_p}{\tau_c}\left(\sx[B_0,B_x-\frac{2}{\pi}B_z]\right.\nonumber\\
&&+\left.\sy[B_0,B_y-\frac{2}{\pi}B_z]+\sz[B_0,B_z+\frac{2}{\pi}B_y]\right)\nonumber\\
\label{eq:Heff-16apFW}
\end{eqnarray}
with an additional error term of $\O(J\beta\tau_d\tau_p)$ along one of the three decoherence channels. The final sequence described above, $RGA_{64c}$, produces a similar effective error Hamiltonian to $RGA_{16a'}$,
\begin{eqnarray}
\H^{RGA_{64c}}_{\text{err}}&\approx&\frac{32i\tau_d\tau_p}{\tau_c}\left(\sx [B_0,B_x+\frac{2}{\pi}B_y]\right.\nonumber\\
&&+\left.\sy[B_0,B_y-\frac{2}{\pi}B_z]+\sz[B_0,B_z+\frac{2}{\pi}B_y]\right),\nonumber\\
\label{eq:Heff-64cFW}
\end{eqnarray}
and further suppresses errors of $\O(J^2\tau^{2}_p)$ to overtake $RGA_{16a'}$ as an optimal sequence for $K=64,256$ in the system-environment interaction-dominant ($J>\beta$) regime.

\begin{table*}[t]
\centering
\begin{tabular}{cccc}
\multicolumn{2}{c}{Sequence} & \multirow{2}{*}{$\tau_p\ll \tau_d$} & \multirow{2}{*}{$\tau_p\gg \tau_d$}\\ \cline{1-2}

Name & Description &  & \\ \hline

$RGA_4$ & $\Pb_2 P_1 \Pb_2 P1$ & $\bb{\O(J\beta\tau^{2}_d,J^2\tau^{2}_d)}$ & $\bb{\O(J\tau_p)}$ \\
$RGA_{4'}$ & $\Pb_2 \Pb_1 \Pb_2 P1$ &  $\O(J\beta\tau^{2}_d,J^2\tau^{2}_d)$ &  $\O(J\tau_p)$\\ \hline

$RGA_{8a'}$ & $I P_1 \Pb_2 P_1 I P_1 \Pb_2 P_1$ & $\bb{\O(J\beta\tau_d\tau_p,J^2\tau_d\tau_p)}$ & $\bb{\O(J\beta\tau^{2}_p,J^2\tau^{2}_p)}$\\ 
$RGA_{8b}$ & $RGA_2[RGA_4]$ & $\O(J\tau_p)$ & $\O(J\tau_p)$\\
$RGA_{8c}$ & $P_1 P_2 P_1 P_2 P_2 P_1 P_2 P_1$ & $\O(J\beta\tau^{2}_d,J^2\tau^{2}_d)$ & $\O(J\beta\tau^{2}_p,J^2\tau^{2}_p)$\\ \hline

$RGA_{16a'}$ & $P_3(RGA_{8a'}) P_3 (RGA_{8a'})$ & $\bb{\O(J\beta\tau_d\tau_p)}$ & $\bb{\O(J\beta\tau_d\tau_p,J^2\tau^{2}_p)}$\\ 
$RGA_{16b'}$ & $RGA_{4}[RGA_{4'}]$ & $\O(J\beta\tau_d\tau_p)$ & $\O(J\beta\tau_d\tau_p,J^2\tau^{2}_p)$\\ \hline

$RGA_{32c}$ & $RGA_{8c}[RGA_4]$ & $\O(J\beta\tau_d\tau_p)$ & $\bb{\O(J\beta\tau^{2}_p)}$\\ \hline

$RGA_{64c}$ & $RGA_{8c}[RGA_{8c}]$ & $\O(J\beta\tau_d\tau_p)$ & $\bb{\O(J\beta\tau^{2}_p)}$\\ \hline

$RGA_{256c}$ & $RGA_{4}[RGA_{64c}]$ & $\O(J\tau_p)$ & $\O(J\tau_p)$\\

\end{tabular}
\caption{Summary of distance measure ($D$) scalings for each optimal $RGA_K$ sequence located by our search algorithm, for DD evolution subjected to finite-width rectangular pulses of duration $\tau_p$ pulses and pulse-interval $\tau_d$. Optimal performance scalings for each $K_{\text{opt}}$ are boxed for each parameter regime (column).}
\label{tbl:scaling-FW}
\end{table*}

A summary of the performance scaling equations for all optimal sequences discussed above is presented in Table~\ref{tbl:scaling-FW}. Note that first order error suppression, in $\tau_p$, is 
achieved for a majority of $K_{\text{opt}}$. However, we are only able to demonstrate the reduction of second order decoherence operators, such as the suppression of $\O(J^2\tau_d\tau_p)$ terms for certain cases, and not complete suppression of $\O(\tau_p\tau_d)$ or $\O(\tau^{2}_p)$ terms. This result is consistent with DD no-go theorems which prove that it is not possible to suppress decoherence operators that are manifested by the second order perturbation expansion for the pulse error evolution operator, i.e. $\O(\tau_p\tau_d)$ and $\O(\tau^{2}_p)$ terms, when rectangular pulse profiles are utilized \cite{PasiniUhrig-NoGo1:08,PasiniUhrig-NoGo2:08}. Our analysis is consistent with these theorems and further conveys the need to utilize pulse shaping techniques in conjunction with optimal sequence construction to 
achieve high order error suppression in the presence of finite-width pulses. Indeed, when liberated from the constraint of rectangular pulse profiles, pulse sequences using DCG and CDCG \cite{DCG,DCG1,CDCG} may be employed when pulse-width errors are the dominant concern.

\begin{figure*}[t]
\centering
\subfigure[\ $K=4$]{\includegraphics[scale=0.024]{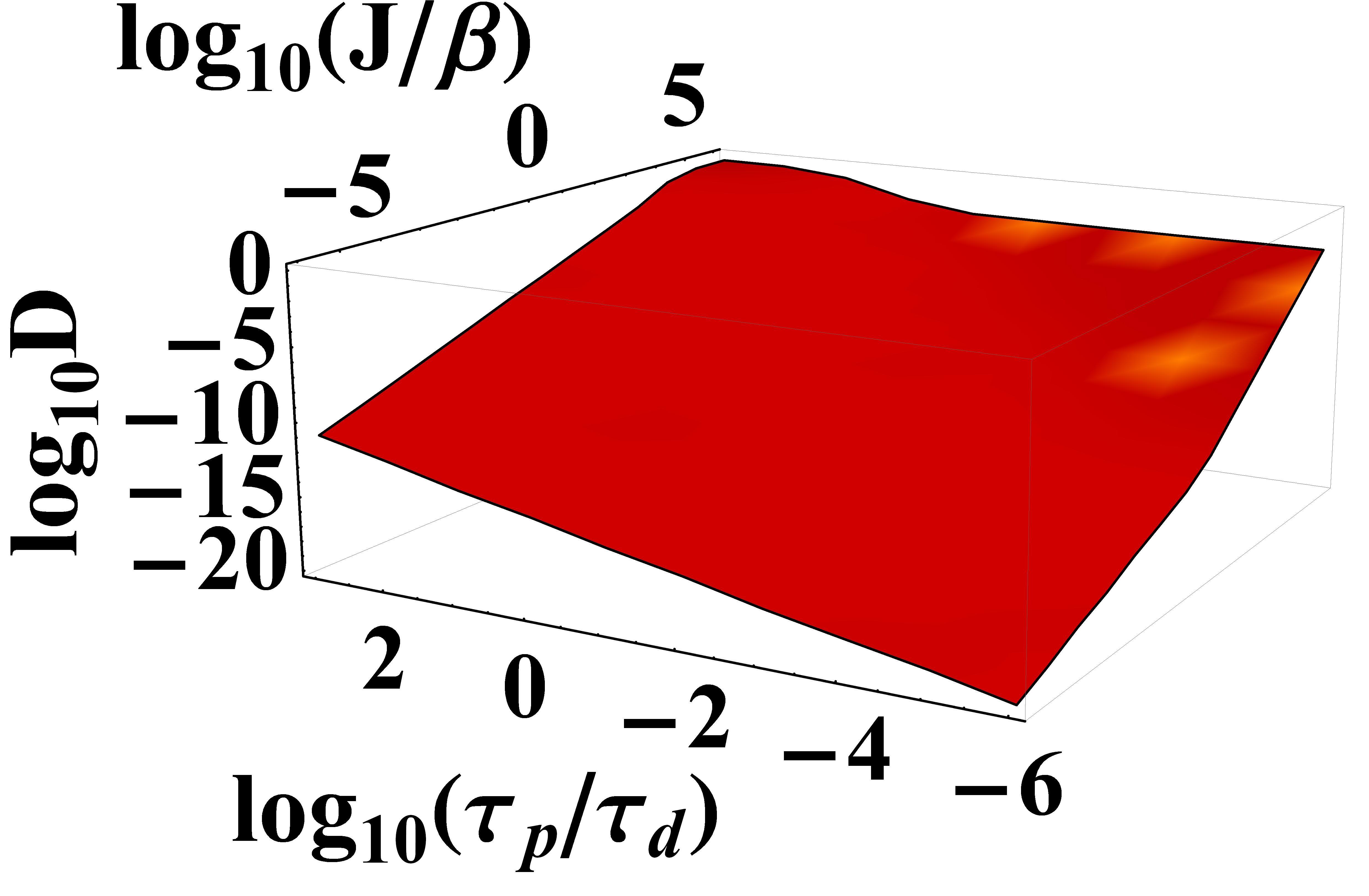}}\hspace{0.1cm}
\subfigure[\ $K=8$]{\includegraphics[scale=0.024]{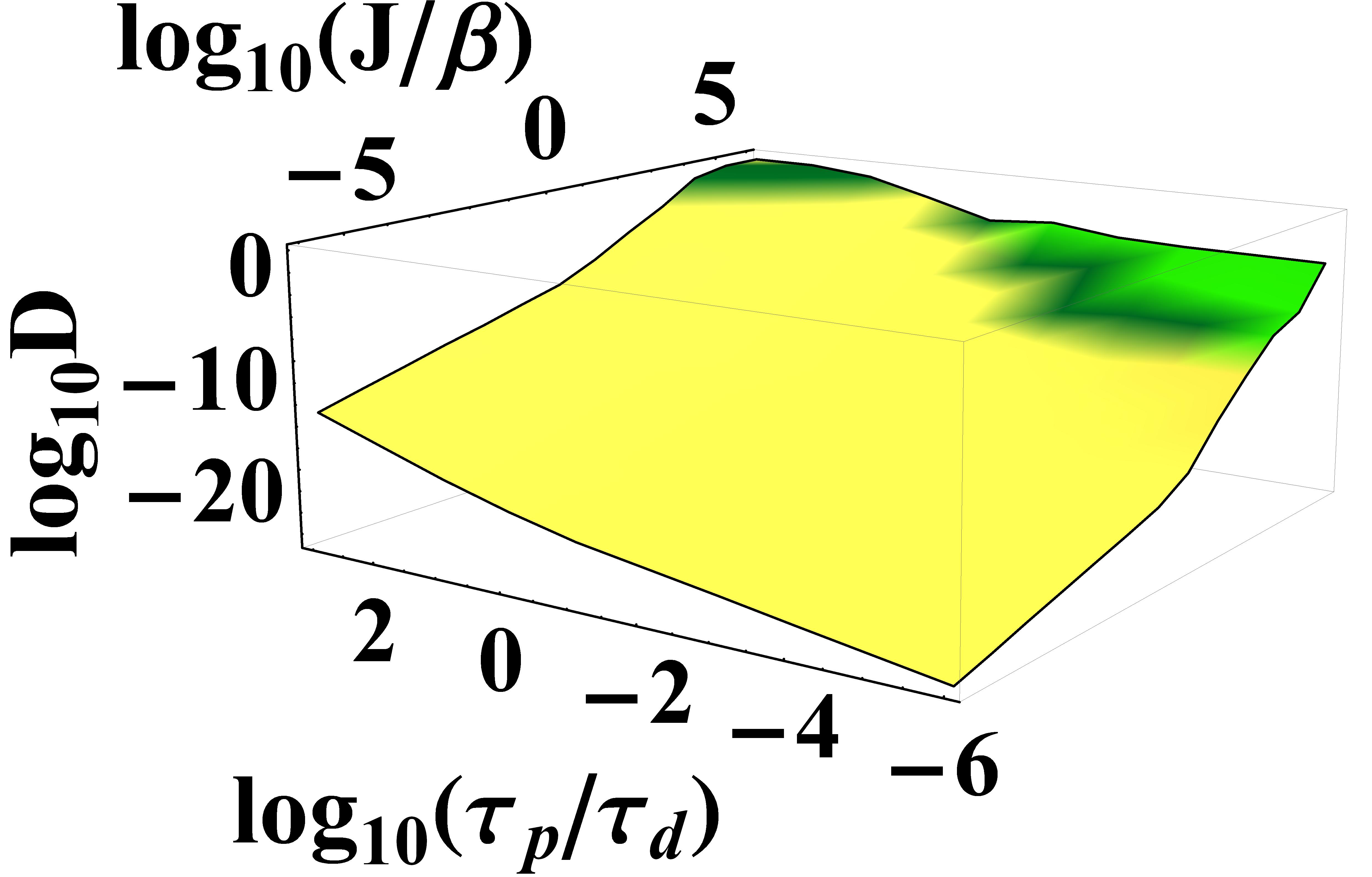}}\hspace{0.1cm}
\subfigure[\ $K=16$]{\includegraphics[scale=0.024]{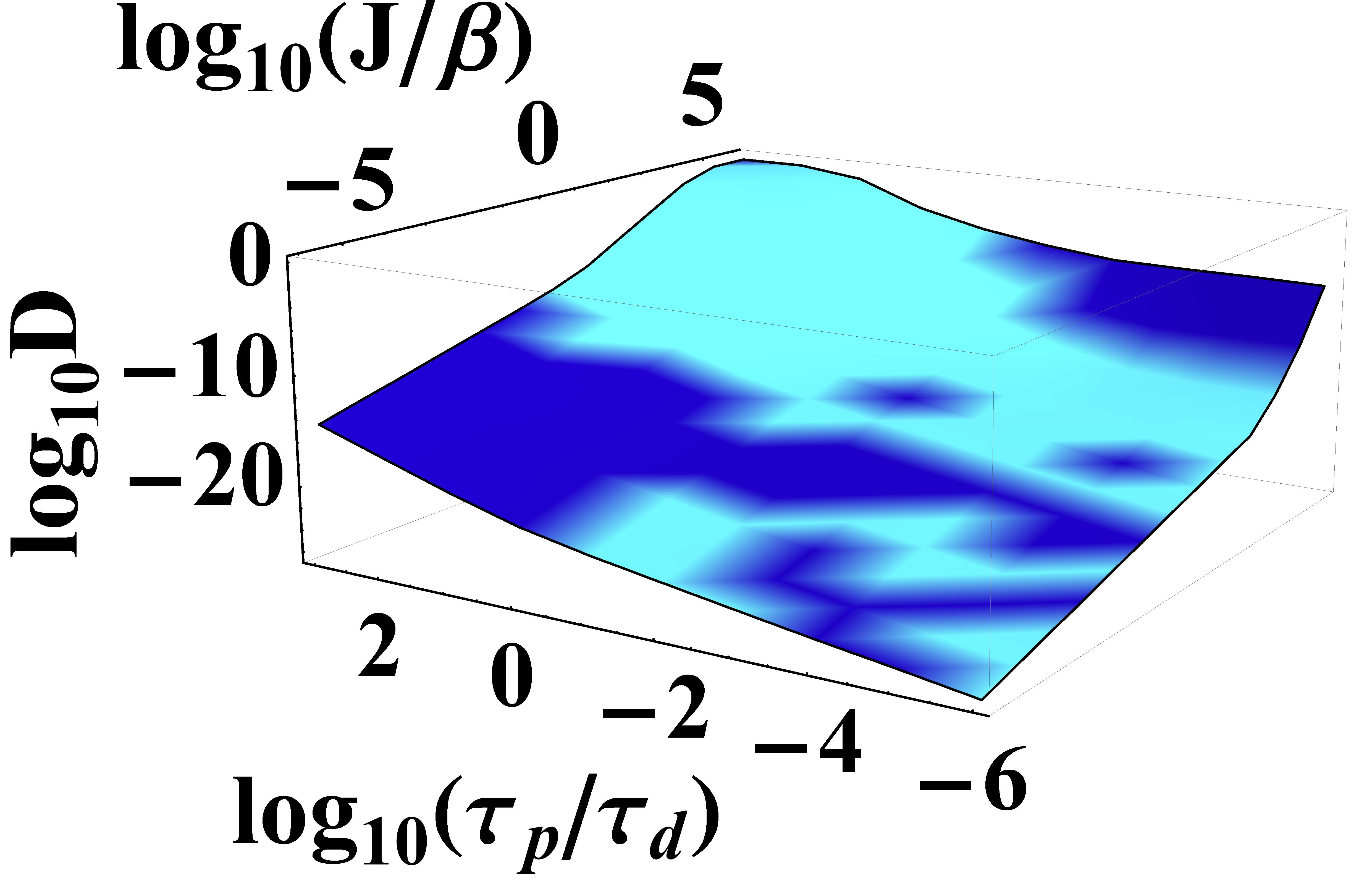}}\\
\subfigure[\ $K=32$]{\includegraphics[scale=0.024]{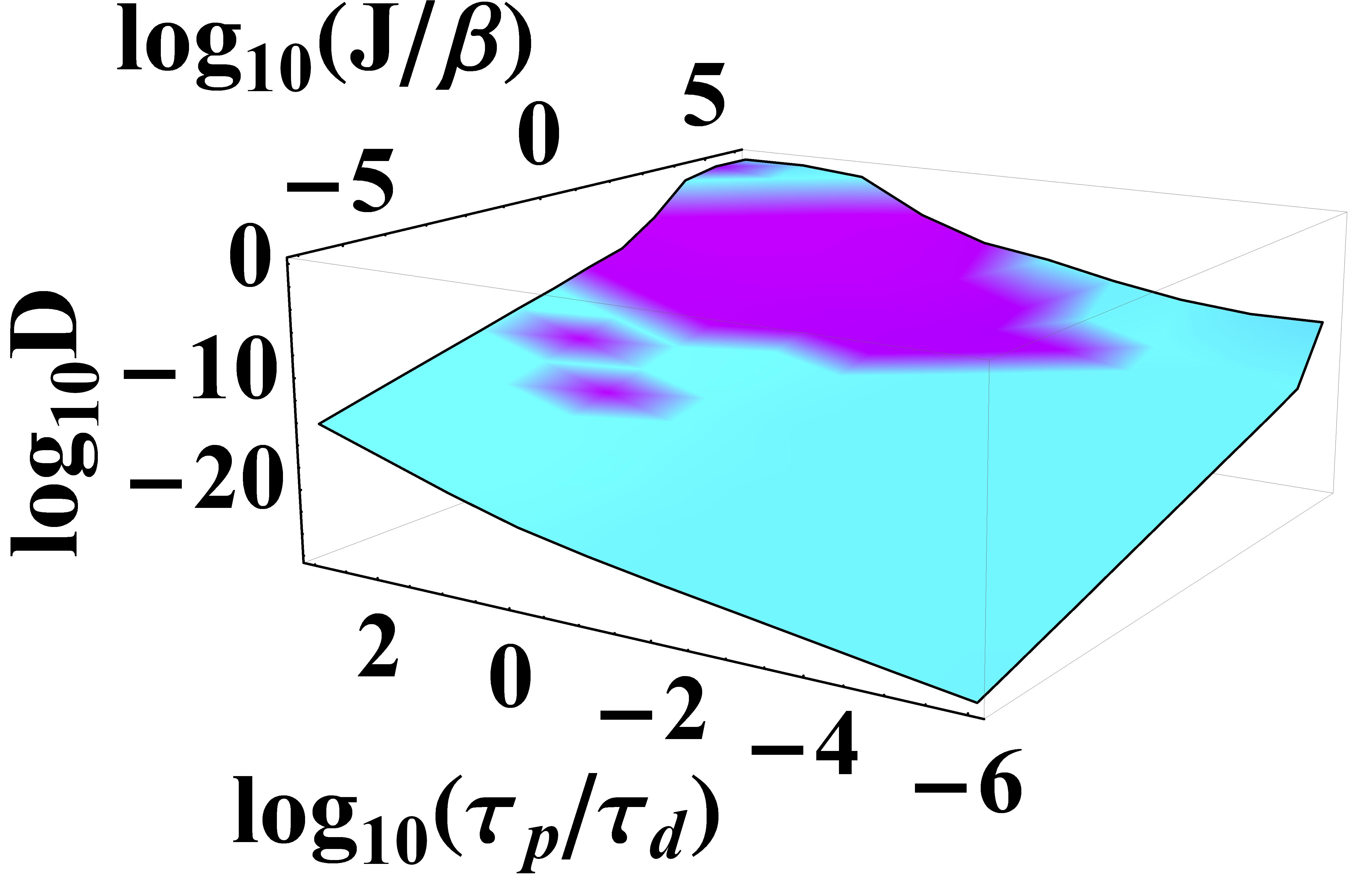}}\hspace{0.1cm}
\subfigure[\ $K=64$]{\includegraphics[scale=0.024]{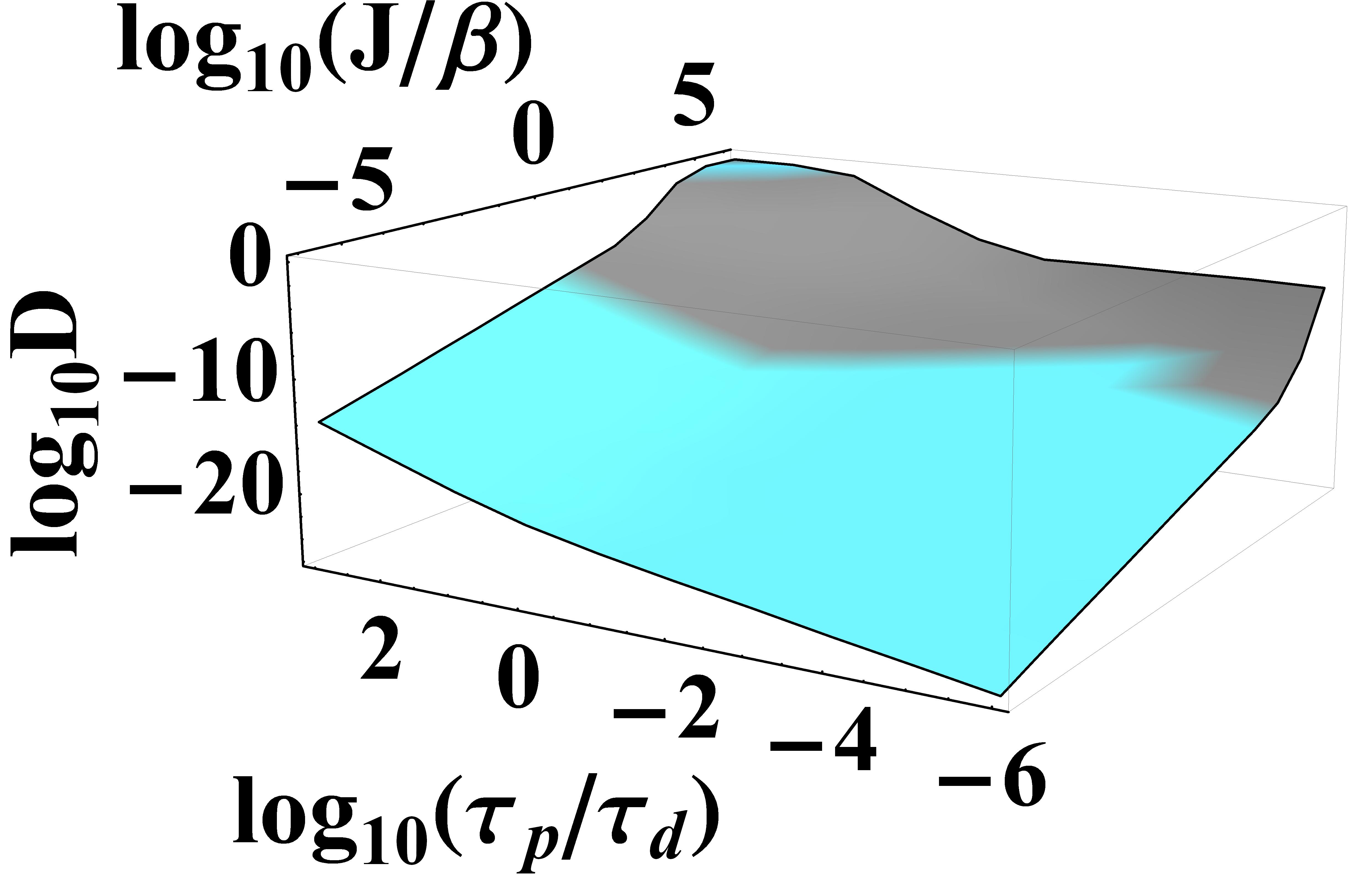}}\hspace{0.1cm}
\subfigure[\ $K=256$]{\includegraphics[scale=0.024]{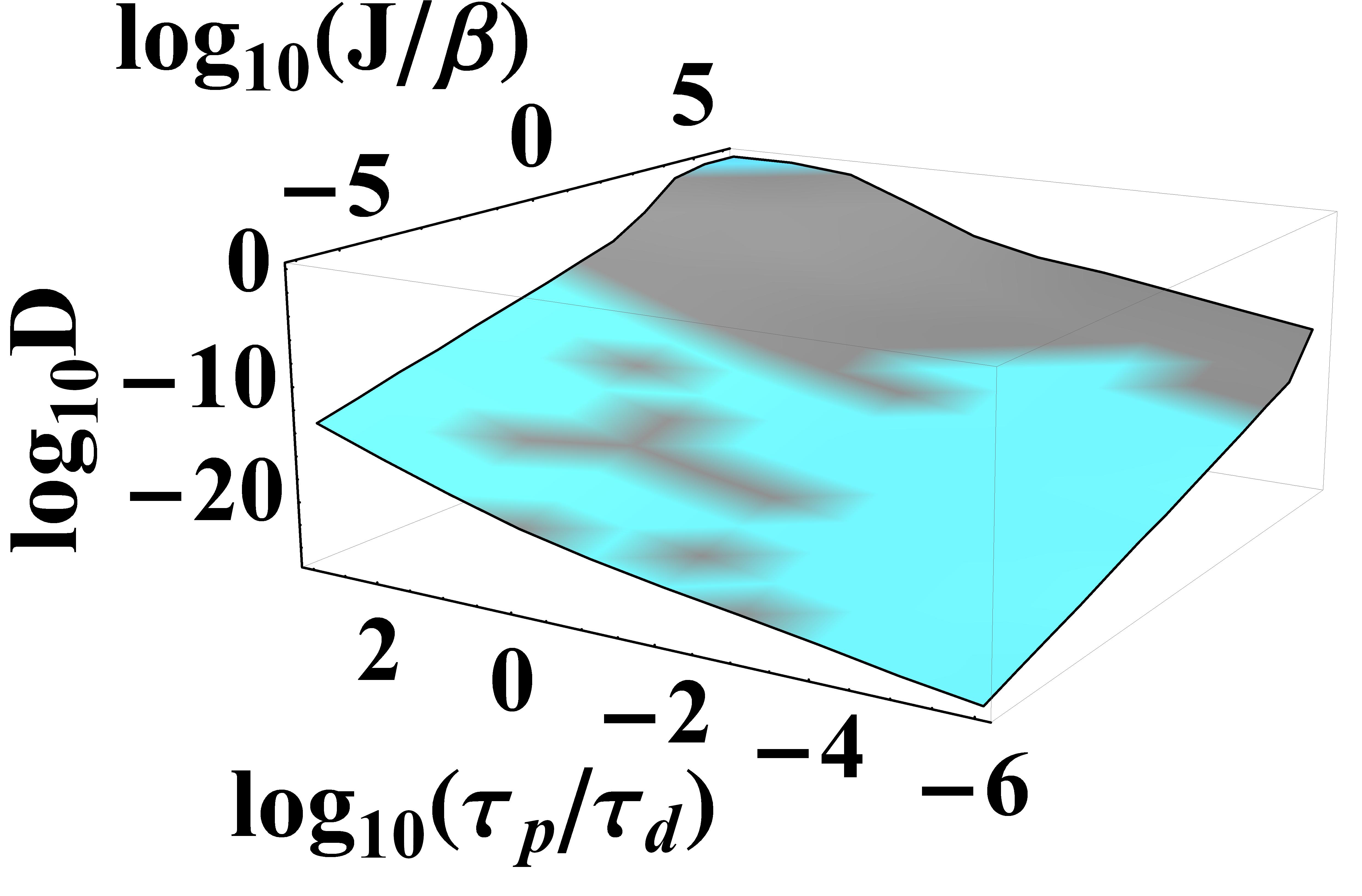}}\\
\subfigure{\includegraphics[scale=0.028]{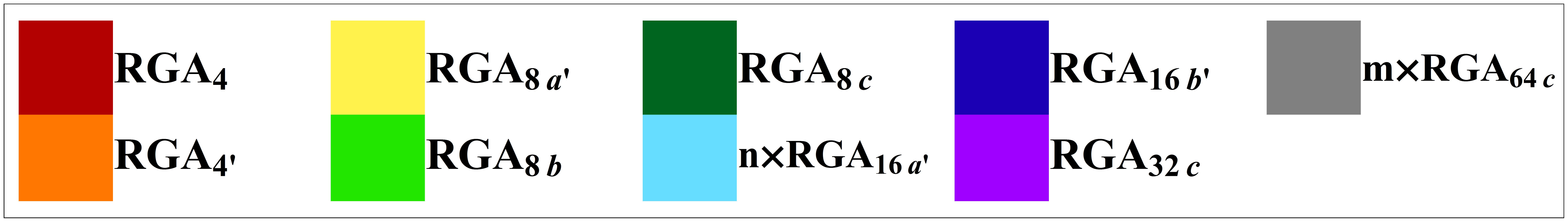}}
\caption{Performance of optimal $RGA_K$ sequences for $K=4,8,16,32,64,256$ shown in (a)-(f), respectively, as a function of $J/\beta$ and $\tau_p/\tau_d$ when DD is subjected to finite pulse duration. The 
norm of the bath Hamiltonian is fixed at $\beta=1$kHz, while $J/\beta\in[10^{-6},10^6]$. The pulse-interval $\tau_d=0.1$ns and the pulse width is varied 
in the range $\tau_p/\tau_d\in[10^3,10^{-6}]$. For a given $K$, the optimal sequence configuration is 
most sensitively dependent upon variations in $J$.
Contrary to the ideal pulse case, concatenated structures composed of {$RGA_{8a^\prime}$ and $RGA_{8c}$ appear to be the most favorable, in particular for $K\geq16$ where $RGA_{16a^\prime}$ and $RGA_{64c}$ repeatedly emerge as optimal sequences.} Sequence performance saturates at $K=16$, while robustness begins to diminish at $K=256$. $n\times RGA_{16b'}$ denotes $n$ cycles of $RGA_{16b'}$; $n=1,4,16$ for $K=16,64,256$ respectively. {The notation is similar for $m\times RGA_{64c}$, where $m=1,4$ cycles are used for $K=64,256$.}}
\label{fig:GALandscapePlotsFW}
\end{figure*}

\begin{figure*}[t]
\centering
\subfigure[]{\includegraphics[width=0.9\columnwidth]{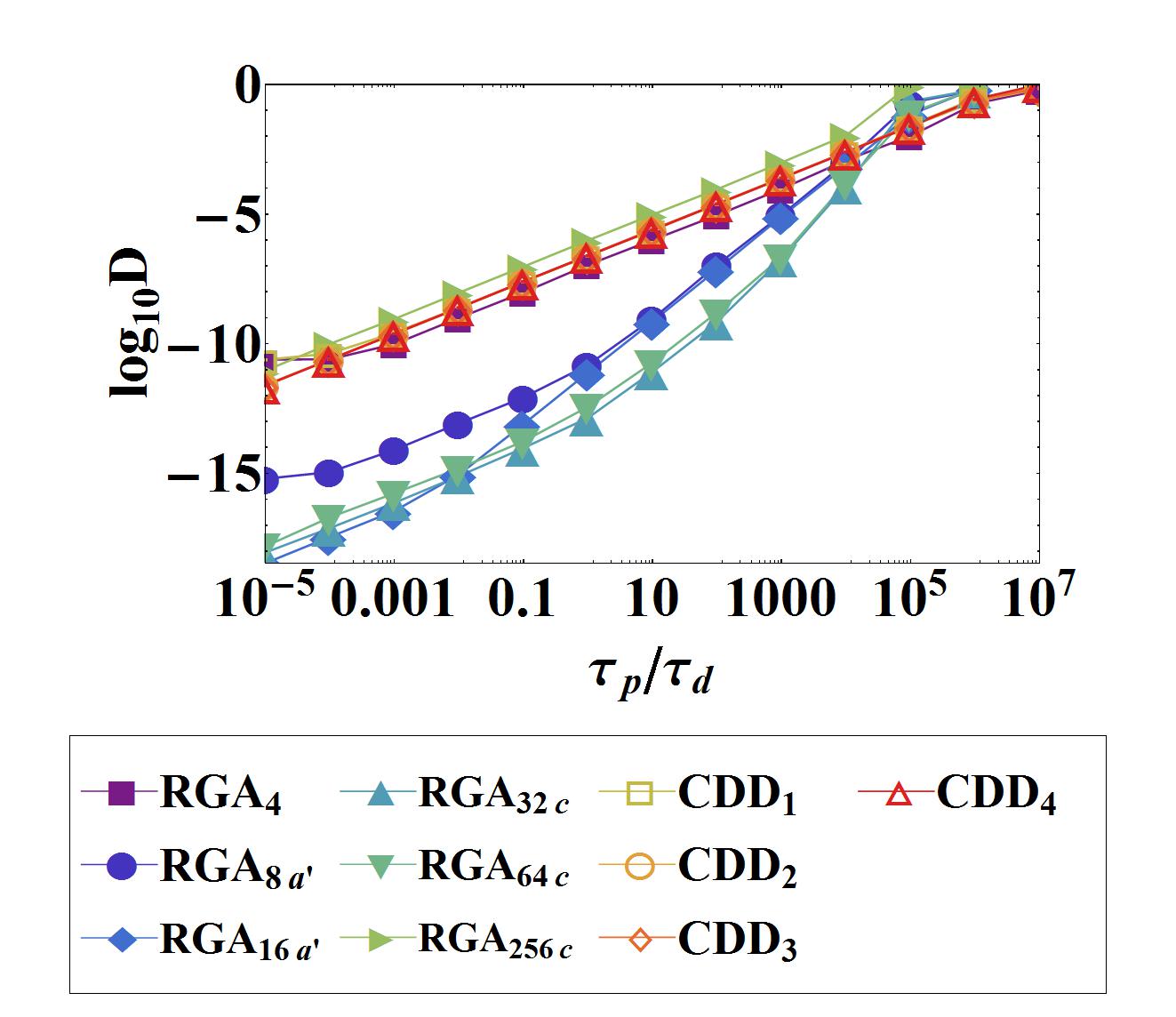}}
\subfigure[]{\includegraphics[width=0.9\columnwidth]{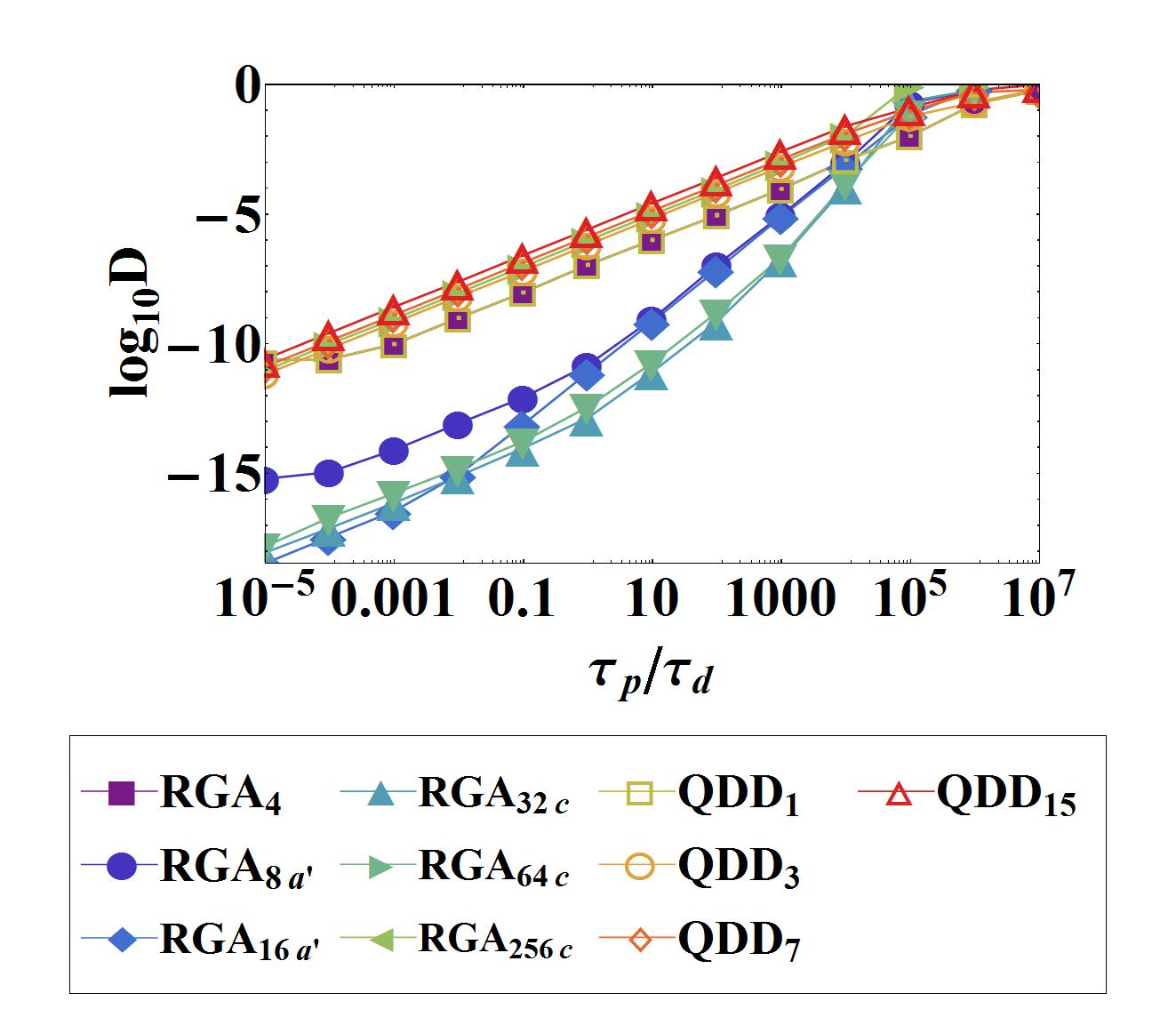}}
\caption{Comparison of performance between $RGA_K$ and (a) CDD$_l$, or (b) QDD$_M$, when subjected to finite pulse duration. Performance is characterized as a function of $\tau_p$, while $\tau_d=0.1$ns. CDD$_l$ performance is essentially the same for all $r$, scaling as $D\sim\mathcal{O}(J\tau_p)$. $RGA_K$ achieves a significant increase in robustness over CDD$_r$ at $K=8a^\prime,16a^\prime,32c,64c$, where the performance surpasses the linear scaling in $\tau_p$. Note the eventual saturation in decoupling order characteristic of the rectangular pulse profile displayed by the nearly equivalent scaling of $K=32c,64c$. QDD$_M$, $M=1,3,7,15$, performance becomes increasingly worse as the sequence order increases due to an accumulation in errors brought about by the finite duration of the pulses. As in the case of CDD$_r$, QDD$_M$ performance maintains $D\sim\mathcal{O}(J\tau_p)$ for all $M$. All results are averaged over 10 realizations of $B_{\mu}$. Error bars are included, but are quite small.}
\label{fig:GAvsCDDFW}
\end{figure*}

\subsubsection{Characterization of $RGA_K$ Sequences in $(\tau_d,\tau_p)$-space}
In the previous section, finite-width pulse error-optimized $RGA_K$ sequences were identified for various values of $K$. In Figure \ref{fig:GALandscapePlotsFW}, we summarize these results using numerical simulations to characterize the regions of optimal performance for each sequence as a function of $J/\beta$ and $\tau_p/\tau_d$. All results are averaged over 10 random realizations of $H_0$ with fixed $\tau_d=0.1$ns and $\beta=1$kHz. The system-environment interaction strength is varied within the range $J/\beta\in[10^{-6},10^{6}]$ and $\tau_p/\tau_d\in[10^3,10^{-6}]$.

Variations in the regions of optimal performance are primarily dependent upon the value of $J$ for a given $K$. Sequences which obtain a favorable performance scaling for a given $J$ tend to maintain their dominance throughout a wide range of $\tau_p$ values extending from the strong pulse to  pulse-width error dominant regimes. As a function of $K$, optimal performance eventually saturates at $K=16$ where $RGA_{16a'}$ maintains regions of optimal performance within $K=32,64,256$ for $J<\beta$. Beyond $K=64$, $RGA_{16a'}$ is accompanied by $RGA_{64c}$, which maintains its region of optimal performance within $J>\beta$ for $K=64,256$. Saturation in optimal sequence configuration and performance clearly agrees with the results of DD no-go theorems related to the achievable order of error suppression for finite-width pulse errors generated by rectangular pulse profiles.

\subsubsection{Comparison with Deterministic Sequences}
In Fig.~\ref{fig:GAvsCDDFW}(a) and (b), CDD$_r$ and QDD$_M$, respectively, are compared to the $RGA_K$ sequences optimized for finite pulse width: $K=4,8a',16a',32c,64c,256c$. The pulse-interval is chosen as $\tau_d=0.1$ns, while $\tau_p/\tau_d\in[10^{-5},10^{7}]$, $J=1$MHz, and $\beta=1$kHz. All results are averaged over $10$ random instances of the Hamiltonian.

Optimal performance for $RGA_K$ is observed predominately for {$K=32c,64c$ with $K=16a^\prime$ exhibiting a more favorable performance only in the strong-pulse regime, namely when $\tau_p/\tau_d<10^{-2}$ in Fig.~\ref{fig:GAvsCDDFW}. As the finite-width pulse errors contribute more substantially,  $\O(J^2\tau^{2}_p)$ terms remaining in the effective error Hamiltonian for $K=16c$ results in a rapid decrease in performance leading to $RGA_{32c}/RGA_{64c}$-dominance.} $K=4$ maintains the lowest performance of all $RGA_K$ sequences due to its inability to suppress the first order contribution in $\tau_p$, $D\sim\mathcal{O}(J\tau_p)$. CDD$_r$ performance is nearly equivalent to $K=4$ for $\tau_p/\tau_d\geq10^{-4}$, scaling as $D\sim\mathcal{O}(J\tau_p)$ for all $r$. The most noticable difference occurs at $\tau_p/\tau_d<10^{-4}$, where CDD$_r$ maintains the linear scaling in $\tau_p$ for $r=2,3,4$ and surpasses $K=4,8c$. In Ref.~\cite{KhodjastehLidar-CDD:07}, an analysis of CDD$_r$ in the presence of finite pulse width is discussed as well. There it was shown that CDD$_r$ can reduce pulse-width errors as the concatenation level increases if $\tau_p\ll\tau_d$. Although the total cycle time was fixed, as opposed to the pulse-interval, the results obtained here are quite similar in the $\tau_p/\tau_d<10^{-4}$ regime and confirm the inherent robustness of CDD$_l$ to finite-width pulse errors.


As discussed in Refs.~\cite{PasiniUhrig-UDDFW:12}, UDD-based schemes are quite susceptible to finite pulse-width errors and must be implemented with specially tailored pulses to regain a portion of the UDD decoupling efficiency. We confirm this result here for the most simplistic pulse shape: the rectangular pulse. Increasing the sequence order does not result in an increase, or sustainability, of performance; rather an accumulation of error results. As in the case of CDD$_r$, the performance maintains a linear scaling in $\tau_p$ throughout the specified range for the higher of the three sequence orders: $M=3,7,15$. The only variation occurs at $M=1$ where the performance becomes dependent upon $\tau_d$ for $\tau_p\ll\tau_d$. Although there exists a regime where both deterministic schemes outperform $K=4,8c$, higher order $RGA_K$ sequences provide a level of robustness that cannot be matched by either CDD$_r$ or QDD$_M$.

\subsection{Flip-angle errors}
\label{subsec:flipErrors}

An additional form of pulse error we consider is that of a flip-angle error. 
The control pulse set $\mathcal{G}=\{X,Y,Z,\Xb,\Yb,\Zb\}$ , with the pulse profile defined in Eq.~(\ref{eq:FAEpulse}). The resulting unitary pulse operators are given by
\begin{equation}
X(Y,Z)=e^{-i\pi/2\,(1+\eps)\sigma^{x(y,z)}}
\end{equation}
and $\{\Xb,\Yb,\Zb\}=\{X^{\dagger},Y^{\dagger},Z^{\dagger}\}$. The analysis is symmetric with respect to over or under-rotations, therefore, our focus on over-rotations does not result in a loss of generality.

\subsubsection{Summary of Numerical Search}
\label{subsubsec:flipErrorResults}
In contrast to the finite-width and finite-width flip-angle error analyses, where eventual saturation in performance was observed, flip-angle error-optimized sequences exhibit an increase in overall decoupling order for a majority of the $K_{opt}$ values. Therefore, manipulation of sequence configuration is sufficient for acquiring robustness to this particular type of pulse imperfection. Advanced pulse shaping techniques may aid in the suppression of additional errors; however, as we will display below, sequence manipulation alone produces a surprisingly high decoupling efficiency.

Robustness against flip-angle errors is completely characterized by the $a$-type $RGA_K$ sequences. For $K=4$, $2\times RGA_2$ is predominately the optimal choice, although $RGA_4$ does appear optimal when $J\gg\beta$. The dominance of $RGA_2$ follows from its $\O(\eps^2)$ decoupling, shown by
\begin{equation}
\H^{RGA_2}_{\text{err}}\approx \sx B_x -\frac{\pi\eps}{2}(\sy B_z -\sz B_y)+\frac{\pi^2\eps^2}{4}(\sy B_y+ \sz B_z).
\label{eq:Heff-FAerrorGA2}
\end{equation}
$RGA_4$ does not achieve such a decoupling,
\begin{eqnarray}
\H^{RGA_4}_{\text{err}}&\approx&-\frac{\pi^2\eps^2}{8\tau_d}\sz-\frac{\pi\eps}{2}\sz B_{x-y}\nonumber\\
&&+\frac{\pi^2\eps^2}{4}[\sx B_{x-y}+\sy(B_y-2B_x)],
\label{eq:Heff-FA4}
\end{eqnarray}
and, therefore, is not the preferred optimal sequence for $K=4$ until the effects of flip-angle errors are negligible compared to the errors generated by free evolution ($J\tau_d\gg\eps$).

We identify $RGA_{8a}$ as the sole optimal sequence for $K=8$. The structure is similar to $GA_{8a}$ and $RGA_{8a^\prime}$, differing only by pulse phases, and identical to a time-symmetrized version of $RGA_4$ sequence, namely $\overline{RGA_4}RGA_4$. Time-symmetrization has 
long been known to be 
beneficial for DD sequence construction since all odd-order terms in the effective error Hamiltonian are averaged out \cite{NLP:11,Wang:72}, even in the case of pulse errors \cite{SouzaAlvarez:12}. The effect of symmetrization is apparent within
\begin{eqnarray}
\H^{RGA_{8a}}&\approx&-\frac{\pi\eps}{2}\sz B_{x-y}+\frac{\pi^2\eps^2}{4}\sy B_{x-y}\nonumber\\
&&-\frac{\pi^2\eps^2}{4}\sy(2B_x-B_y),
\label{eq:Heff-PEGA8a}
\end{eqnarray}
where the dominant error term scales as $\O(\eps J\tau_d)$. Comparing $RGA_{8a}$ to $RGA_2$, the primary difference is the second order suppression of  $\O(J\tau_d)$ terms acquired by $RGA_{8a}$. While $RGA_2$ achieves a similar robustness against flip-angle errors, it fails to address errors created by free evolution.

In the case of $K=16,32$, all optimal sequences can be characterized by $RGA_{16a}$. As one may notice from the definition of the sequence given in Table~\ref{tbl:scaling-FWPE}, there is some freedom in the choice of the $P_3$ pulse. Considering the usual case of $\{P_1,P_2\}=\{X,Y\}$ discussed so far, the following effective error Hamiltonians emerge for each $P_3\neq I$:
\begin{eqnarray}
\H^{RGA_{16a}}_{\text{err},P_3=X}&\approx&\frac{\pi^2\eps^2}{4}[\sx B_{x-y}-\sy B_{x-y}],\\
\H^{RGA_{16a}}_{\text{err},P_3=Y}&\approx&\frac{\pi^2\eps^2}{4}[\sx B_{x-y}-\sy(2B_x-B_y)],\\
\H^{RGA_{16a}}_{\text{err},P_3=Z}&\approx&\frac{\pi\eps}{2}\sz B_{x-y}.
\end{eqnarray}
Interestingly, the decoupling order for terms proportional to $\eps$ is $P_3$-dependent. Choosing $P_3$ to be orthogonal to $P_1,P_2$ is clearly the least favorable choice, with $P_3=P_1$ being the optimal choice, most notably in the case of uniform decoherence in the xy-plane. From the effective error Hamiltonians above, we find that optimal $RGA_{16a}$ performance is determined by $\O(\eps^2J\tau_d)$ terms. Additional sequence structures such as $RGA_{32a}$ do not achieve similar performance and, in fact, suffer from the presence of $\O(\eps^2)$ terms. 

\begin{table}[t]
\centering
\begin{tabular*}{\linewidth}{@{\extracolsep{\fill}}cccc}
\multicolumn{2}{c}{Sequence} & \multirow{2}{*}{$\eps\ll J\tau_d$} & \multirow{2}{*}{$\eps\gg J\tau_d$}\\ \cline{1-2}

Name & Description &  & \\ \hline

$RGA_2$		&  $\Pb P$				&  $\O(J\tau_d)$	&  $\bb{\O(\eps J\tau_d)}$\\

$RGA_4$		&  $\Pb_2P_1\Pb_2P_1$	& $\O(J\beta\tau^{2}_d,J^2\tau^{2}_d)$	&  $\O(\eps^2)$\\

$RGA_{8a}$	&  $I\Pb_1 P_2 \Pb_1 I P_1 \Pb_2 P_1$  &  $\bb{\O(\eps J\tau_d)}$	&  $\bb{\O(\eps J\tau_d)}$\\

$RGA_{16a}$	&  $\Pb_3(RGA_{8a}) P_3(RGA_{8a})$	  &  $\bb{\O(\eps^2 J\tau_d)}$	&  $\bb{\O(\eps^2 J\tau_d)}$\\

$RGA_{32a}$	&  $RGA_4[RGA_{8a}]$	& $\O(\eps^2)$	&  $\O(\eps^2)$\\

$RGA_{64a}$	&  $RGA_{8a}[RGA_{8a}]$	&  $\bb{\O(\eps^3 J\tau_d)}$	&  $\bb{\O(\eps^3 J\tau_d)}$\\

$RGA_{256a}$	&  $RGA_4[RGA_{64a}]$	&  $\O(\eps^2)$	&  $\O(\eps^2)$

\end{tabular*}
\caption{Summary of distance measure $D$ scalings for each optimal $RGA_K$ sequence located for DD pulses subjected to flip-angle errors with rotation error $\eps$, with fixed pulse-interval $\tau_d$. Boxed performance scalings highlight optimal performance scaling for various $K_{\text{opt}}$.}
\label{tbl:scaling-PE}
\end{table}

\begin{figure*}[t]
\centering
\subfigure[\ $K=4$]{\includegraphics[scale=0.024]{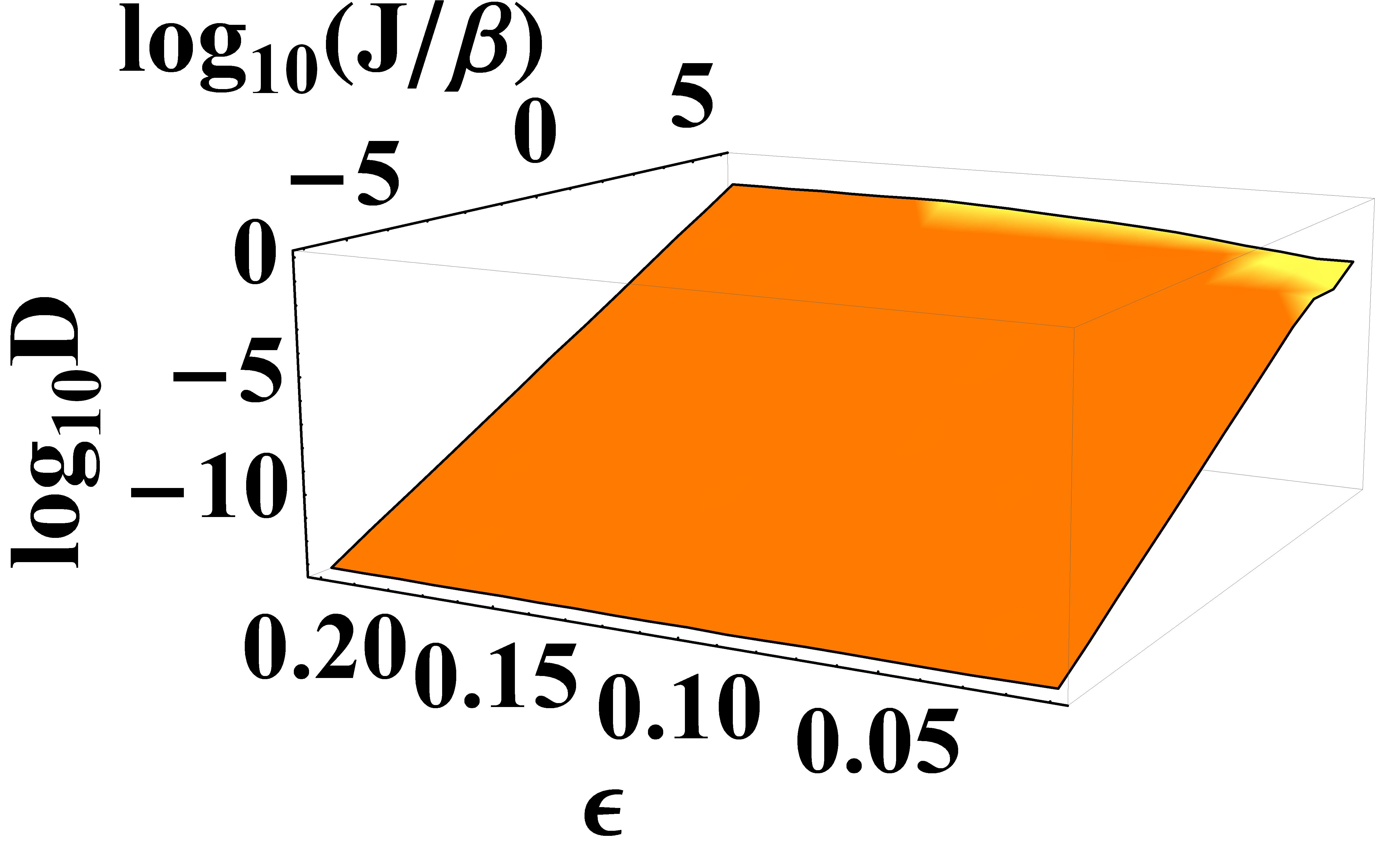}}\hspace{0.1cm}
\subfigure[\ $K=8$]{\includegraphics[scale=0.024]{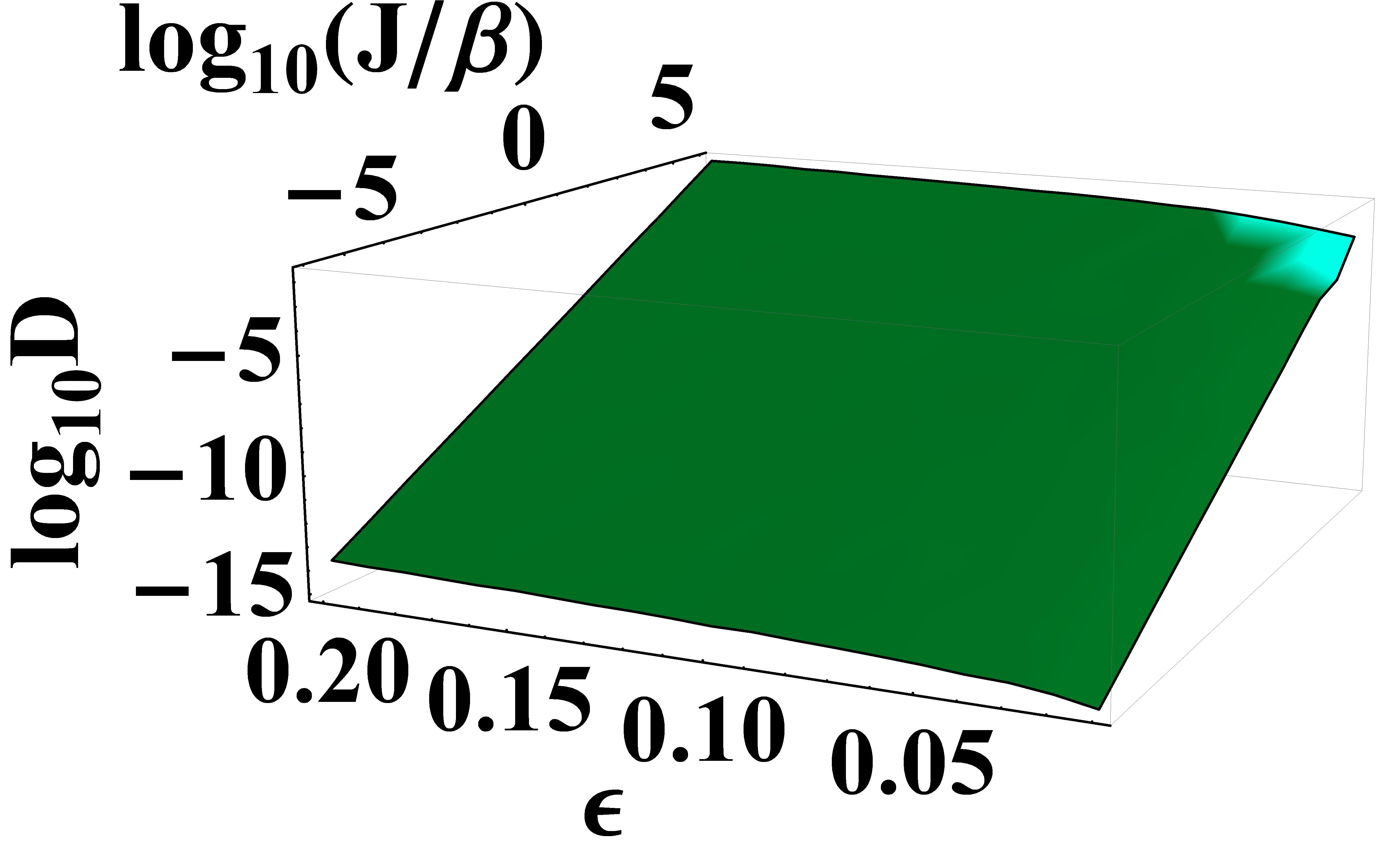}}\hspace{0.1cm}
\subfigure[\ $K=16$]{\includegraphics[scale=0.024]{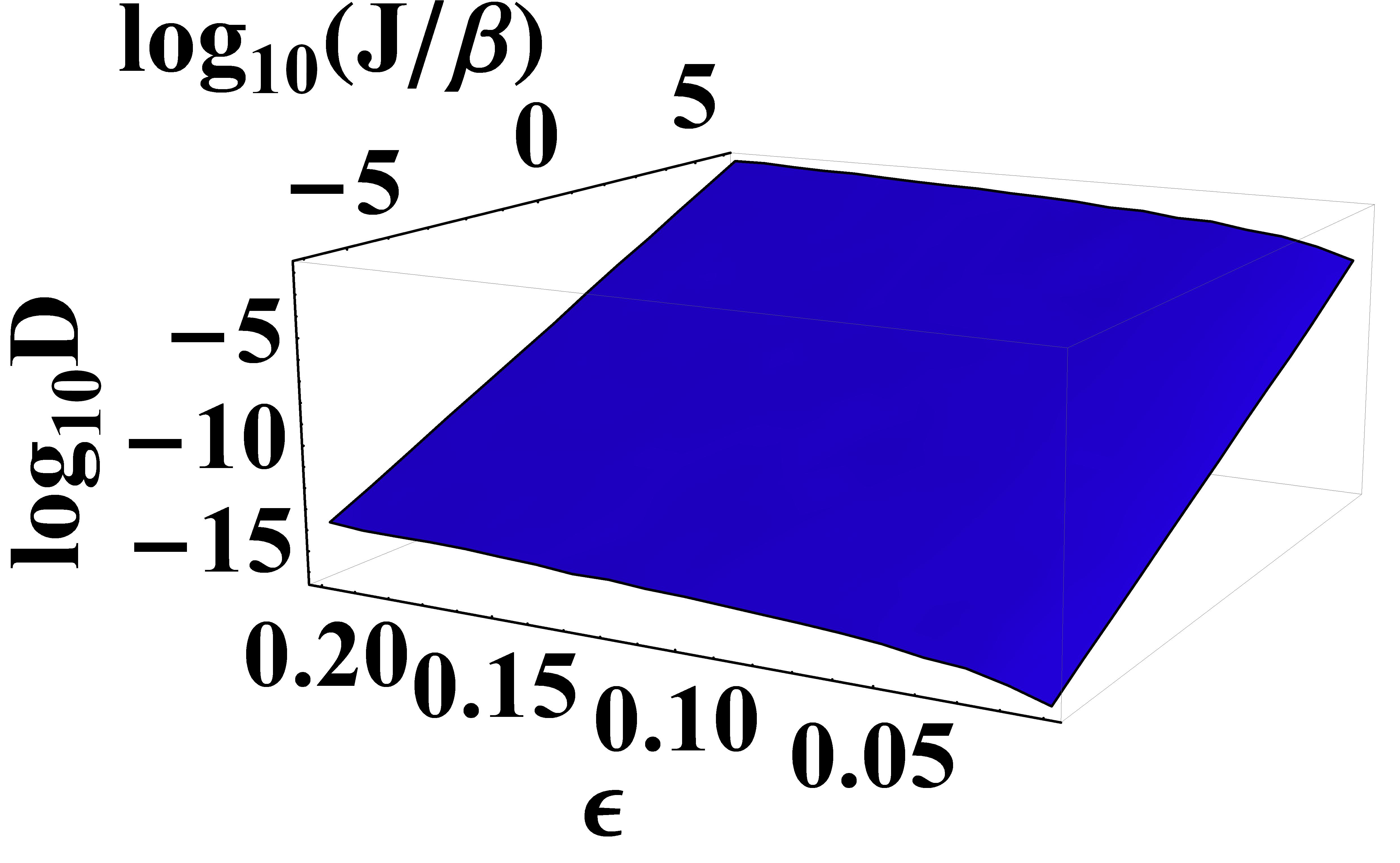}}\\
\subfigure[\ $K=32$]{\includegraphics[scale=0.024]{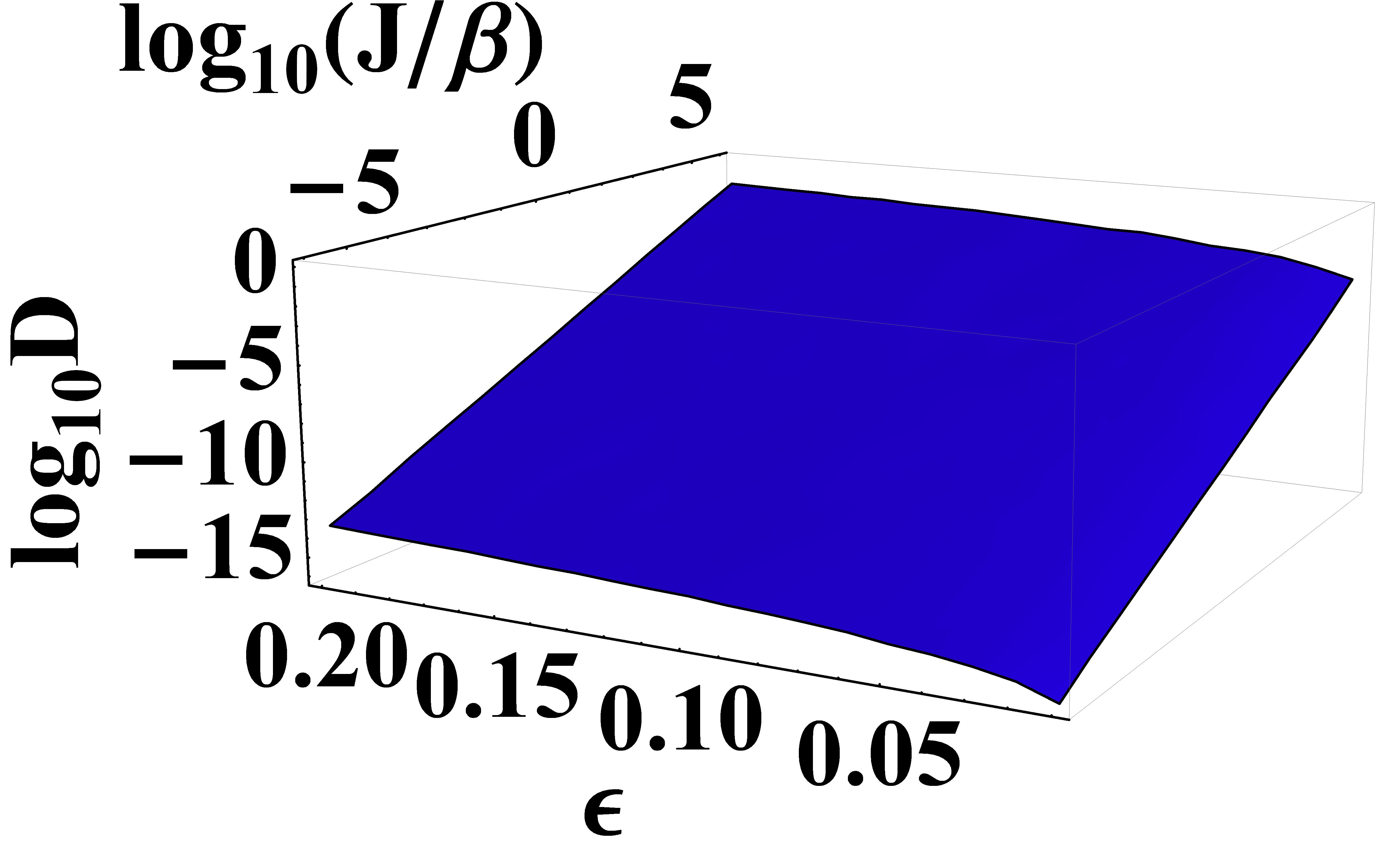}}\hspace{0.1cm}
\subfigure[\ $K=64$]{\includegraphics[scale=0.024]{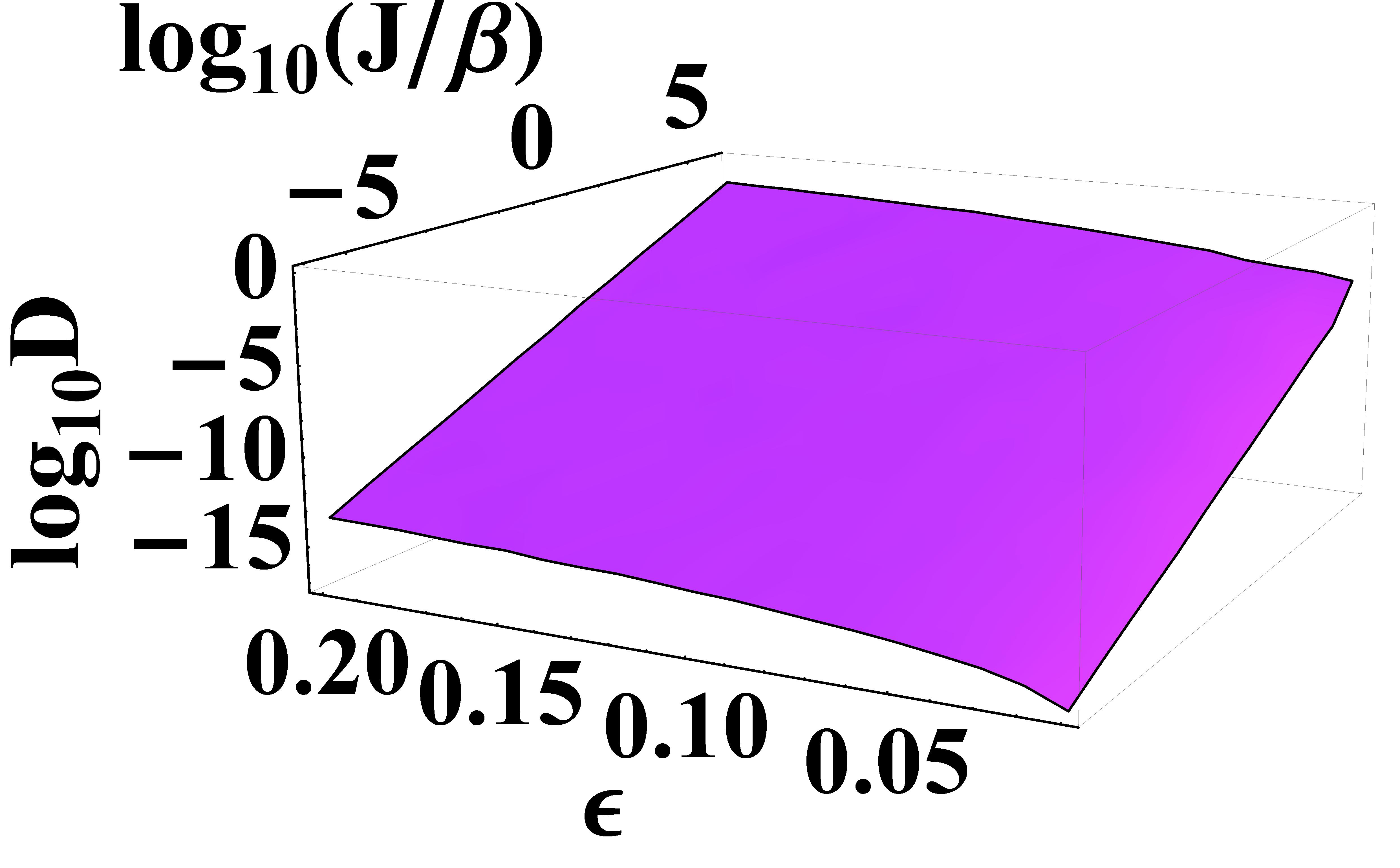}}\hspace{0.1cm}
\subfigure[\ $K=256$]{\includegraphics[scale=0.024]{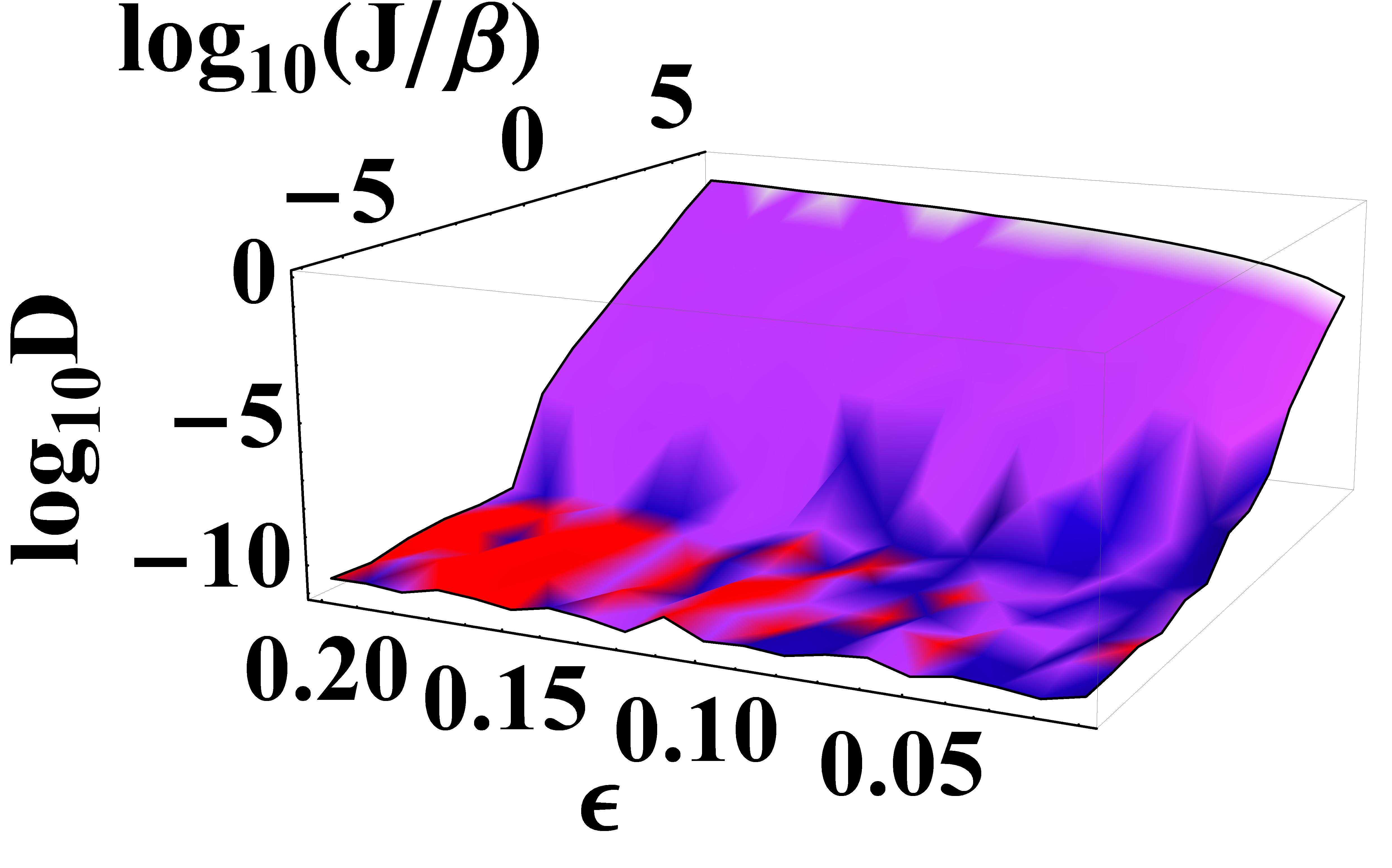}}\\
\subfigure{\includegraphics[scale=0.028]{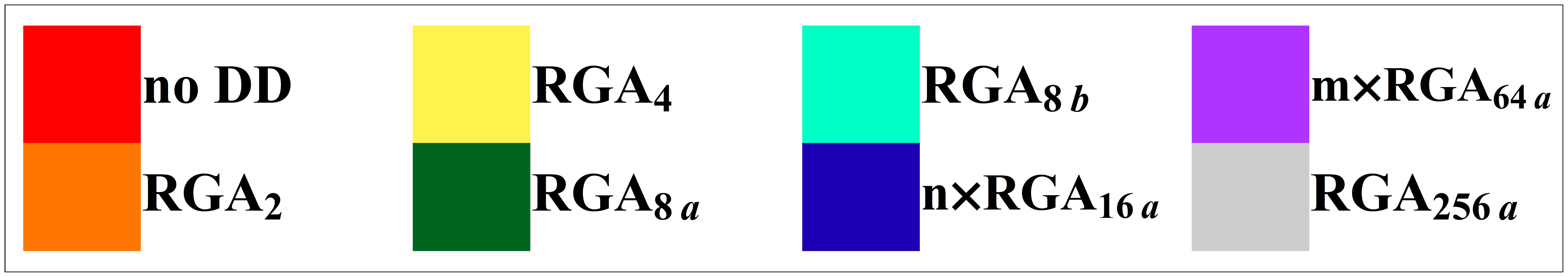}}
\caption{Performance of $RGA_K$ sequences for $K=4,8,16,32,64,256$, as shown in (a)-(f), respectively, as a function of $J/\beta$ and $\eps$. The minimum pulse interval is fixed at $\tau_d=0.1$ns, $J/\beta\in[10^{-6},10^{6}]$, and $\beta=1$kHz, while $\eps$ is varied from one to twenty percent rotation error. Results are averaged over 10 realizations of $B_{\mu}$. Sequence performance mostly increases from $K=4$ to $K=64$, indicating a reduction in the error terms proportional to $\eps$ in the effective error Hamiltonian. Successive error suppression is achieved for $K=4,8,16,64$, where the maximum error suppression yields $D\sim\mathcal{O}(\eps^3 J\tau_d)$. for $RGA_{64a}$. Multiple cycles of $RGA_{16a}$ and $RGA_{64a}$ appear as optimal sequences for various $K_{\text{opt}}$. The number of cycles for each sequence is given as follows: (d) $n=2$, (e) $m=1$, and (f) $n=16$, $m=4$.}
\label{fig:GALandscapePlotsPE}
\end{figure*}

The decoupling order again increases at $K=64$, where $RGA_{64a}$ attains suppression of all errors up to $\O(\eps^3 J\tau_d)$ terms, confirmed by
\begin{equation}
\H^{RGA_{64a}}\approx\frac{\pi^3\eps^3}{4}\sz B_y-\frac{3\pi^3\eps^3}{8}\sz B_x.
\end{equation}
Comparing $RGA_{8a}$ and its first level concatenation, $RGA_{64a}$, we find that an additional \emph{two} orders of error suppression are achieved simply by a single level concatenation. An obvious question that arises from this result is whether additional orders of decoupling, and possibly arbitrary order decoupling, is attainable by continuing the concatenation procedure. While the question of arbitrary order decoupling will not be addressed here, we are confident that such a scheme exists due to the $\O(\eps^5 J\tau_d)$ scaling acquired by $RGA_{512a}=RGA_{8a}[RGA_{8a}[RGA_{8a}]]$.

The final sequence length considered, $K=256$, is the first instance of complete breakdown in performance. Optimal sequences consist of cycles of $RGA_{16a}$ and $RGA_{64a}$ with various regions where free evolution reigns supreme. More sophisticated sequence structures, such as $RGA_{256a}$, are not found to be optimal due to remaining $\O(\eps^2)$ terms.  In Table~\ref{tbl:scaling-PE}, we display the performance scaling for $K=256$ along with the scalings for the remaining values of $K_{\text{opt}}$ discussed above.

In summary, the results presented for $K$$=$$K_{\text{opt}}$ in the presence of flip-angle errors demonstrate that successive error suppression is achievable by concatenation only if the outer sequence maintains the same decoupling order as the inner sequence(s). Supplying a lower order decoupling outer sequence ultimately leads to an effective error Hamiltonian that possesses dominant error terms that are intrinsic to the low-order sequence. Provided this condition is satisfied flip-angle error-optimized sequences exhibit a continual increase in decoupling order with an increasing number of pulses. Uninhibited, due to the absence of no-go theorems for flip-angle errors, we believe that extending 
the search beyond $K
=256$ will result in additional sequence configurations that utilize concatenations of $RGA_{8a}$, or even a more robust construction such as $RGA_{16a}$, to achieve higher order decoupling. This conclusion is supported by the increase in decoupling order found for $RGA_{512a}$, which suggests that arbitrary order error suppression using $\ell$ concatenations of $RGA_{8a}$ can be used to achieve $D\sim\O(\eps^{2\ell-1}J\tau_d)$.

\subsubsection{Characterization of $RGA_K$ Sequences in $(\eps,J\tau_d)$-space}
In this section, we illustrate the results obtained for the flip-angle error-optimized sequence search using numerical simulations to correctly identify the regions of optimal performance as a function of $J/\beta$ and $\eps$ [see Fig.~\ref{fig:GALandscapePlotsPE}]. The pulse delay and the strength of the bath dynamics are fixed at $\tau_d=0.1$ns and $\beta=1$kHz, respectively. The strength of the system-environment interaction is varied within the range $J/\beta\in[10^{-6},10^6]$ and the flip-angle error $\eps\in[0,0.2]$, corresponding to a $0\%$ to $20\%$ error in pulse rotation. All results are averaged over 10 random realizations of the bath operators $B_\mu$.

In contrast to the results obtained for finite-width pulse errors, optimal sequence configuration and performance generally increases as a function of $K$, up to $K=64$, indicating an increase in the suppression of error terms proportional to $\eps$. Saturation in performance is first observed at $K=32$, where two cycles of $RGA_{16a}$ is the optimal configuration $\forall J,\eps$ considered. This effect is actually quite brief, as an increase in error suppression returns at $K=64$ via $RGA_{64a}$. The most significant attenuation in performance is found for $K=256$. Optimal sequence configurations either consist of complete free evolution or cycles of previously located sequences. As discussed in the previous section, we expect additional increases in performance using more sophisticated sequencs beyond $K=256$, perhaps most obviously for concatenations of $RGA_{8a}$. We leave this analysis for future studies.

\begin{figure*}[t]
\centering
\subfigure[]{\includegraphics[width=0.41\textwidth]{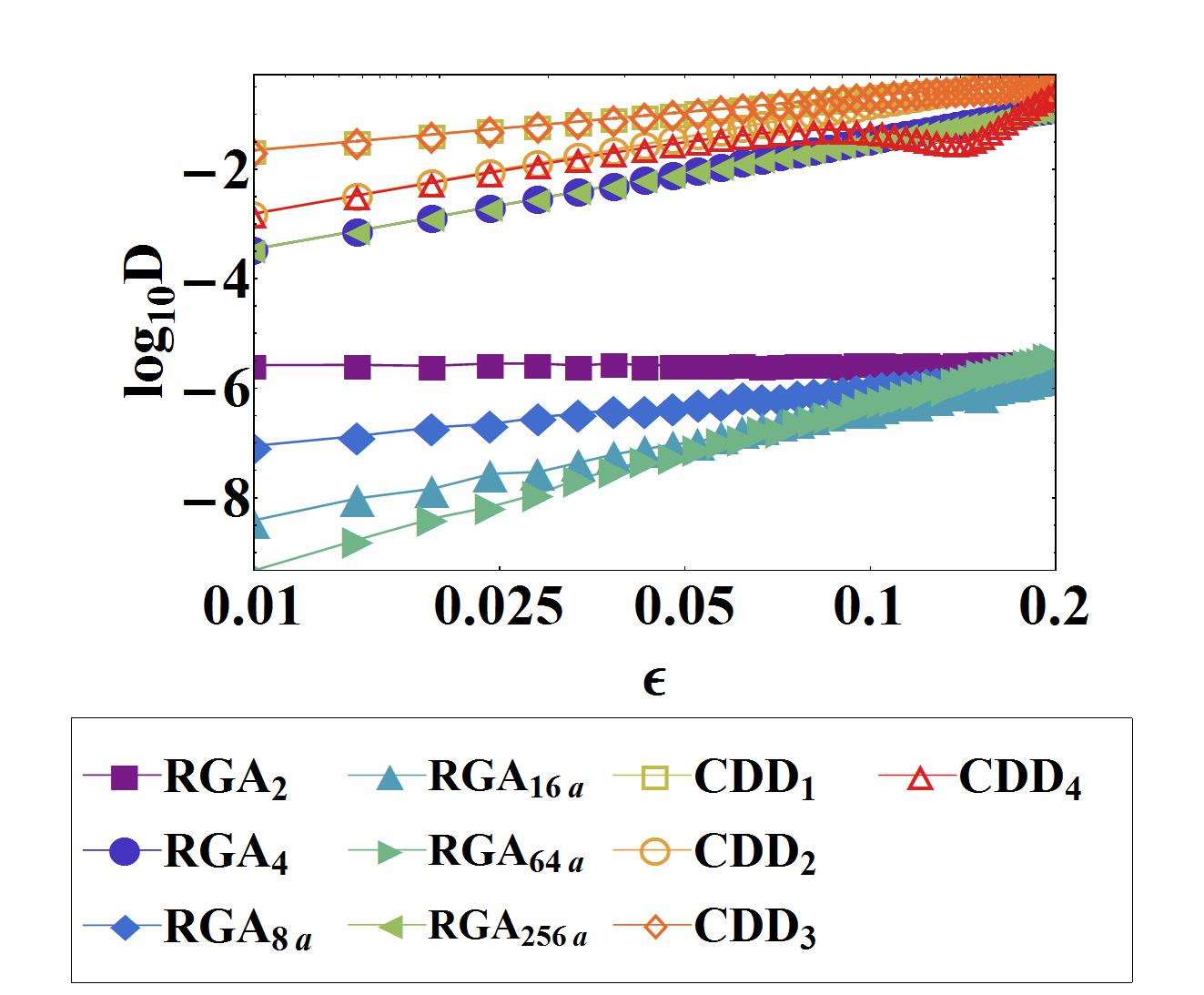}}
\subfigure[]{\includegraphics[width=0.41\textwidth]{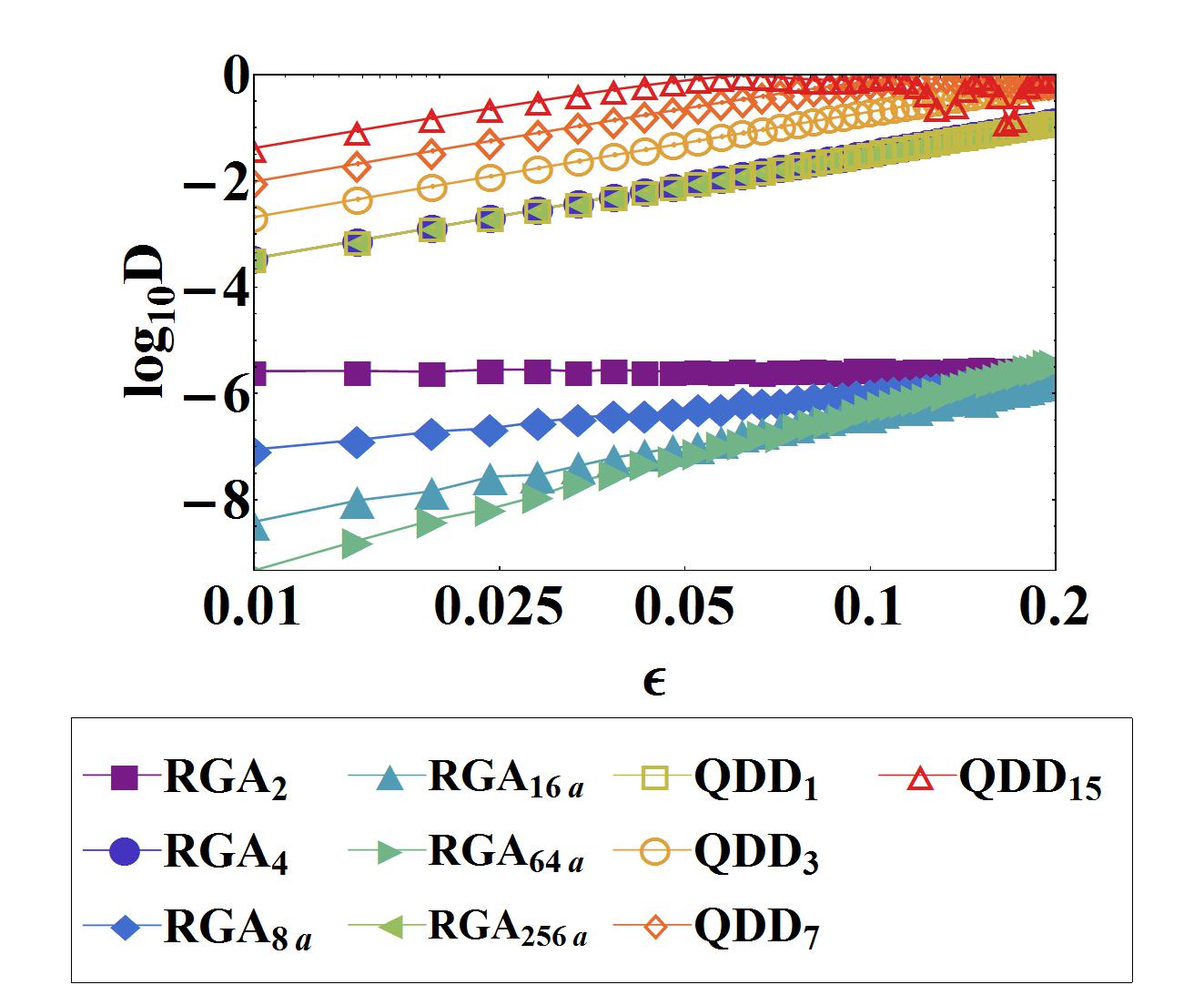}}
\caption{Performance of $RGA_K$ sequences versus (a) CDD$_r$ and (b) QDD$_M$ as a function of flip-angle error $\eps\in[0.01,0.2]$ averaged over $20$ realizations of $B_\mu$. The relevant parameters are chosen as $J=1$MHz, $\beta=1$KHz, and $\tau_d=0.1$ns. Numerically optimal sequences are found to be highly robust against flip-angle errors, significantly outperforming CDD$_r$ for $r=1,2,3,4$. For QDD$_M$, the sequence orders are chosen as $M=3,7,15$ and directly correspond to $K=16,64,256$. QDD is shown to be highly sensitive to flip-angle errors, decreasing in performance as $M$ grows. Again, robust GA sequences achieve optimal performance.}
\label{fig:GAvsCDDPE}
\end{figure*}

\subsubsection{Comparison with Deterministic Schemes}

In \ref{subsubsec:flipErrorResults}, robust sequences were identified for control pulses subjected to flip-angle errors. Here, we compare the numerically optimal $RGA_K$ sequences to CDD$_r$ and QDD$_M$ as a function of $\eps$. We consider the case of interaction-dominated dynamics and set the strengths of the environment dynamics and system-bath interaction to $\beta=1$kHz and $J=1$MHz, respectively. The pulse delay is chosen as $\tau_d=0.1$ns and all data is averaged over 20 realizations of the bath operators $\{B_\mu\}$. In addition to $K=2,8a,16a,64a$, which exhibit an increase in decoupling order for terms proportional to $\eps$, $K=4,32a,256a$ are included in the comparison as well to fully characterize $RGA_K$ performance with respect to both deterministic schemes.

We first focus on $RGA_K$ and CDD$_r$, where $RGA_K$ superiority is clearly evident for all values of $K$ shown in Fig.~\ref{fig:GAvsCDDPE}(a). Optimal performance is observed for $K=2,8a,16a,64a$, as expected, with $K=64c$ providing the highest level of robustness to flip-angle errors using the 
smallest number of pulses. Although the lowest level of performance for $RGA_K$ occurs at $K=4,32a,256a$, a considerable improvement over {the corresponding} CDD$_r$, $r=1,2,3,4$, is seen. Optimal CDD$_r$ performance {is achieved at} $r=2,4$, where the performance can be shown to scale as $D\sim\mathcal{O}(\eps^2)$. The remaining levels of concatenation, $r=1,3$ do not achieve first order suppression, therefore, $D\sim\mathcal{O}(\eps)$. The comparison clearly indicates that the numerically optimized sequences are highly robust to flip-angle errors and capable of dramatically outperforming CDD$_r$.

Analyzing QDD$_M$, for $M=1,3,7,15$, as a function of $\eps$, we find that the performance maintains a $D\sim\mathcal{O}(\eps^2)$ scaling for all $M$; see Fig.~\ref{fig:GAvsCDDPE}(b). 
The performance for QDD$_M$ diminishes with increasing $M$, indicating an accumulation of error rather than a reduction, or sustainability as in the case of CDD$_r$. Although variations in performance exist between the deterministic schemes, their robustness to flip-angle errors is clearly not comparable to the optimized $RGA_K$ sequences.


%

\end{document}